\documentclass[ALICE,manyauthors]{cernphprep}

\usepackage{rotating}
\usepackage[normalem]{ulem}
\usepackage{amssymb,amsmath}
\usepackage{graphicx}
\usepackage{subfigure}
\usepackage{amssymb}
\usepackage{epstopdf}
\usepackage{cite}
\usepackage{multirow}
\usepackage{color}
\usepackage{amsmath,bm}
\usepackage{amsbsy}
\usepackage{hyperref}
\usepackage{datetime}
\usdate
\usepackage{lineno}
\usepackage[toc,page]{appendix}
\usepackage[section] {placeins}

\newcommand{\CommentBlock}[1]{}

\newcommand{\PbPb}{Pb--Pb}

\newcommand{\pp}{pp}

\newcommand{\sqrtsNN}{\ensuremath{\sqrt{s_{\mathrm {NN}}}}}
\newcommand{\sqrts}{\ensuremath{\sqrt{s}}}
\newcommand{\rr}{\ensuremath{R}}

\newcommand{\gev}{\ensuremath{\mathrm{GeV/}c}}

\newcommand{\kT}{\ensuremath{k_\mathrm{T}}}
\newcommand{\antikT}{anti-\ensuremath{k_\mathrm{T}}}

\newcommand{\pT}{\ensuremath{p_\mathrm{T}}}

\newcommand{\pTjet}{\ensuremath{p_{\mathrm{T,jet}}}}
\newcommand{\pTjetch}{\ensuremath{p_\mathrm{T,jet}^\mathrm{ch}}}
\newcommand{\pTtrig}{\ensuremath{p_{\mathrm{T,trig}}}}

\newcommand{\pTh}{\ensuremath{p_{\mathrm{T,h}}}}

\newcommand{\dNjetdpT}{\ensuremath{\frac{\rm{d}^{2}N^\mathrm{AA}_{jet}}{\mathrm{d}\pTjetch\mathrm{d}\eta_\mathrm{jet}}}}
\newcommand{\dNjetdpTdphi}{\ensuremath{\frac{\rm{d}^{2}N_{jet}}{\mathrm{d}\pTjetch\mathrm{d}\dphi}}}

\newcommand{\dpT}{\ensuremath{\delta{\pT}}}

\newcommand{\qhat}{\ensuremath{\hat{q}}}
\newcommand{\vtwo}{\ensuremath{v_2}}

\newcommand{\Ntrig}{\ensuremath{\mathrm{N}^\mathrm{AA}_{\rm{trig}}}}

\newcommand{\RAA}{\ensuremath{R_\mathrm{AA}}}

\newcommand{\Drecoil}{\ensuremath{\Delta_\mathrm{recoil}}}
\newcommand{\cRef}{\ensuremath{c_\mathrm{Ref}}}

\newcommand{\IAA}{\ensuremath{I_{\rm{AA}}}}
\newcommand{\DIAA}{\ensuremath{\Delta\IAA}}

\newcommand{\pTconst}{\ensuremath{p_\mathrm{T,const}}}

\newcommand{\pTrawi}{\ensuremath{p_\mathrm{T,jet}^\mathrm{raw,i}}}

\newcommand{\pTreco}{\ensuremath{p_\mathrm{T,jet}^\mathrm{reco,ch}}}

\newcommand{\pTpart}{\ensuremath{p_\mathrm{T,jet}^\mathrm{part}}}

\newcommand{\pTrecoi}{\ensuremath{p_\mathrm{T,jet}^\mathrm{reco,i}}}

\newcommand{\Ajeti}{\ensuremath{A_\mathrm{jet}^\mathrm{i}}}

\newcommand{\rhoAi}{\ensuremath{\rho\cdot{A_\mathrm{jet}^\mathrm{i}}}}

\newcommand{\phiTrig}{\ensuremath{\varphi_\mathrm{trig}}}
\newcommand{\phiJet}{\ensuremath{\varphi_\mathrm{jet}}}
\newcommand{\dphi}{\ensuremath{\Delta\varphi}}

\newcommand{\TTSig}{\ensuremath{\mathrm{TT}_{\mathrm{Sig}}}}
\newcommand{\TTRef}{\ensuremath{\mathrm{TT}_{\mathrm{Ref}}}}

\newcommand{\Rdet}{\ensuremath{R_{\mathrm{det}}}}
\newcommand{\Rtot}{\ensuremath{R_{\mathrm{tot}}}}

\newcommand{\pTrec}{\ensuremath{p_{\mathrm{T,jet}}^{\mathrm{det}}}}
\newcommand{\pTgen}{\ensuremath{p_{\mathrm{T,jet}}^{\mathrm{part}}}}

\newcommand{\EffKin}{\ensuremath{\epsilon_{\rm{kin}}}}

\newcommand{\Rargs}{\ensuremath{\left(\pTrec,\pTgen\right)}}
\newcommand{\Drecoilphi}{\ensuremath{\Phi(\dphi)}}

\newcommand{\phithresh}{\ensuremath{\Delta\varphi_{\mathrm{thresh}}}}
\newcommand{\Dyieldthresh}{\ensuremath{\Sigma\left(\phithresh\right)}}

\newcommand{\AAtohjet}{\mathrm{AA}\rightarrow\rm{h}+{jet}+X}
\newcommand{\AAtoh}{\mathrm{AA}\rightarrow\rm{h}+X}

\begin{document}

\begin{titlepage}
\PHnumber{136}                 
\PHdate{03 Jun}              
\PHyear{2015}              

\title{Measurement of jet quenching with semi-inclusive hadron-jet
  distributions in central \PbPb\ collisions at ${\sqrt{\bf{s}_{\mathrm {\bf{NN}}}}}$ = 2.76 TeV}
\ShortTitle{Semi-inclusive hadron-jet measurement in central \PbPb} 

\Collaboration{ALICE Collaboration%
         \thanks{See Appendix~\ref*{app:collab} for the list of collaboration
                      members}}
\ShortAuthor{ALICE Collaboration}      

\begin{abstract}
  We report the measurement of a new observable
  of jet quenching in central \PbPb\ collisions at \sqrtsNN\ = 2.76
  TeV, based on the semi-inclusive rate of charged jets recoiling from
  a high transverse momentum (high-\pT) charged hadron trigger. Jets
  are measured using collinear-safe jet reconstruction with infrared
  cutoff for jet constituents of 0.15 \gev, for jet resolution
  parameters \rr\ = 0.2, 0.4 and 0.5. Underlying event background is
  corrected at the event-ensemble level, without imposing bias on the
  jet population. Recoil jet spectra are reported in the range
  $20<\pTjetch<100$ \gev. Reference distributions for \pp\ collisions
  at \sqrts\ = 2.76 TeV are calculated using Monte Carlo and NLO pQCD
  methods, which are validated by comparing with measurements in \pp\
  collisions at \sqrts\ = 7 TeV. The recoil jet yield in central
  \PbPb\ collisions is found to be suppressed relative to that in \pp\
  collisions. No significant medium-induced broadening of the
  intra-jet energy profile is observed within 0.5 radians relative to
  the recoil jet axis. The angular distribution of the recoil jet
  yield relative to the trigger axis is found to be similar in central
  \PbPb\ and \pp\ collisions, with no significant medium-induced
  acoplanarity observed. Large-angle jet deflection, which may provide
  a direct probe of the nature of the quasi-particles in hot QCD
  matter, is explored.
\end{abstract}

\maketitle
\end{titlepage}
\setcounter{page}{2} 

\section{Introduction}
\label{sect:intro}

Hadronic jets are unique probes of the hot Quantum Chromodynamic
(QCD) matter generated in nuclear collisions at collider
energies. Interactions of hard-scattered partons with colored matter may
modify intra-jet structure, softening and broadening the distribution
of hadronic jet fragments relative to jets generated in vacuum, and may
deflect jets by large angles. These phenomena, known as jet quenching
\cite{Majumder:2010qh}, can probe dynamical properties of the hot QCD
medium \cite{Burke:2013yra} and the nature of quasi-particles in the
Quark-Gluon Plasma (QGP) \cite{D'Eramo:2012jh}.

Jet quenching generates marked, experimentally observable effects.
Measurements of inclusive distributions and correlations of high
transverse momentum (high-\pT) hadrons have revealed significant yield
suppression in nuclear collisions relative to vacuum
\cite{Adcox:2001jp,Adare:2012wg,Adare:2010ry,Adler:2002xw,Adler:2002tq,
  Adams:2003kv,Adams:2006yt,Adamczyk:2013jei,Adare:2012qi,
  Abelev:2012hxa, CMS:2012aa, Aamodt:2011vg, Chatrchyan:2012wg}.
Suppression of the inclusive yield of reconstructed jets
\cite{Abelev:2013kqa,Aad:2014bxa,CMS:2012rba,Adam:2015ewa} and
enhancement in the rate of energy-imbalanced back-to-back di-jet pairs
\cite{Chatrchyan:2012nia,Aad:2010bu} have also been observed in
nuclear collisions.  A measurement of event-averaged missing \pT\
suggests that the radiated energy induced by the interaction of an
energetic parton with the medium is carried to a
significant extent by soft particles at large angles relative to the
jet axis \cite{Chatrchyan:2011sx}. 

The measurement of reconstructed jets over a wide range in jet energy
and jet resolution parameter (\rr) is required for comprehensive
understanding of jet quenching in heavy-ion collisions. Such
measurements are challenging, however, due to the presence of 
complex, uncorrelated background to the jet signal, and the
need to minimize biases in the selected jet population imposed by
background suppression techniques. Multiple, complementary measurement
approaches, differing both in instrumentation and in analysis
algorithm, are therefore important to elucidate the physics of jet
quenching using reconstructed jets.

In this article we present a new approach to the measurement of jet
quenching, based on the semi-inclusive distribution of charged jets
recoiling from a high-\pT\ charged hadron trigger (``h-jet''
coincidence) in central (0-10\%) \PbPb\ collisions at \sqrtsNN\ = 2.76
TeV. Jets are reconstructed using charged particle tracks with the
\kT\ \cite{Cacciari:2011ma} and \antikT\ algorithms
\cite{FastJetAntikt}, with infrared cutoff for tracks $\pTconst>0.15$
\gev. Uncorrelated background to the recoil jet signal is corrected
solely at the level of ensemble-averaged distributions, without
event-by-event discrimination of jet signal from background, using a
technique that exploits the phenomenology of jet production in
QCD. The correction is carried out using an unfolding technique. This
approach enables the collinear-safe measurement in heavy-ion
collisions of reconstructed jets with low infrared cutoff over a wide
range of jet energy and \rr. Recoil jet distributions, which are
differential in \pTjet\ and in azimuthal angle relative to the trigger
axis, are reported for \rr\ = 0.2, 0.4 and 0.5, over the range
$20<\pTjetch<100$ \gev.

Suppression of the recoil jet yield due to quenching is measured by
comparison to the yield in \pp\ collisions. However, our current data
for \pp\ collisions at \sqrts\ = 2.76 TeV do not have sufficient
statistical precision to provide a reference for the \PbPb\
measurements reported here. The reference distribution is therefore
calculated using the PYTHIA event generator \cite{Sjostrand:2006za}
and perturbative QCD (pQCD) calculations at Next-to-Leading Order
(NLO) \cite{deFlorian:2009fw}, which are validated by comparison with
ALICE measurements of \pp\ collisions at \sqrts\ = 7 TeV.  Angular
broadening of the internal jet structure due to quenching is
investigated by comparing the differential recoil jet distributions
for different values of \rr. Acoplanarity between the trigger hadron
and recoil jet directions is measured to explore the deflection of the
jet axis induced by quenching. The rate of large angular deviations is
measured; this rate may be dominated by single hard (Moli{\`e}re)
scattering, which could potentially probe the quasi-particle nature of
the hot QCD medium \cite{D'Eramo:2012jh,Wang:2013cia}.

These observables are directly comparable to theoretical
calculations, without the need to model the heavy-ion collision
background, due to utilization of a hadron trigger, the
semi-inclusive nature of the observables, and the background
suppression technique. The only non-perturbative component required
to calculate the hard-process bias is the inclusive charged hadron
fragmentation function (in-vacuum or quenched) for the trigger hadron.

The paper is organized as follows: Sect. \ref{sect:Dataset}, dataset,
event selection, and simulations; Sect. \ref{Sect:JetReco}, jet
reconstruction; Sect. \ref{sect:Observables}, discussion of
observables; Sect. \ref{sect:RawDistr}, raw distributions; Sect.
\ref{sect:DrecoilCorr}, corrections; Sect. \ref{sect:SysUncert},
systematic uncertainties; Sect. \ref{sect:ppReference}, reference
distributions for \pp\ collisions; Sect.  \ref{sect:Results}, results;
and Sect.  \ref{sect:Summary}, summary.

\section{Data set, offline event selection, and simulations}
\label{sect:Dataset}

The ALICE detector and its performance are described in
\cite{Aamodt:2008zz,Abelev:2014ffa}. 

The \PbPb\ collision data were recorded during the 2011 LHC \PbPb\ run
at \sqrtsNN\ = 2.76 TeV. This analysis uses the 0\,--\,10\% most-central
Pb-Pb collisions selected by the online trigger based on the hit
multiplicity measured in the forward V0
detectors. The online trigger had 100\% efficiency for the 0\,--\,7\% interval in
centrality percentile, and 80\% efficiency for the 8\,--\,10\%
interval.

Events are reconstructed offline as described in
Ref. \cite{Abelev:2012hxa}. Charged tracks are measured in the ALICE
central barrel, with acceptance $|\eta|<0.9$ over the full
azimuth. Accepted tracks are required to have $0.15<\pT< 100$~\gev,
with at least 70 Time Projection Chamber (TPC) space-points and at
least 80\% of the geometrically findable space-points in the TPC.  To
account for the azimuthally non-uniform response of the Inner Tracking
System (ITS) in this dataset, two exclusive classes of tracks are used
\cite{Abelev:2014ffa}: tracks with Silicon Pixel Detector (SPD) hits
(70\% of all tracks in central \PbPb\ collisions, and 95\% in \pp\
collisions); and tracks without SPD hits but with a primary vertex
constraint. The primary vertex is required to lie within 10~cm of the
nominal center of the detector along the beam axis, and within 1~cm of
it in the transverse plane.  After offline event selection, the \PbPb\
dataset consists of 17M events in the 0\,--\,10\% centrality
percentile interval.

The \pp\ collision data used to validate PYTHIA and pQCD  calculations
were recorded during the 2010 low-luminosity \pp\ run at \sqrts\ = 7
TeV, using a MB trigger. The MB trigger configuration, offline event
selection, and tracking are the same as described in
\cite{ALICE:2014dla}. After event selection cuts, the \pp\ dataset
consists of 168M events. There is negligible
  difference in the inclusive jet cross section for events selected by
  the ALICE online trigger, and for a non-single diffractive event population.

Simulations of \pp\ collisions were carried out
  using PYTHIA 6.425, with the Perugia 0, Perugia 2010, and Perugia
  2011 tunes \cite{Skands:2010ak}. Instrumental effects are calculated
  using the Perugia 0 and Perugia 2010 tunes for \pp\ and \PbPb\
  collisions respectively, with a detailed detector model implemented
  using GEANT3 \cite{GEANT3}. In addition, a simulation based on
  HIJING \cite{Wang:1991hta} is used to evaluate the detector response
  in the high multiplicity environment of \PbPb\ collisions.  Perugia
2011, which has been tuned to other LHC data, is used as an
alternative to compare with the new data presented here. Simulated
events, which include primary particles and the daughters of strong
and electromagnetic decays but not instrumental effects or the
daughters of weak decays, are denoted ``particle level''. Simulated
events also including instrumental effects and weak decay daughters
where reconstructed tracks are selected using the experimental cuts
are denoted ``detector level''.

For central \PbPb\ collisions, tracking efficiency is 80\% for $\pT>1$
\gev, decreasing to 56\% at 0.15 \gev. Track momentum resolution is
1\% at \pT\ = 1 \gev\ and 3\% at \pT\ = 50 \gev. For \pp\ collisions,
tracking efficiency is 2\%-3\% higher than in central \PbPb\
collisions. Track momentum resolution is 1\% at \pT\ = 1 \gev\ for all
reconstructed tracks; 4\% at \pT\ = 40 \gev\ for tracks with SPD hits;
and 7\% at \pT\ = 40 \gev\ for tracks without SPD hits
\cite{Abelev:2014ffa,ALICE:2014dla}.

\section{Jet reconstruction}
\label{Sect:JetReco}

Jet reconstruction for both the \pp\ and \PbPb\ analyses is carried
out using the \kT\ \cite{Cacciari:2011ma} and
\antikT\cite{FastJetAntikt} algorithms applied to all accepted charged
tracks.  The boost-invariant \pT-recombination scheme is used
\cite{Cacciari:2011ma}. Jet area is calculated by the Fastjet
algorithm using ghost particles with area 0.005\cite{FastJetArea}.

Charged jets are not safe in perturbation theory, because radiation
carried by neutral particles is not included. However, infrared-safe
calculations of charged-jet observables can be performed using
non-perturbative track functions, which absorb infrared divergences
and describe the energy fraction of a parton carried by charged tracks
\cite{Chang:2013rca}. Track functions are analogous to fragmentation
functions, with DGLAP-like evolution, and perturbative calculations
using them are in good agreement with PYTHIA
calculations\cite{Chang:2013rca}. Track functions can provide the basis
for rigorous comparison of the charged-jet measurements reported
here with both analytic and Monte Carlo QCD calculations.

For the \PbPb\ analysis, adjustment of jet energy for the presence of
large background utilizes the FastJet procedure \cite{FastJetPileup},
in which jet reconstruction is carried out twice for each event. The
first pass applies the \kT\ algorithm with \rr\ = 0.4 to estimate
$\rho$, the density of jet-like transverse-momentum due to background,
which is defined as

\begin{equation}
\rho=\mathrm{median}\left\{ \frac{\pTrawi}{\Ajeti} \right\},
\label{eq:rho}
\end{equation}

\noindent
where index i runs over all jet candidates in an event, and \pTrawi\
and \Ajeti\ are the transverse momentum and area of the
i$^{\textup{th}}$ reconstructed jet. Further details are presented in
\cite{Abelev:2012ej}.  The central data points in this analysis are
determined by excluding the two jets with highest \pTrawi\ from
calculation of the median, with a variant used to study systematic
sensitivity to this choice.

The second pass, which generates jet candidates
for the reported distributions, applies the \antikT\
algorithm with resolution parameter \rr\ = 0.2, 0.4, and 0.5. The
value of \pTrawi\ is corrected according to \cite{FastJetPileup},

\begin{equation}
\pTrecoi=\pTrawi - \rhoAi,
\label{eq:pTraw}
\end{equation}

\noindent
where \pTrawi\ and \Ajeti\ are measured for the i$^\mathrm{th}$ jet in an
event, and $\rho$ is a scalar value common to all jets in each event,
but varies from event to event. 

Jet candidates are accepted if $|\eta_{\rm{jet}}|<0.5$ for \rr\ = 0.2
and 0.4 and $|\eta_{\rm{jet}}|<0.4$ for \rr\ = 0.5, where
$\eta_{\mathrm{jet}}$ is the pseudo-rapidity of the jet candidate
centroid. The azimuthal acceptance of the recoil yield measurement is
$\pi - \dphi < 0.6$, where $\dphi=|\phiTrig - \phiJet|$ is the
difference between the azimuthal angle of the trigger hadron
(\phiTrig) and the jet candidate centroid (\phiJet), and
$0\leq\dphi\leq\pi$.

A cut on jet area is applied to suppress combinatorial jets while
preserving high efficiency for true hard jets
\cite{deBarros:2011ph,Jacobs:2010wq}. Jet candidates are rejected if
$\Ajeti<0.07$ for \rr\ = 0.2; $\Ajeti<0.4$ for \rr\ = 0.4; and
$\Ajeti<0.6$ for \rr\ = 0.5.

Similar procedures are followed for the \pp\ data analysis. Jets are
reconstructed with the \antikT\ algorithm for \rr\ = 0.2, 0.4 and 0.5.
Reconstructed \pTreco\ is adjusted using Eq. \ref{eq:pTraw}, where
$\rho$ is estimated in this case by the summed \pT\ in two cones of
radius \rr\ = 0.4, with centroids at the same $\eta$ but perpendicular
in azimuth to the leading jet in the event.

The instrumental jet energy resolution (JER), which characterizes the
detector response relative to charged jets at the particle level,
varies from 20\% at \pTjet\ = 20 \gev\ to 25\% at \pTjet\ = 100 \gev,
for both \PbPb\ and \pp\ collisions, with negligible dependence on
\rr. The jet energy scale (JES) uncertainty, which is dominated by the
uncertainty of tracking efficiency, is approximately 5\% for both
\PbPb\ and \pp\ collisions, with negligible dependence on \pTjetch\
and \rr. However, the instrumental response is significantly
non-Gaussian\cite{ Abelev:2013kqa} and unfolding of the full response
matrix is used for corrections.

\section{Discussion of observables}
\label{sect:Observables}

\subsection{General considerations}
\label{sect:ObservablesGeneral}

Energetic jets that arise from high momentum transfer (high-$Q^2$)
scattering of partons are readily visible in event displays of high
multiplicity heavy-ion collisions
\cite{Aad:2010bu,Chatrchyan:2011sx}. However, accurate measurement of
jet energy in such events, and unbiased measurement of jet
distributions, are more difficult. Application of a jet reconstruction
algorithm to high multiplicity events will cluster hadrons arising from
multiple incoherent sources into each reconstructed jet,
resulting in significant smearing of the true hard jet energy
distribution. It will also generate a large population of
``combinatorial'' background jets comprising solely hadrons generated
by soft production processes ($Q^2$ below a few GeV$^2$), which cannot
be identified as hard jets with smeared energy.

Current heavy-ion jet analyses select the hard jet population on a
jet-by-jet basis by several different approaches: removal of an
estimated background component of transverse energy prior to jet
reconstruction \cite{Kodolova:2007hd}; or imposition of a
fragmentation bias requiring a cluster of high-\pT\ tracks or a single
high-\pT\ track in the jet, and imposition of a jet \pT\ threshold
\cite{Aad:2012vca,Abelev:2013kqa,CMS:2012rba,Adam:2015ewa}.  These
rejection techniques may bias towards certain fragmentation patterns
in the accepted hard jet population.

This analysis takes a different approach, in which corrections for
background and instrumental effects are applied solely at the level of
ensemble-averaged distributions, without rejection of individual jet
candidates or removal of event components. The analysis is based on
the semi-inclusive differential distribution of charged jets recoiling
from a high-\pT\ trigger hadron, with the trigger hadron selected
within a limited \pTtrig\ interval (Trigger Track, or TT, class). This
distribution, which is the number of jets measured in the recoil
acceptance normalized by the number of trigger hadrons, is equivalent
to the ratio of inclusive production cross sections,

\begin{equation}
\frac{1}{\Ntrig}\dNjetdpT\Bigg\vert_{\pTtrig\in{\mathrm{TT}}} = \left(
\frac{1}{\sigma^{\AAtoh}} \cdot
\frac{\rm{d}^2\sigma^{\AAtohjet}}{\mathrm{d}\pTjetch\rm{d}\eta_\mathrm{jet}}\right) 
\Bigg\vert_{\pTh\in{\mathrm{TT}}},
\label{eq:hJetDefinition}
\end{equation}

\noindent
where AA denotes \pp\ or \PbPb\ collisions, $\sigma^{\AAtoh}$ is the
cross section to generate a hadron within the \pT\ interval of the
selected TT class,
$\rm{d}^2\sigma^{\AAtohjet}/\rm{d}\pTjetch\rm{d}\eta$ is the
differential cross section for coincidence production of a hadron in
the TT interval and a recoil jet, and \pTjetch\ and
$\eta_\mathrm{jet}$ are the charged jet transverse momentum and
pseudo-rapidity.

Because the observable in Eq.~\ref{eq:hJetDefinition} is
semi-inclusive, the selection of events containing a hard process
(``hard-process selection'') is based solely on the presence of a
high-\pT\ hadron trigger. In particular, there is no requirement that
a jet satisfying certain criteria be found in the recoil acceptance.
Rather, all jet candidates in the recoil acceptance are counted in
Eq.~\ref{eq:hJetDefinition}, regardless of their specific
properties. Events with no hard jet candidates (however defined)
falling within the acceptance are not rejected, and contribute to the
normalization. This observable thereby measures the absolutely
normalized rate of recoil jets observed per trigger. Correction for
the contribution of uncorrelated background jets in
Eq.~\ref{eq:hJetDefinition} is carried out at a later step in the
analysis, as discussed below.

Other jet correlation measurements in heavy-ion collisions have been
carried out, in which hard-process selection utilizes a compound
condition that requires the presence of both a trigger object (jet or
photon) and a recoil jet satisfying certain criteria
\cite{Aad:2010bu,Chatrchyan:2011sx,Chatrchyan:2012gt}. The jet
correlation distributions in these analyses are normalized per
trigger--recoil pair; absolute normalization requires scaling by the
inclusive trigger yield, together with selection of the recoil jet
population using the semi-inclusive procedure described above. The
role of normalization in the measurement of in-medium large-angle
scattering is discussed in Sect.~\ref{sect:AngDep}.

\subsection{Trigger hadrons and hard-process bias}
\label{sect:HadronTrigger}

The use of high-\pT\ hadron triggers for hard-process
selection in this analysis is based on the following considerations.

Hadrons with \pT\ larger than about 5-7 \gev\ are expected to originate
primarily from fragmentation of energetic jets, in both \pp\ and
\PbPb\ collisions at \sqrtsNN\ = 2.76 TeV (see
e.g. \cite{Renk:2011gj}). They provide experimentally clean triggers,
without the need for correction for uncorrelated background. Selection
of events by requiring the presence of a high-\pT\ hadron biases
towards events containing a high-$Q^2$ partonic interaction, with jets
in the final state.

Inclusive distributions of high-\pT\ hadrons have been measured and
calculated theoretically in both \pp\ and heavy-ion collisions at
collider energies. For \pp\ collisions at the LHC, agreement within a
factor two is found between NLO calculations and data for $\pT>10$
\gev, with the discrepancies attributable to poorly constrained gluon
fragmentation functions that can be improved by fitting to LHC data
\cite{d'Enterria:2013vba}. For inclusive hadron production in
heavy-ion collisions \cite{Adams:2003kv, Adare:2012wg, Abelev:2012hxa,
  CMS:2012aa}, the medium-induced modification and evolution of
fragmentation functions have been calculated in several frameworks,
showing good agreement with data (\cite{Armesto:2007dt,Chang:2014fba}
and references therein). Suppression of inclusive hadron production in
heavy-ion collisions has been used to determine the jet transport
parameter \qhat\cite{Burke:2013yra}.

Any hard-process selection procedure imposes bias on the accepted
event population, and accurate comparison of theory calculations with
such measurements requires calculation of this selection bias. In this
analysis, hard-process selection uses the same cuts as those used for
high-\pT\ inclusive hadron measurements. Since inclusive hadron
production is calculable in both \pp\ and \PbPb\ collisions, the
selection bias in this analysis is likewise calculable using current
theoretical approaches.

\subsection{Hadron--jet coincidences}
\label{sect:hJetDiscussion}

There are additional considerations for interpreting hadron-triggered
recoil distributions in Eq.~\ref{eq:hJetDefinition} and comparing such
measurements to theoretical calculations, as follows.

The h-jet coincidence cross section in \pp\ collisions has been
calculated in a pQCD framework \cite{deFlorian:2009fw}. In this
process at LO, a pair of final-state partons is generated with
opposing transverse momenta, with one of the pair fragmenting into a
hadron which carries momentum fraction $z$ of the recoiling
jet. Since $z=\pTtrig/\pTjet\leq1$ at LO, the requirement of a
high-\pTtrig\ hadron above threshold therefore biases against
coincident recoil jets with $\pTjet<\pTtrig$, but does not impose a
kinematic constraint on recoil jets with $\pTjet>\pTtrig$.

The inclusive hadron distribution at high-\pT\ is biased towards
high-$z$ jet fragments, due to interplay between the shape of the
inclusive jet \pT\ spectrum and the shape of the inclusive
fragmentation function, with
$\langle{z_\mathrm{incl}}\rangle\approx0.6$ at LHC energies
\cite{Alessandro:2006yt}. However, in the semi-inclusive measurement
based on Eq. \ref{eq:hJetDefinition}, the trigger-normalized rate of
recoil jets is measured as a function of \pTjetch. At LO this
corresponds to $z=\pTtrig/\pTjet$, which can differ significantly from
$\langle{z_\mathrm{incl}}\rangle$ \cite{deFlorian:2009fw}. For fixed
\pTtrig\ the $z$-bias therefore varies as a function of recoil \pTjet,
with stronger bias than the inclusive case for $\pTjet\approx\pTtrig$,
and weaker bias for $\pTjet\gg\pTtrig$. The $z$-bias has been
calculated for \pp\ collisions at \sqrts\ = 2.76 and 7 TeV using the
approach of \cite{deFlorian:2009fw} and found to be similar at LO and
NLO. This $z$-bias, which is kinematic in origin, likewise occurs in
nuclear collisions in which jets experience quenching. This effect is
intrinsic to any theoretical framework based on pQCD, both in-vacuum
and quenched, and will be properly accounted for in such calculations.

For quenched jets in nuclear collisions, high-\pT\ hadron selection
may generate two additional, related biases. The first is a bias
towards high-$z$ fragments of jets that have lost relatively little
energy in the medium \cite{Baier:2002tc}. The second is a geometric
bias in which small energy loss corresponds to small path length in
matter. In the latter case, jets generating high-\pT\ trigger hadrons
are generated predominantly on the surface of the collision region and headed
outward
\cite{Drees:2003zh,Dainese:2004te,Eskola:2004cr,Renk:2006nd,Loizides:2006cs,Zhang:2007ja,Renk:2012cb},
with the corresponding recoil jet population biased towards longer
path length in matter than the unbiased, fully inclusive jet
population.

The degree to which high-\pT\ hadron selection biases towards small
energy loss of its parent jet determines the degree of similarity in
the underlying distribution of high-$Q^2$ processes in \pp\ and \PbPb\
collisions, for the same hadron trigger cuts. In the following, we
refer to potential differences in such $Q^2$ distributions as being
due to ``trigger-jet'' energy loss. Quantitative assessment of these
effects can be carried out using theoretical calculations of inclusive
charged hadron production.

\subsection{Semi-inclusive recoil jet measurements}
\label{sect:RecoilJetDistributions}

Trigger hadrons lie within the charged-track
  acceptance $|\eta|<0.9$ and are selected in the intervals
$8<\pTtrig<9$ \gev, denoted by TT\{8,9\} and referred to as the Reference TT
class; and $20<\pTtrig<50$ \gev, denoted by TT\{20,50\} and referred to as
the Signal TT class.

\begin{figure}[tbh!f]
\centering
\includegraphics[width=0.5\textwidth]{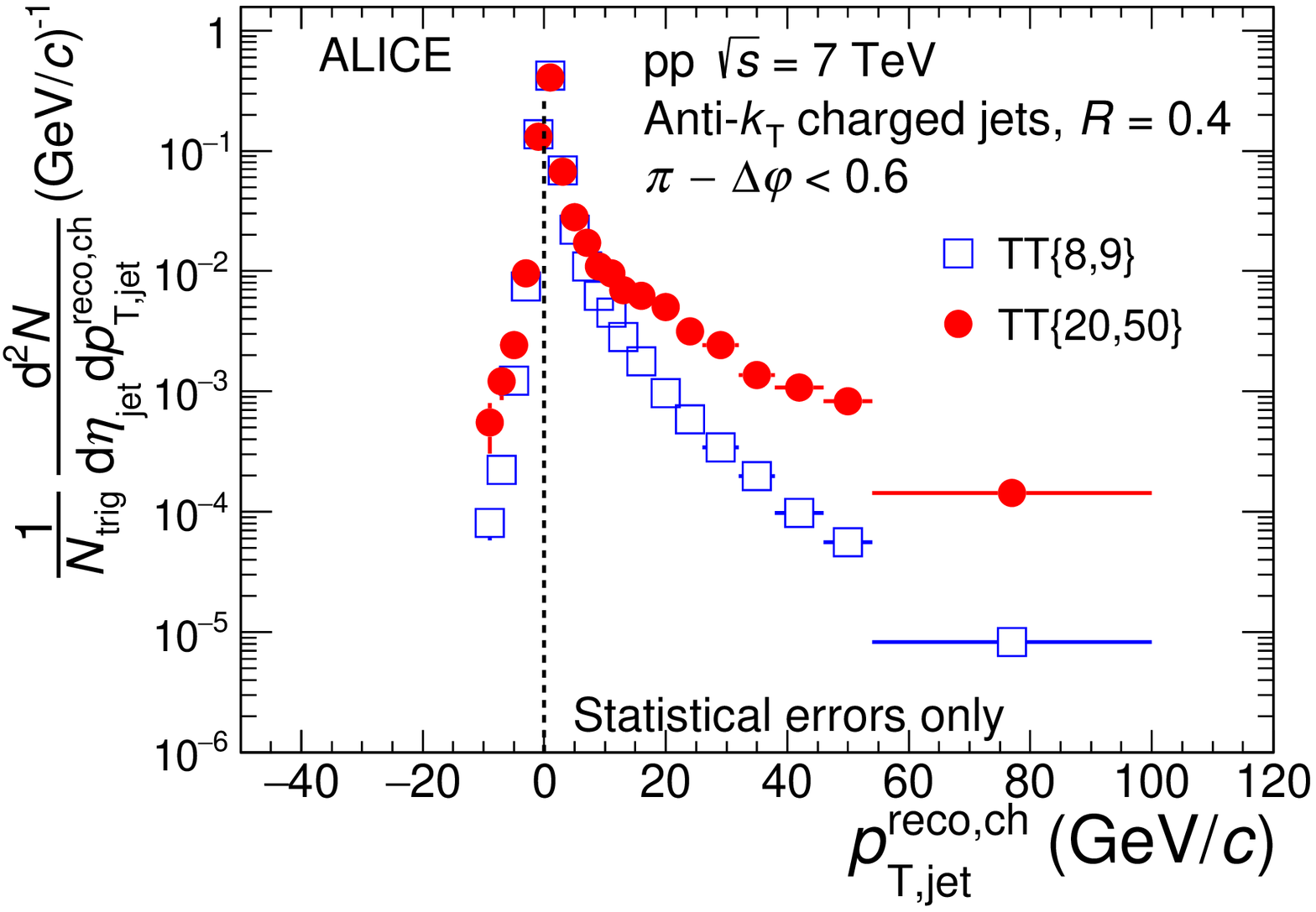}
\includegraphics[width=0.5\textwidth]{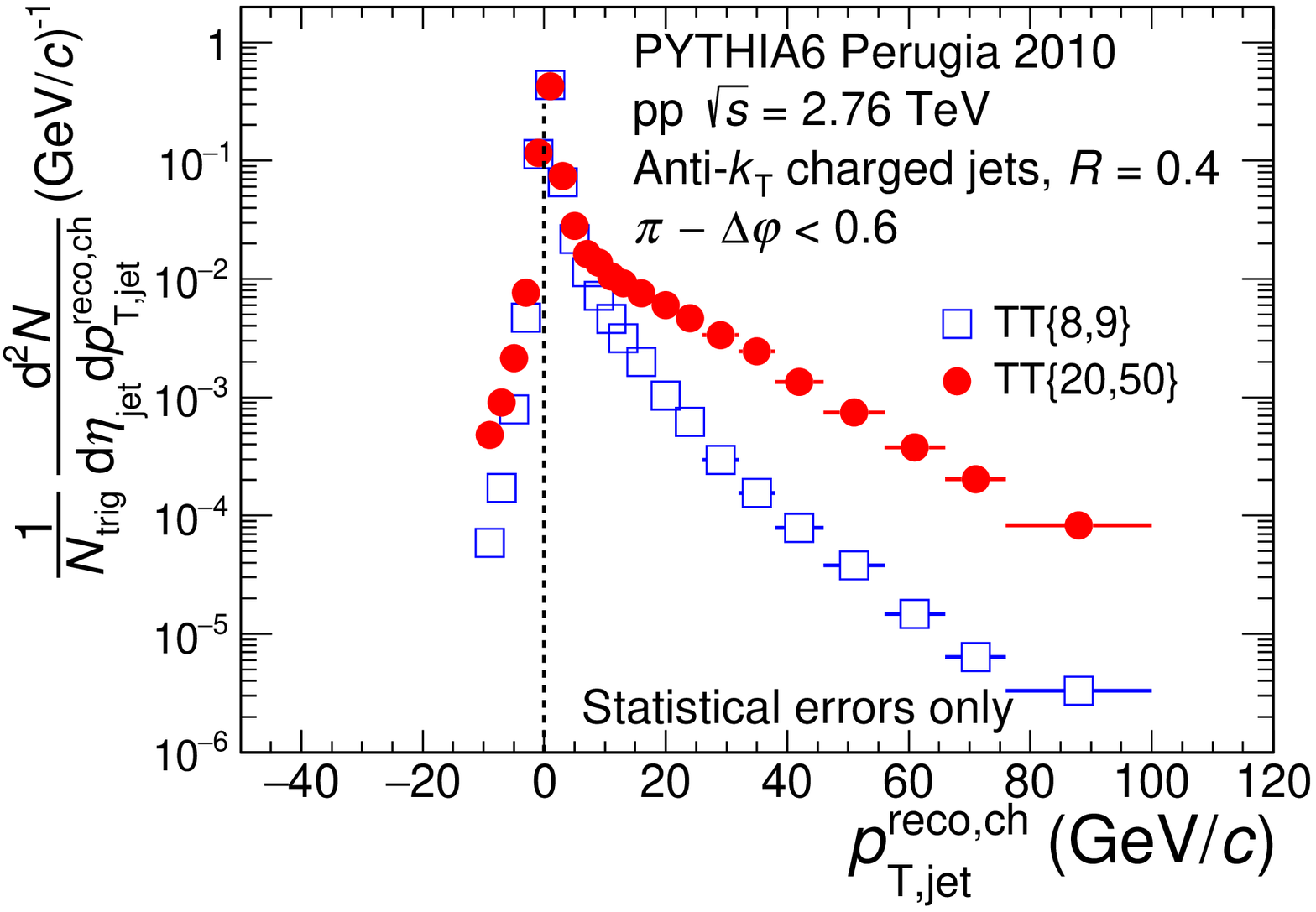}
\includegraphics[width=0.5\textwidth]{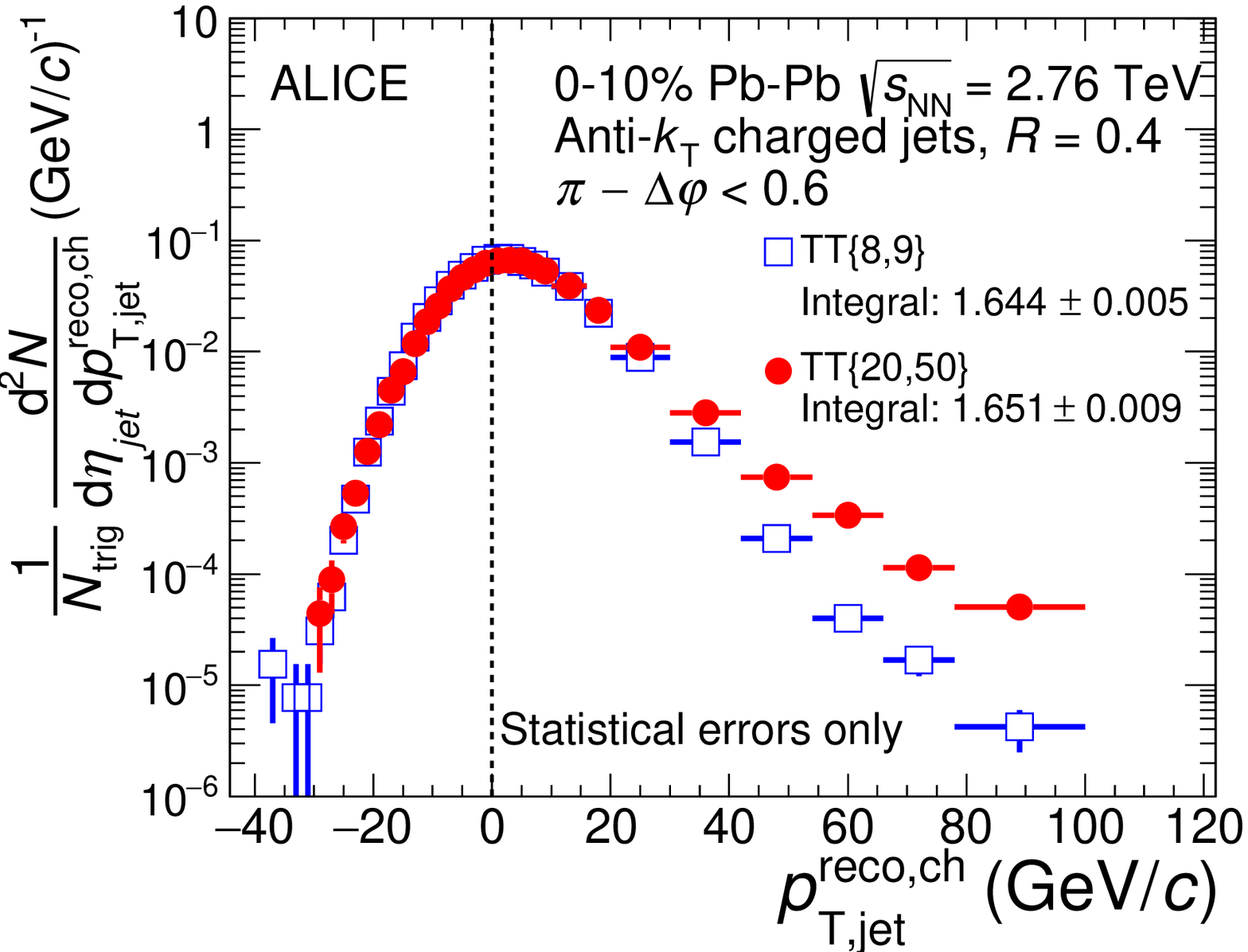}
\caption{Semi-inclusive distributions of jets recoiling from a hadron
  trigger for two exclusive TT classes (Eq. \ref{eq:hJetDefinition}),
  for \pp\ collisions at \sqrts\ = 7 TeV from ALICE data (top), \pp\
  collisions at \sqrts\ = 2.76 TeV from particle-level PYTHIA simulations
  (center), and central \PbPb\ collisions at \sqrtsNN\ = 2.76 TeV from
  ALICE data (bottom). All distributions are a function of \pTreco
  (Eq. \ref{eq:pTraw}). Distributions from data are not corrected for
  background fluctuations and instrumental effects. }
\label{fig:hJetRecoil}
\end{figure}

Figure \ref{fig:hJetRecoil} shows semi-inclusive distributions
(Eq. \ref{eq:hJetDefinition}) for recoil jets with R = 0.4, for the
Signal and Reference TT classes in \pp\ collisions at \sqrtsNN\ = 7
TeV and central \PbPb\ collisions measured by ALICE, and in \pp\
collisions at \sqrtsNN\ = 2.76 TeV simulated by PYTHIA. The
distributions include all jet candidates in the recoil acceptance.

Since $\rho$ is the median energy density in an event, there must be
jet candidates with energy density less than $\rho$, and which
consequently have $\pTreco<0$.  The recoil jet distribution in the
region $\pTreco<0$ is seen to be largely uncorrelated with TT class in
all cases, indicating that the yield in this region is dominated by
combinatorial jets. In \pp\ collisions the distribution in this region
is narrow, indicating only small background density fluctuations. The
predominant feature of the \pp\ distributions is the strong dependence
on TT class for $\pTreco>0$, with a harder recoil jet spectrum for
higher \pTtrig, as expected from the systematics of jet production in
QCD. For \PbPb\ collisions the distribution in the region $\pTreco<0$
is much broader, indicating significantly larger background density
fluctuations than in \pp\ collisions.  For large and positive \pTreco,
the recoil jet distribution in \PbPb\ is strongly correlated with TT
class, similar to \pp\ collisions, showing that this region has
significant contribution from the true coincident recoil jet yield.

The integrals of the \PbPb\ distributions in Fig.
\ref{fig:hJetRecoil} are $1.645\pm{0.005}$(stat) for TT\{8,9\} and
$1.647\pm{0.009}$(stat) for TT\{20,50\}.  This integral represents the
average number of jet candidates per trigger hadron, both correlated
and uncorrelated, and is seen to be consistent, within errors of a few
per mil, for the two TT classes. Similar features have been observed
in model calculations \cite{deBarros:2012ws}. Since central \PbPb\
events have high multiplicity, the recoil acceptance in each event is
fully populated by jet candidates. Invariance of the integral with TT
class therefore indicates that the number of jet candidates per
trigger hadron is due largely to geometric factors, specifically the
acceptance and jet resolution parameter \rr. This behavior is
consistent with the robustness of the \antikT\ algorithm against
modification of jet structure by soft particles from the underlying
event \cite{FastJetAntikt}.  Jet candidate distributions reconstructed
using the \antikT\ algorithm for different trigger hadron kinematics
appear to differ most significantly in the shape of the distribution
as a function of \pTjetch, not in the total number of jet candidates
per event.

Based on these considerations we define a new observable, \Drecoil,
which suppresses the uncorrelated jet yield in a purely data-driven
way. \Drecoil\ is the difference between two semi-inclusive recoil jet
distributions (Eq. \ref{eq:hJetDefinition}) for the Signal and Reference TT
classes \cite{deBarros:2012ws},

\begin{equation}
\Drecoil = 
\frac{1}{\Ntrig}\dNjetdpT\Bigg\vert_{\pTtrig\in{\TTSig}} -
\cRef\cdot \frac{1}{\Ntrig}\dNjetdpT\Bigg\vert_{\pTtrig\in{\TTRef}}.
\label{eq:DRecoil}
\end{equation}

\noindent
The scale factor \cRef, which is within a few percent of unity, is
discussed in Sect. \ref{sect:DrecoilRaw}.

The raw \Drecoil\ distribution must be corrected for instrumental
effects and for smearing of coincident recoil jet energy by
fluctuations of energy density in the underlying event. After
corrections, \Drecoil\ represents the change in the distribution of
jets recoiling in coincidence with a trigger hadron, as the trigger
hadron \pT\ changes from the Reference to Signal TT
interval. While this differential coincidence observable has not been
reported previously, it is nevertheless well-defined in terms of
perturbative QCD.

We also extend Eq. \ref{eq:DRecoil} to measure the angular
distribution of recoil jet yield with respect to the axis defined by the
trigger hadron momentum, in order to investigate medium-induced acoplanarity
\cite{D'Eramo:2012jh,Wang:2013cia} (``inter-jet broadening''). The
azimuthal correlation between the trigger hadron and coincident recoil
charged jets is measured via

\begin{equation}
\Drecoilphi = 
\frac{1}{\Ntrig}\dNjetdpTdphi\Bigg\vert_{\pTtrig\in{\TTSig}} -
\cRef\cdot \frac{1}{\Ntrig}\dNjetdpTdphi\Bigg\vert_{\pTtrig\in{\TTRef}},
\label{eq:DRecoildphi}
\end{equation}

\noindent
where the recoil acceptance for this observable is
$\pi/2<\dphi<\pi$. Normalization to unit $\eta$ is omitted from the
notation for clarity.

We quantify the rate of medium-induced large-angle scattering by
measuring the integrated recoil yield at large angular deflection
relative to $\dphi=\pi$, defined as
\begin{equation}
\Dyieldthresh = \int^{\pi-\phithresh}_{\pi/2} {\rm{d}\dphi} \left[
  \Drecoilphi \right], 
\label{eq:Dyield}
\end{equation}

\noindent
where the lower limit of the integration is set arbitrarily to
$\pi/2$. The upper limit excludes the main peak of the \Drecoilphi\
distribution, $|\dphi-\pi| < \phithresh$, in order to measure the
yield in the tail of the distribution.  \Dyieldthresh\ is measured as
a function of \phithresh.

The distributions \Drecoilphi\ and \Dyieldthresh\ likewise represent the
change in the angular distribution of recoil jet yield, as the trigger hadron
\pT\ changes from the Reference to Signal TT interval.

\section{Raw distributions}
\label{sect:RawDistr}

In order to ensure statistical independence of the recoil jet
distributions for the Signal and Reference TT classes, each event is
assigned randomly to one of the TT classes and is used only for its
assigned TT class.  The statistical reach of the analysis is optimized
by assigning 80\% of the events to the Signal TT subset and 20\% to
the Reference TT subset. This choice balances retention of the
high-\pTreco\ component of the Signal recoil jet distribution with
statistical precision of the Reference distribution in the region
$\pTreco<0$, with the latter condition required to provide accurate
normalization of the combinatorial background jet distribution.

Events within each subset are then selected for further analysis if
they contain at least one hadron within the \pTtrig\ interval of their
assigned TT class. If more than one hadron satisfying this criterion
is found, one hadron is chosen randomly as the trigger hadron. For the
\PbPb\ analysis there are 65k events with trigger hadrons in the
Reference TT class and 22k in the Signal TT class. For the \pp\
analysis there are 74k events with trigger hadrons in the Reference TT
class and 5k in the Signal TT class.

\subsection{Distributions of \Drecoil}
\label{sect:DrecoilRaw}

\begin{figure}[tbh!f]
\centering
\includegraphics[width=0.45\textwidth]{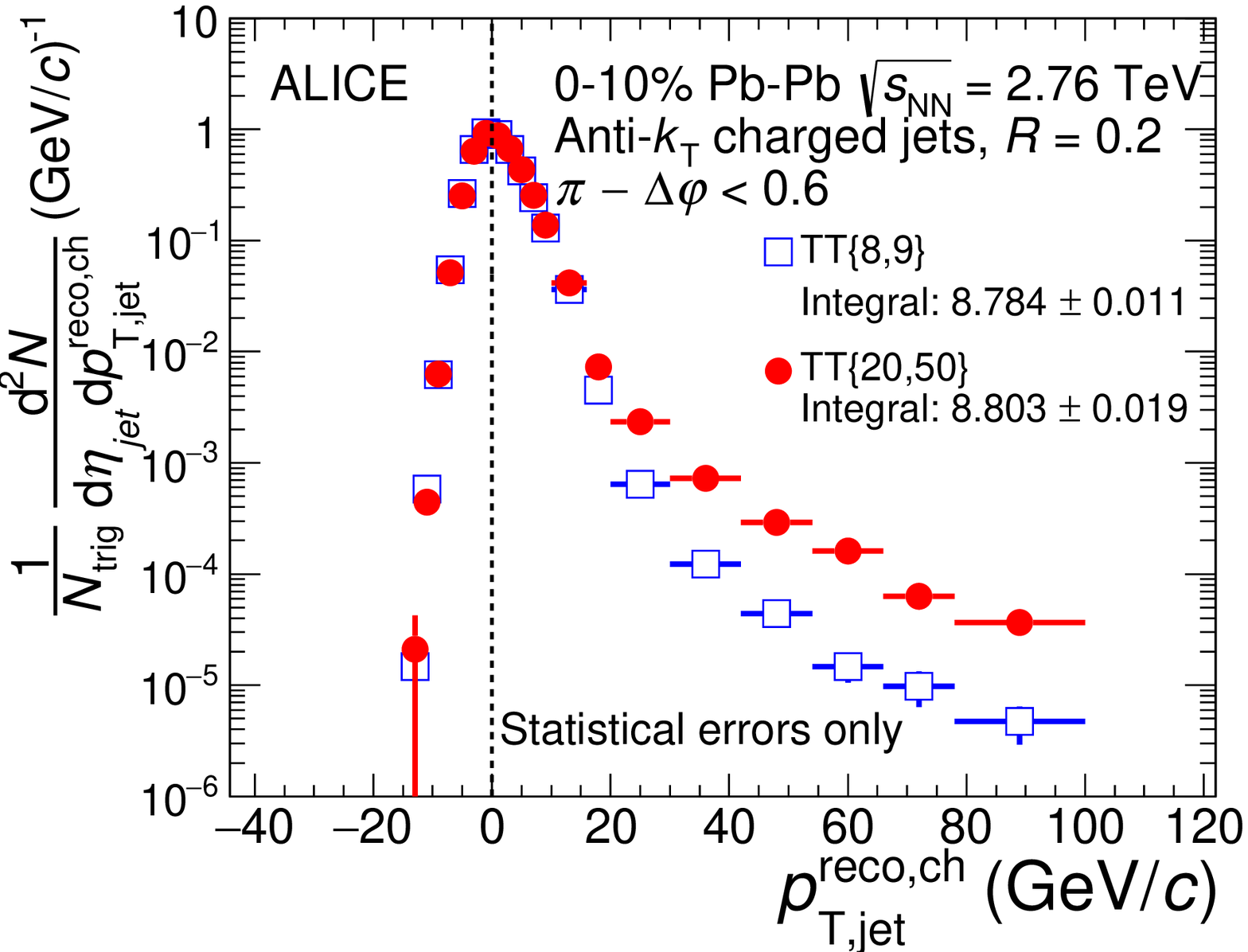}
\includegraphics[width=0.45\textwidth]{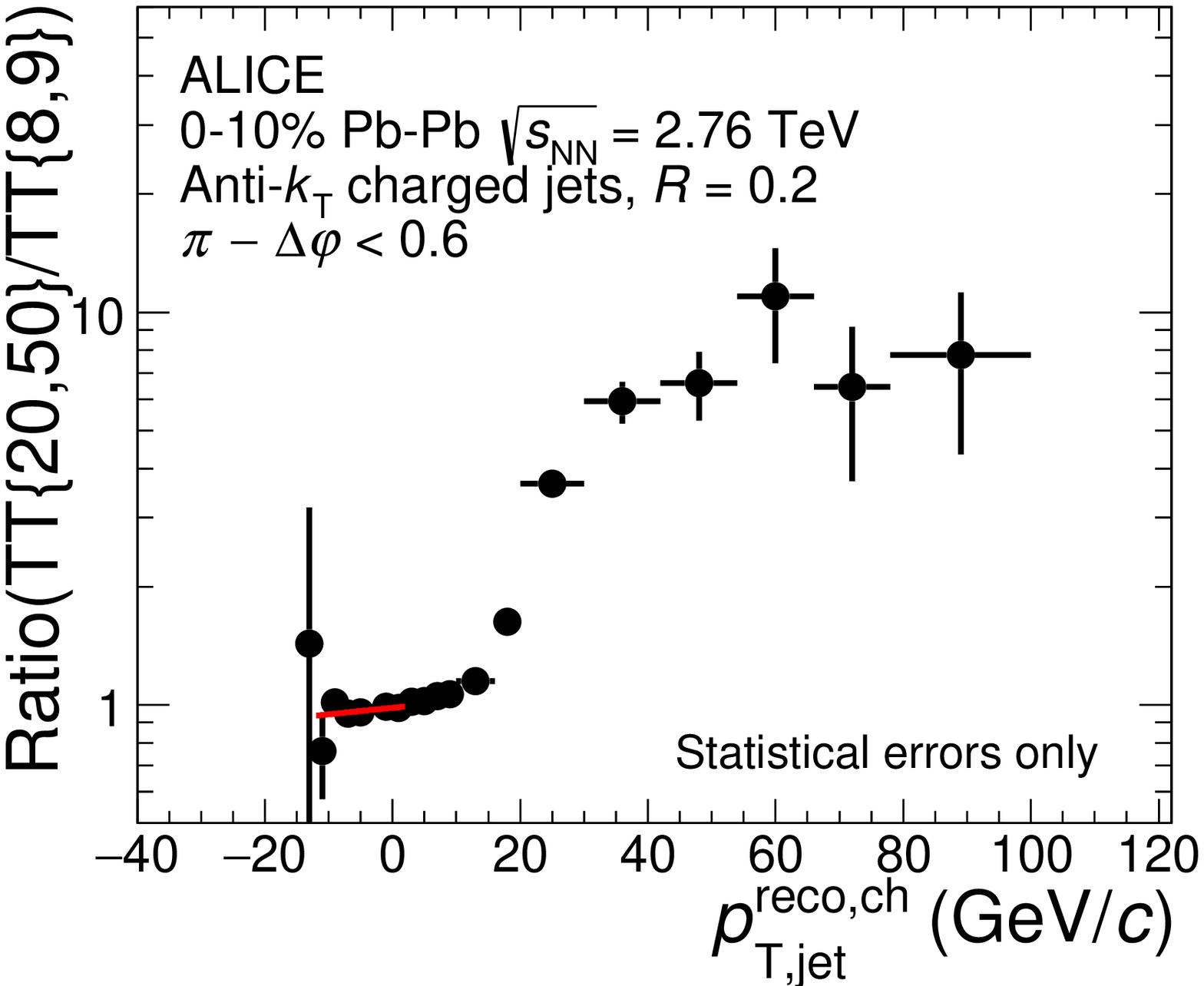}
\includegraphics[width=0.45\textwidth]{./Figs/AllInRawFinalR04TT205089}
\includegraphics[width=0.45\textwidth]{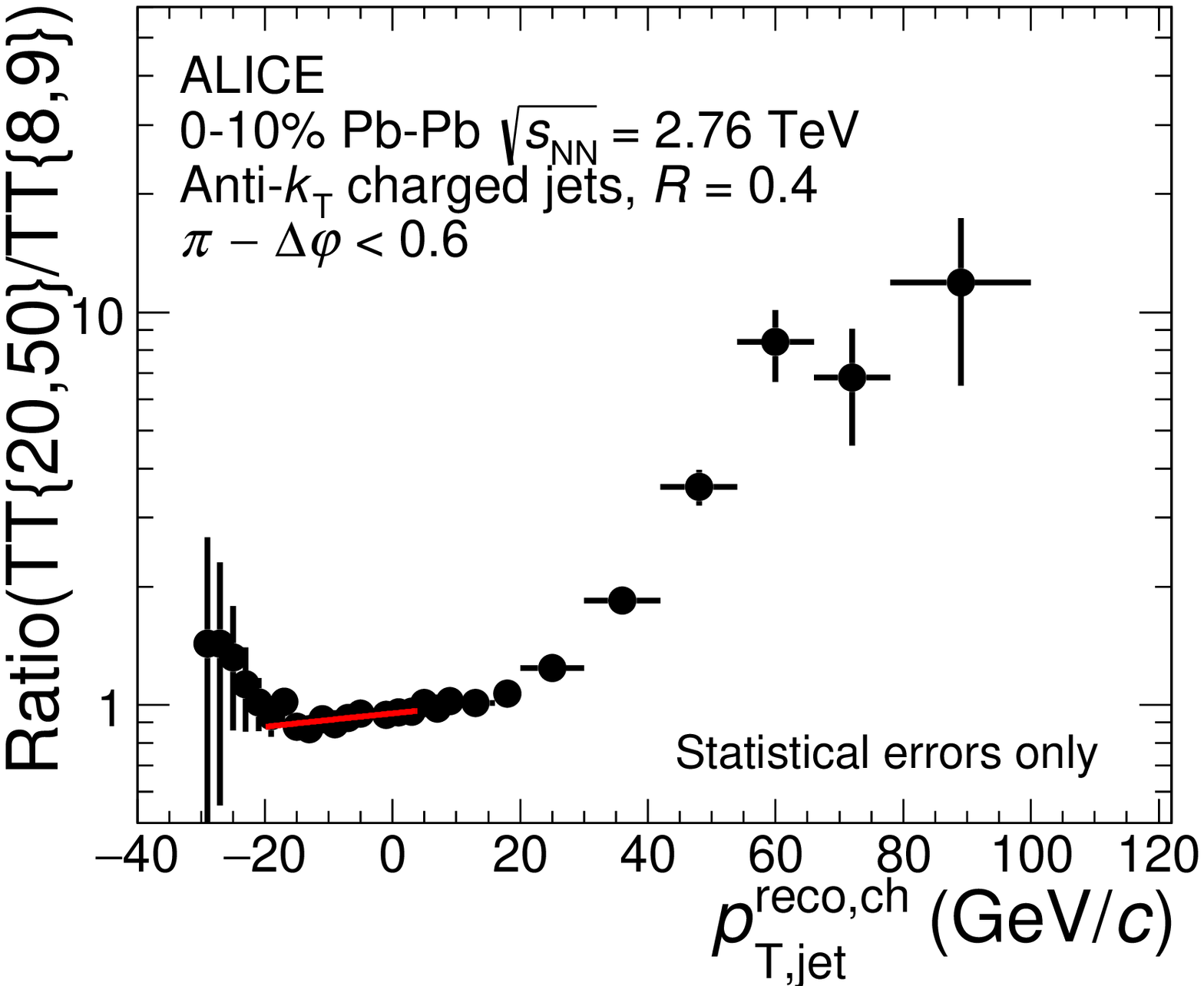}
\includegraphics[width=0.45\textwidth]{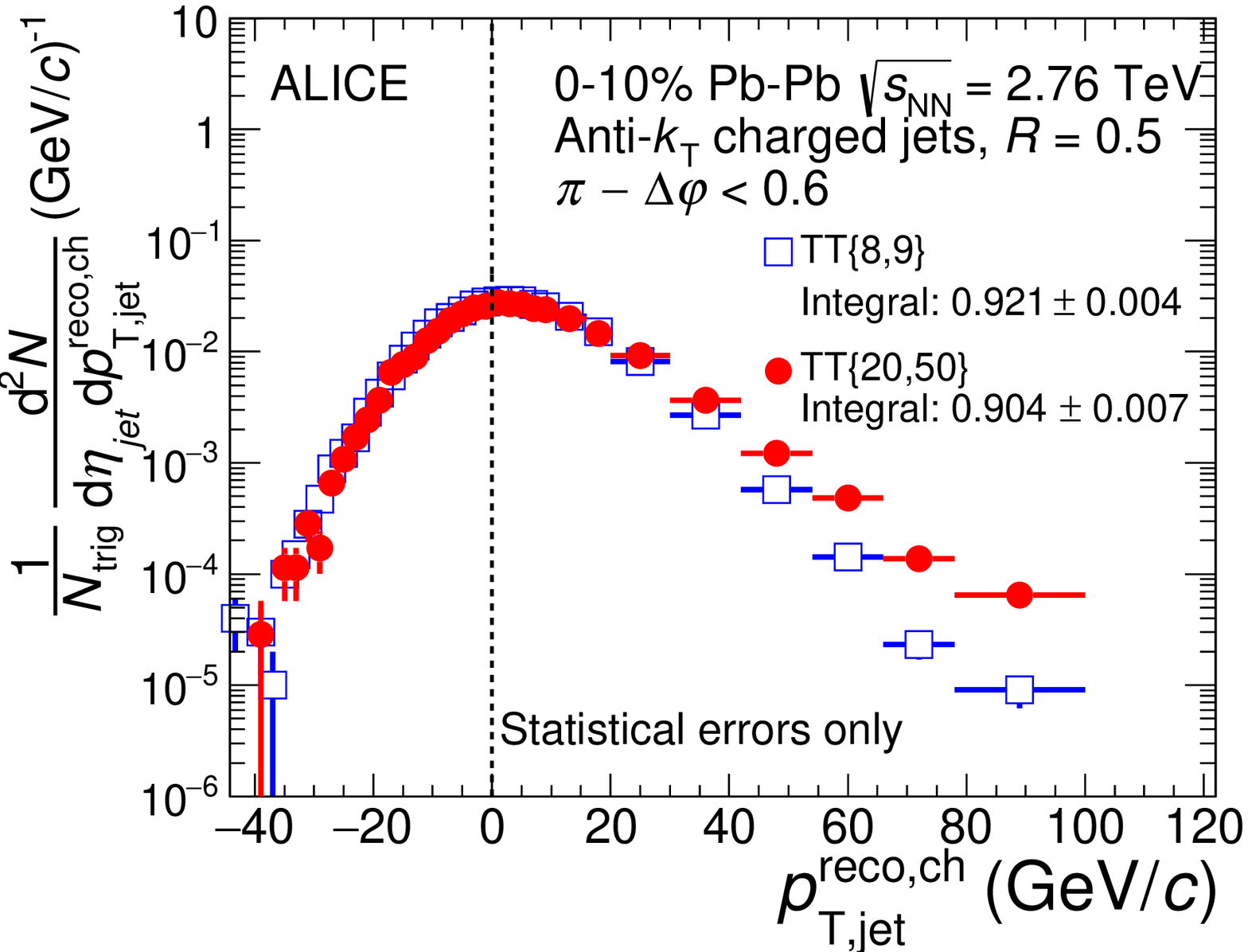}
\includegraphics[width=0.45\textwidth]{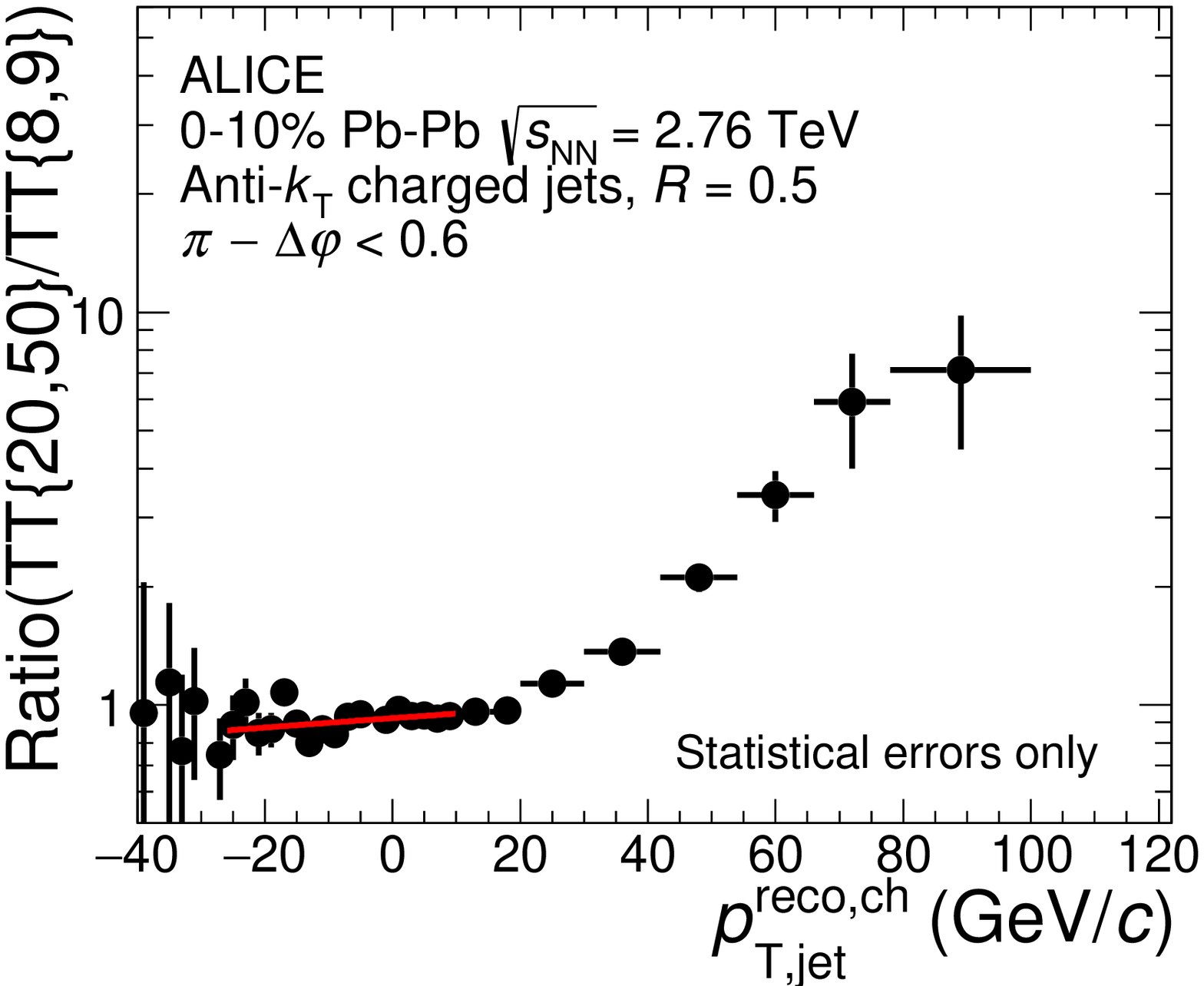}
\caption{Uncorrected trigger-normalized recoil charged jet
  distributions for central \PbPb\ collisions, with Signal TT\{20,50\}
  and Reference TT\{8,9\}. Jets are reconstructed with the \antikT\
  algorithm, constituent $\pTconst>0.15$ \gev, and \rr\ = 0.2, 0.4,
  and 0.5. Left: individual spectra. Right: their ratios. The red line
  shows a linear fit in the indicated region. Error bars show
  statistical errors only. Left-middle panel (\rr\ = 0.4) is identical
  to right panel in Fig.\ \ref{fig:hJetRecoil}.}
\label{fig:hJetRecoil2050}
\end{figure}

Figure \ref{fig:hJetRecoil2050}, left panels, show uncorrected
trigger-normalized recoil jet distributions for \rr\ = 0.2, 0.4 and
0.5 for both Signal and Reference TT classes. The right panels show
the ratio of Signal and Reference distributions for each value of
\rr. The error bars in Fig.\ \ref{fig:hJetRecoil2050} are statistical
only, and are dominated by the statistics of the recoil jet yield in
all cases. The statistical error due to trigger hadron yield is
negligible.

Figure \ref{fig:hJetRecoil2050}, left panels, also show the integrals
of the distributions. The integrals are seen to be insensitive to TT
class for a given \rr, with variations at the percent level or
smaller, while the value of the integral depends strongly on
\rr. These features are consistent with the geometric interpretation
of the integral given in Sect. \ref{sect:RecoilJetDistributions}.

\begin{center} 
\begin{table}[h]
\centering
\begin{tabular}{ |c|c|c|c| } \hline
 \rr & \pTreco\ fit range (\gev) & Constant
 (\cRef\ in Eq. \ref{eq:DRecoil}) &
 Slope  (\gev)$^{-1}$\\ \hline \hline
0.2  & [-12, 2] & $0.99\pm0.01$ & $0.004\pm0.002$ \\ 
0.4 & [-20, 4] & $0.96\pm0.01$ & $0.004\pm0.001$ \\
0.5 & [-26, 10] & $0.93\pm0.01$ & $0.002\pm0.001$ \\ \hline
\end{tabular}
\caption{Parameters from linear fits to ratios shown in right panels
  of Fig.\ \ref{fig:hJetRecoil2050}.}
\label{tab:SpecRatioFit}
\end{table}
\end{center}

Table \ref{tab:SpecRatioFit} shows the parameters resulting from the
fit of a linear function to the ratios in the right panels of Fig.\
\ref{fig:hJetRecoil2050}, in the region of \pTreco\ where the
distributions are largely uncorrelated with TT class. The constant
term of the fit, \cRef, is less than unity by a few percent, while the
slopes exceed zero by about 2 $\sigma$. The individual distributions
vary by three orders of magnitude in this region.

A value of \cRef\ below unity arises because the higher TT class has a
larger rate of true coincident recoil jets, and the integrals of the
distributions are largely uncorrelated with TT class. Larger yield at
positive \pTreco\ consequently depletes the yield at negative and small
positive values of \pTreco.

Accurate subtraction of the uncorrelated component from the Signal TT
distribution therefore requires scaling of the Reference TT
distribution by \cRef, as indicated in Eq. \ref{eq:DRecoil}\ and
Eq. \ref{eq:DRecoildphi}. Scaling of the Reference TT distributions in
the \pp\ analysis by \cRef\ has negligible effect. 

\begin{figure}[tbh!f]
\centering
\includegraphics[width=0.45\textwidth]{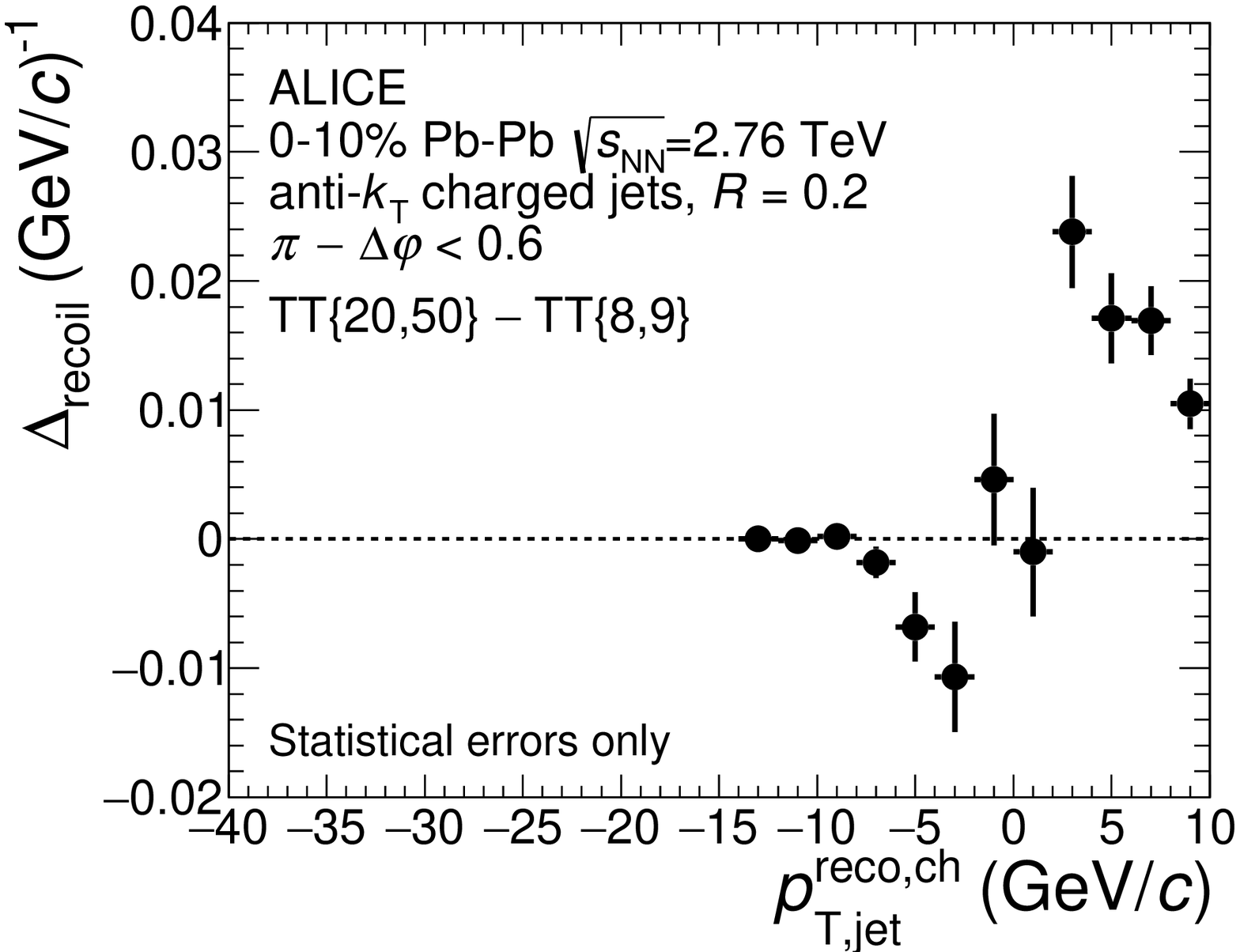}
\includegraphics[width=0.45\textwidth]{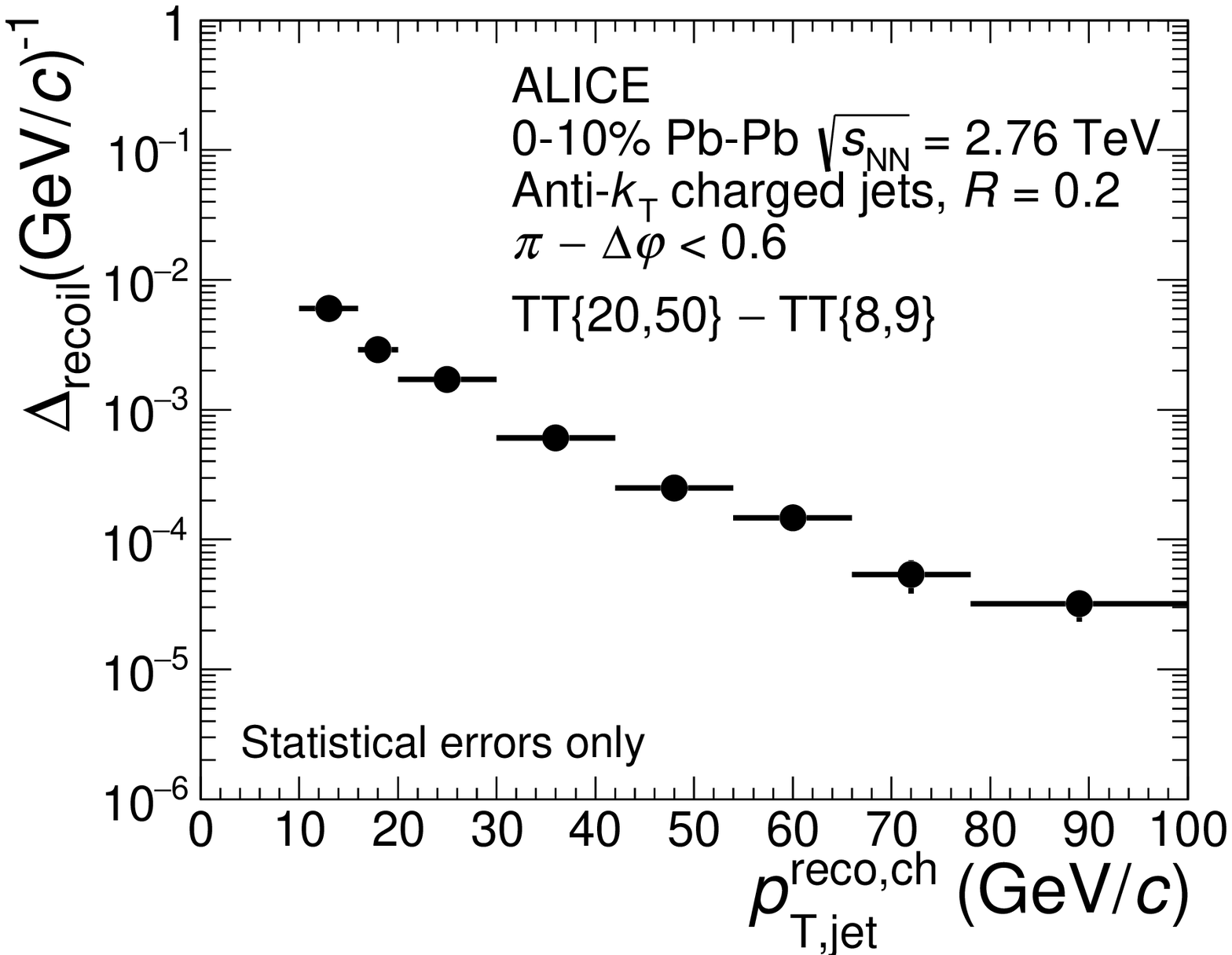}
\includegraphics[width=0.45\textwidth]{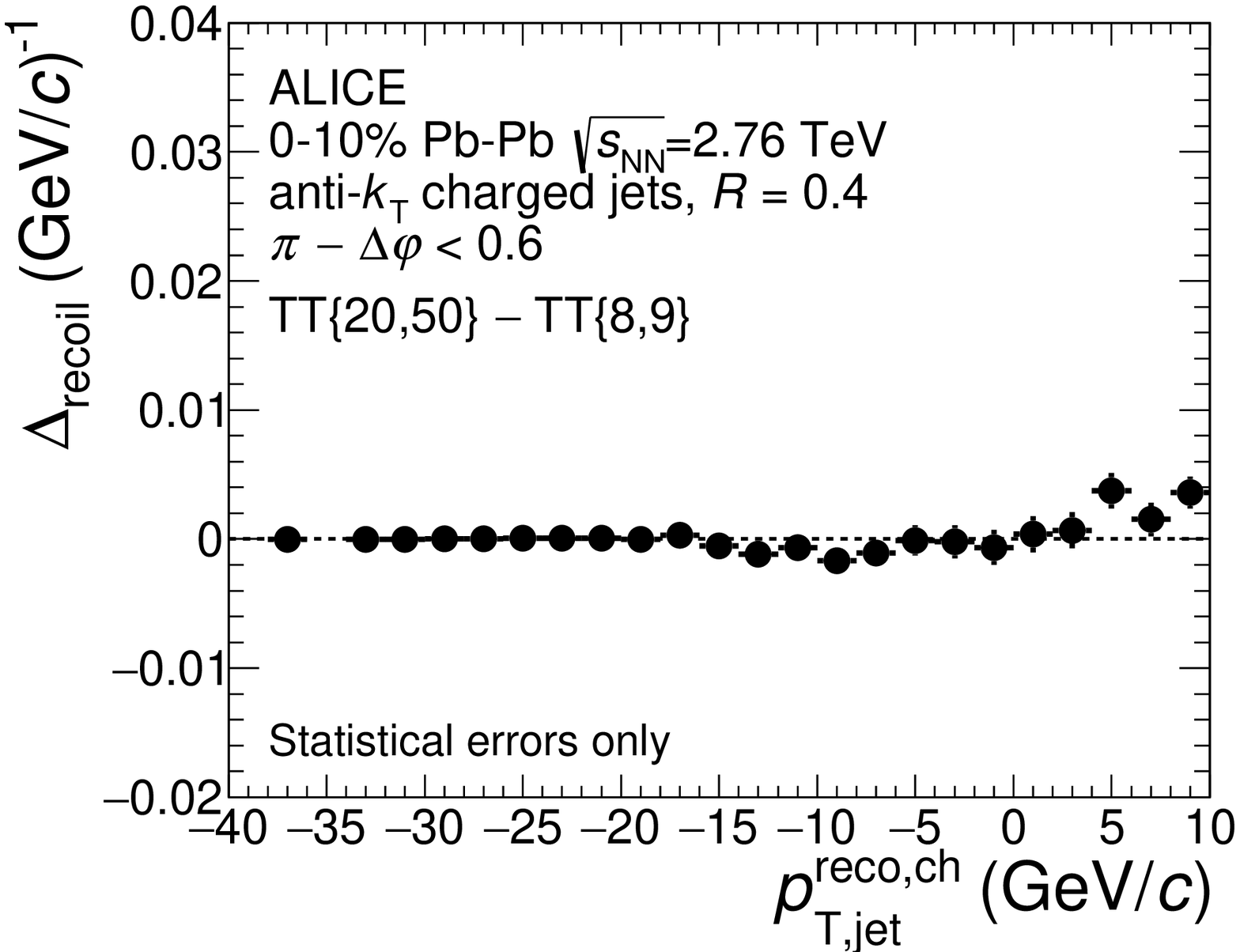}
\includegraphics[width=0.45\textwidth]{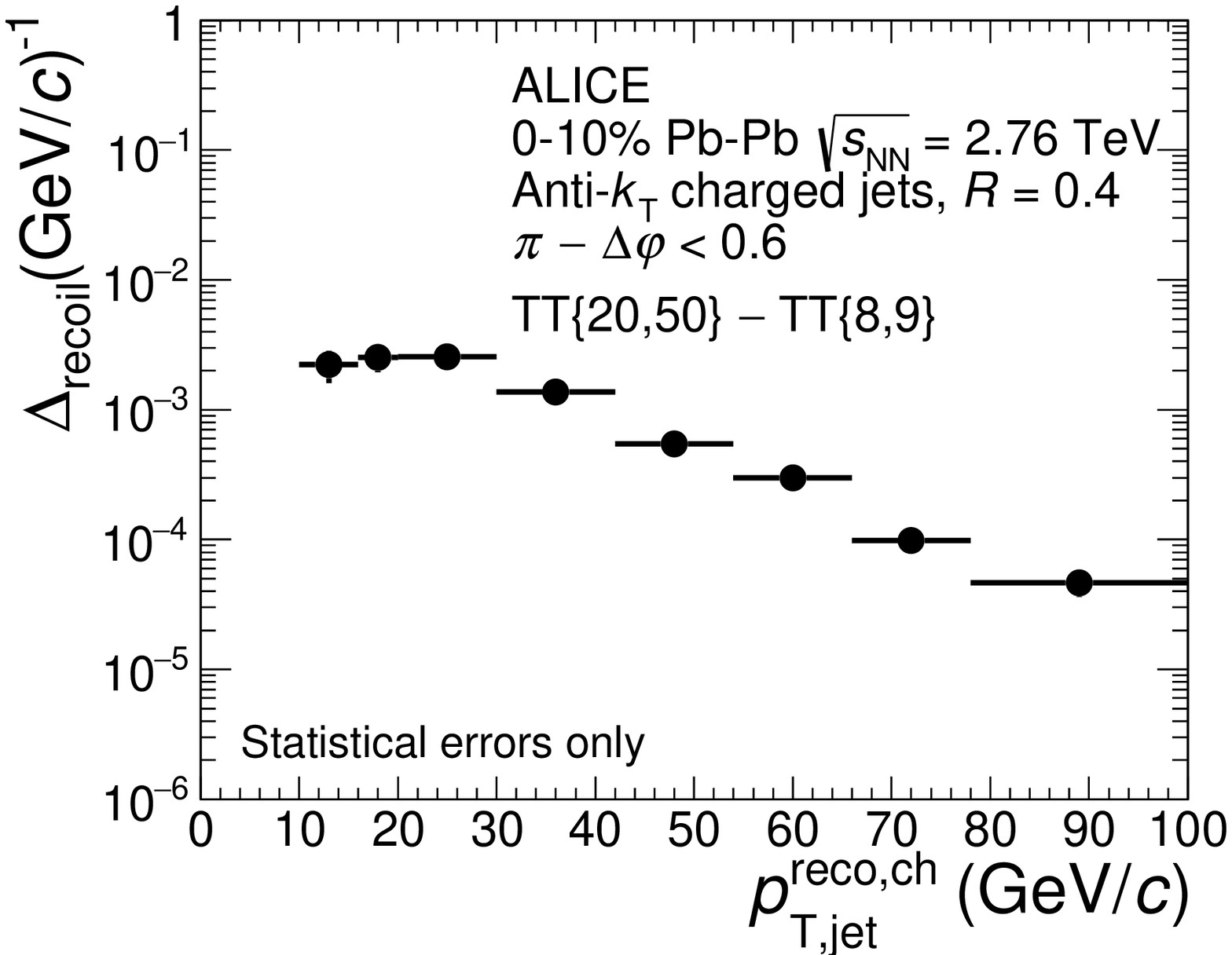}
\includegraphics[width=0.45\textwidth]{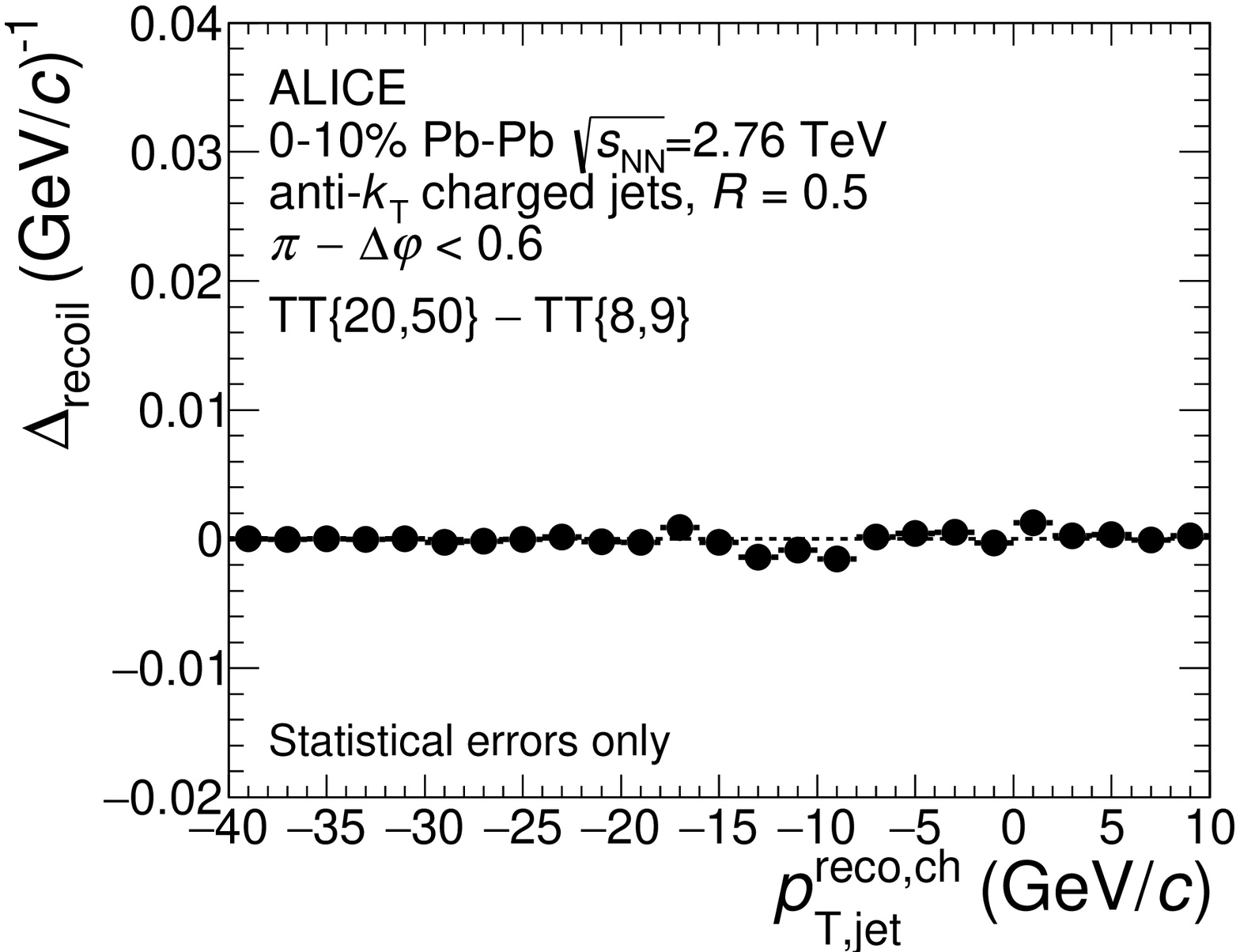}
\includegraphics[width=0.45\textwidth]{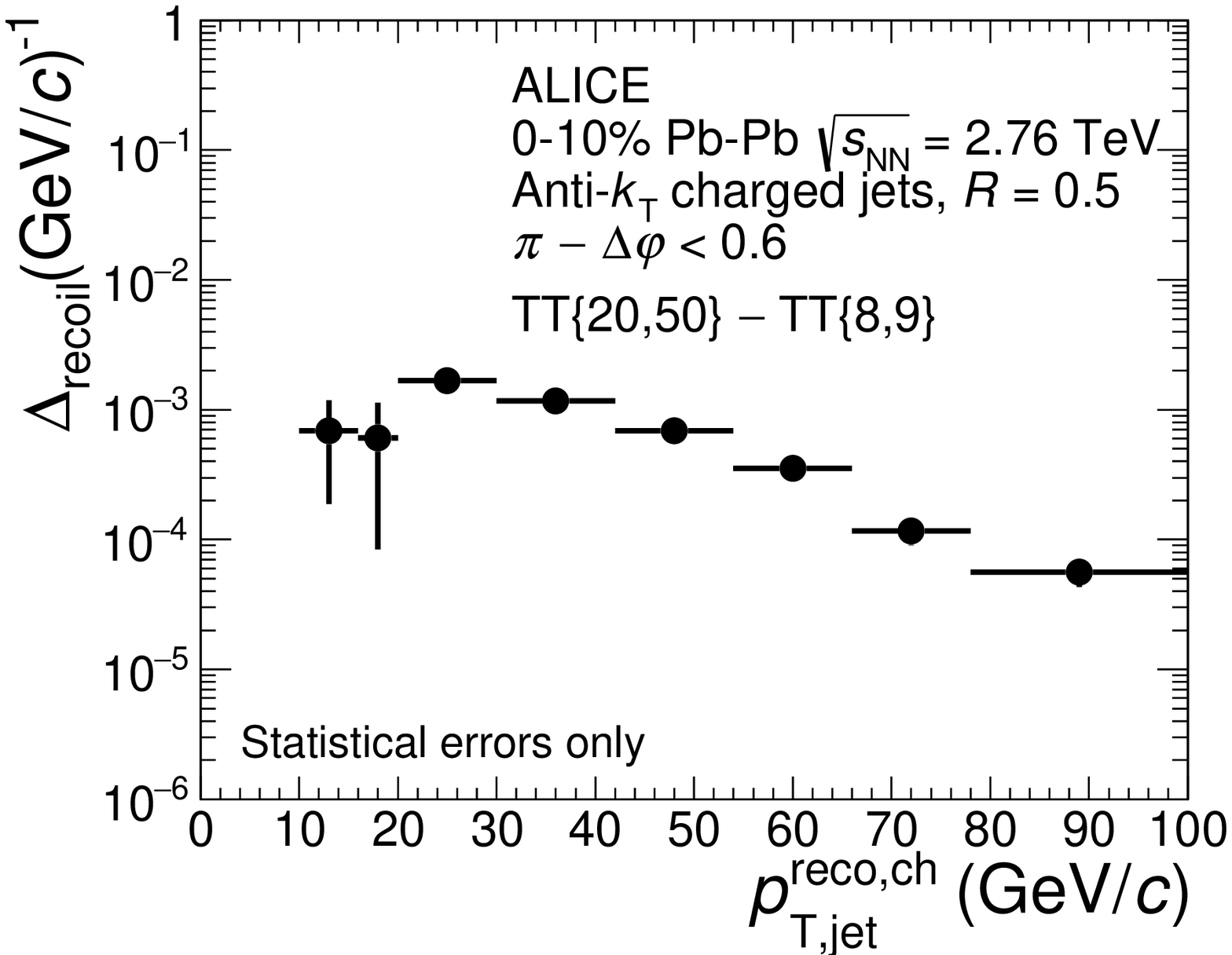}
\caption{Distribution of raw \Drecoil\ for \rr\ = 0.2, 0.4, and 0.5,
  measured in central \PbPb\ collisions for Signal TT class \{20,50\}
  and Reference TT class \{8,9\}. Left panels: \pTreco\ range of fit
  to extract \cRef, with linear vertical scale. Right panels: \pTreco\
  range above fit region, with logarithmic vertical scale. Error bars
  show statistical errors only.}
\label{fig:DRecoilbR5TT2050}
\end{figure}

Figure \ref{fig:DRecoilbR5TT2050} shows \Drecoil\ distributions for
\rr\ = 0.2, 0.4 and 0.5. The left panels, which have linear vertical
scale, show \Drecoil\ in the region of \pTreco\ in which the scale
factor \cRef\ is determined. \Drecoil\ is seen to be consistent with
zero over the entire fitting range. These panels also show the close
similarity of the shapes of the Signal and Reference distributions in
this region.

Figure \ref{fig:DRecoilbR5TT2050}, right panels, show \Drecoil\ at
positive \pTreco, where the Signal and Reference
distributions diverge. This is the ensemble-averaged distribution of
the trigger-correlated differential jet yield, but with measured
\pTreco\ not yet corrected for instrumental effects and
fluctuations of the underlying event background. 

\subsection{\Drecoilphi\ and \Dyieldthresh}
\label{sect:AngleDepRaw}

The analysis of \Drecoilphi\ (Eq. \ref{eq:DRecoildphi}) and
\Dyieldthresh\ (Eq. \ref{eq:Dyield}) is the same as that for \Drecoil\
in terms of event selection, track cuts, jet reconstruction, and jet
candidate selection. For this analysis we only consider jets
with \rr\ = 0.4 and $40<\pTreco<60$ \gev.

\begin{figure}[tbh!f]
\centering
\includegraphics[width=0.49\textwidth]{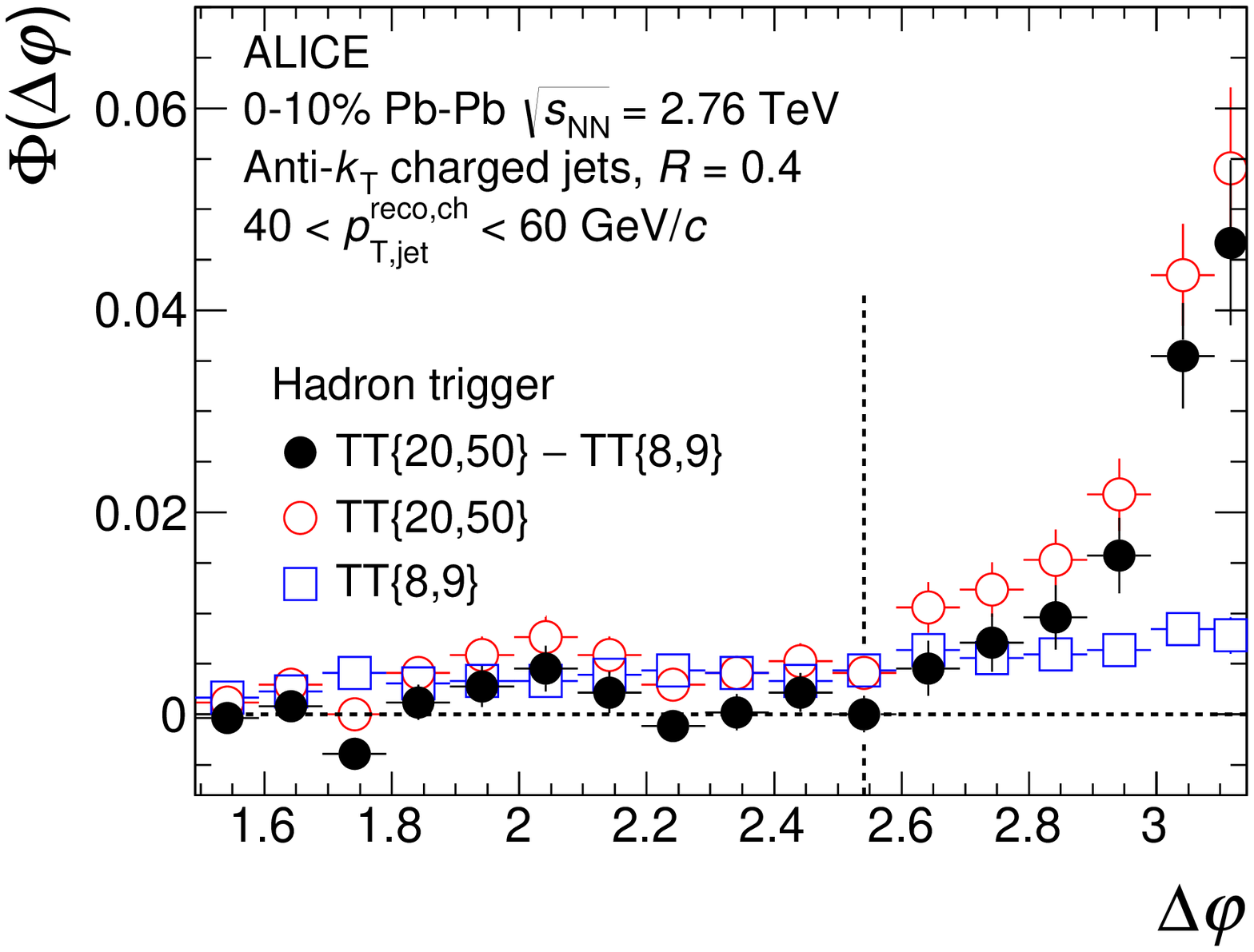}
\includegraphics[width=0.49\textwidth]{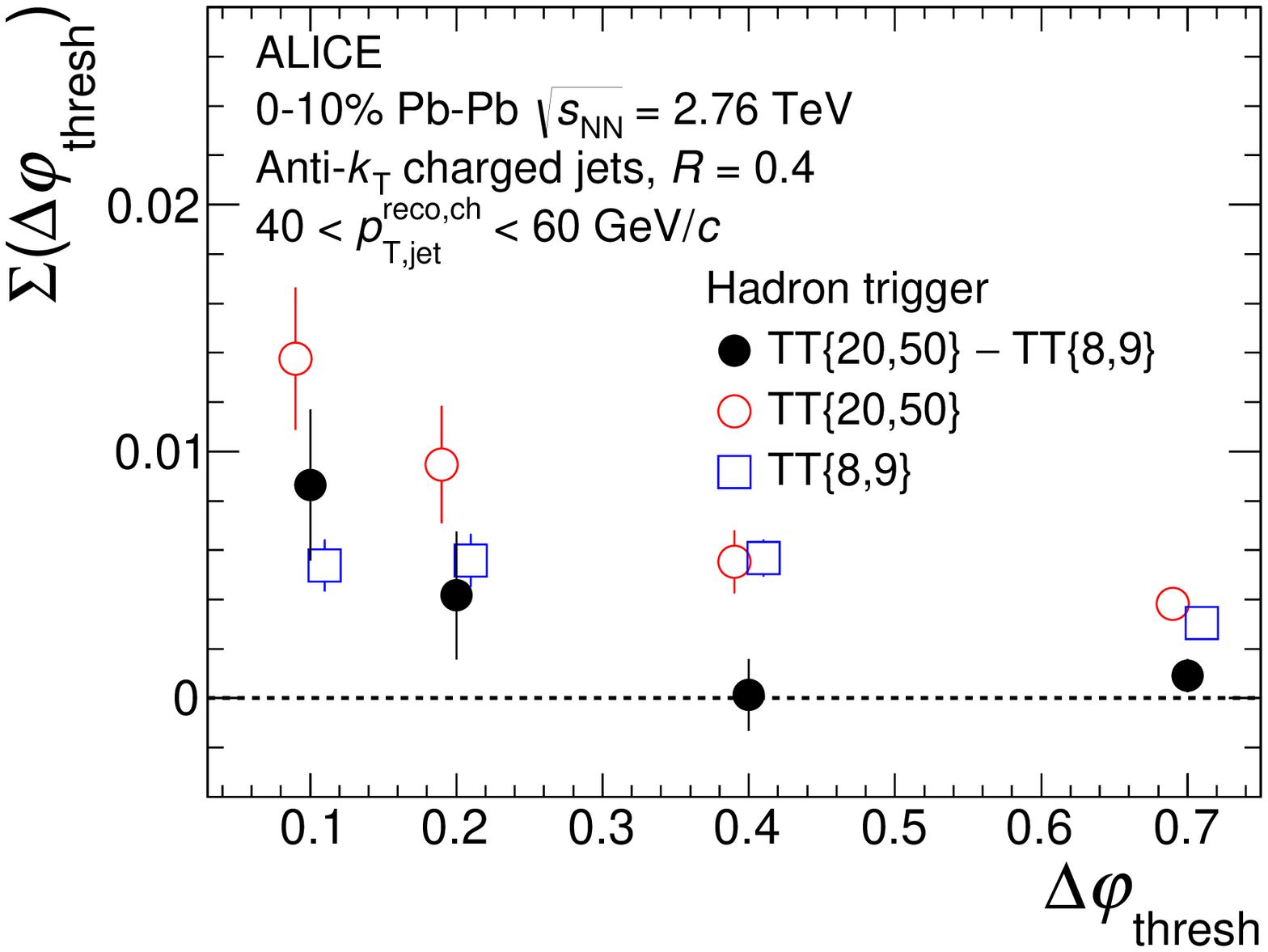}
\caption{\Drecoilphi\ (left) and \Dyieldthresh\ (right) distributions
  in central \PbPb\ collisions for TT\{20,50\} (open red circles),
  TT\{8,9\} (open blue boxes), and TT\{20,50\}-TT\{8,9\} (filled black
  circles), for jets with $40<\pTreco<60$ \gev. All error bars are
  statistical only. The vertical dashed line in the left panel indicates the
  acceptance limit for the \Drecoil\ measurement. Points in the right
  panel are displaced horizontally for clarity.}
\label{fig:data-dphi-raw}
\end{figure}

Figure \ref{fig:data-dphi-raw}, left panel, shows the distributions
of \Drecoilphi\ for TT\{8,9\} and TT\{20,50\} individually, and for
TT\{20,50\}-TT\{8,9\}, illustrating the effect of the subtraction. 

Figure \ref{fig:data-dphi-raw}, right panel, shows the raw
distribution of \Dyieldthresh, likewise for TT\{8,9\} and TT\{20,50\}
individually, and for TT\{20,50\}-TT\{8,9\}. Since \Dyieldthresh\ is an
integral over \dphi\ beyond a specified threshold, care must be taken
to ensure statistical independence of measurements for different
values of the threshold. Each point in Fig.\ \ref{fig:data-dphi-raw},
right panel, is therefore generated from an exclusive subset of the
data, with 10\% of the data used for threshold values 0.1 and 0.2,
20\% for 0.4, and 60\% for 0.7.  Subsets of unequal size are chosen to
optimize the statistical errors.

Due to the limited statistical precision of the data, correction of
the raw distributions in Fig.\ \ref{fig:data-dphi-raw} via unfolding
for background fluctuations and instrumental effects is not
possible. In order to compare the \PbPb\ distributions with a reference
distribution for \pp\ collisions, we therefore impose the effects of
instrumental response and \PbPb\ background fluctuations on the
distribution calculated by PYTHIA for \pp\ collisions at \sqrts\ =
2.76 TeV. The instrumental response, modeled by GEANT, is dominated by
tracking efficiency and momentum resolution. The effects of background
fluctuations are modeled by embedding detector-level PYTHIA events
into real \PbPb\ events. Recoil jets are reconstructed from these
hybrid events, using the same procedures as real data analysis.

\section{Corrections to {\bf\Drecoil} distributions}
\label{sect:DrecoilCorr}

Corrections to the raw \Drecoil\ distributions for underlying event
fluctuations and instrumental response are carried out using unfolding
methods \cite{Cowan:2002in,Hocker:1995kb}, in which
the true jet distribution $T$ is determined from the measured
distribution $M$ using a response matrix. We denote by \Rtot\ the
response matrix that incorporates all corrections, due  to underlying
event fluctuations and to instrumental response.  \Rtot\ maps
$T(\pTgen)$ to $M(\pTrec)$,

\begin{equation}
M(\pTrec) = \Rtot\Rargs \times T(\pTgen),
\label{eq:foldequation}
\end{equation}

\noindent
where \pTgen\ is the particle-level charged-jet \pT\ and \pTrec\ is
the detector-level or reconstructed jet \pT.

Precise inversion of Eq. \ref{eq:foldequation} for non-singular
\Rtot\ may result in a solution with large fluctuations in central
values and large variance, arising from statistical noise in $M(\pTrec)$
\cite{Cowan:2002in}. Inversion of Eq. \ref{eq:foldequation} to obtain
a physically interpretable solution is achieved via regularized
unfolding, which imposes the additional constraint of smoothness on
the solution.

Input to the unfolding procedure uses jets in the range
$20<\pTrec\ <100$ \gev. The distribution in Eq. \ref{eq:DRecoil}
provides a natural cutoff at low \pTrec, where the difference between
central values of Signal and Reference distributions is smaller than
the statistical error of the difference, so that imposition of a lower
bound in this range is strictly speaking not required. However, in
practice it was found that imposition of a lower bound at \pTrec\ = 20
\gev, which is above the LO cutoff in terms of charged jet \pTrec, is
needed for stable unfolding. This bound was kept as low as possible,
to retain as much correlated signal as possible. The upper bound is
set by the requirement that the highest \pTrec\ bin has at least 10
counts.

Correction for loss of jet yield in the excluded regions is carried
out by applying a \pTgen-dependent efficiency \EffKin, which is
determined using PYTHIA simulations. \EffKin\ is close to unity for
all \rr\ in the analysis, over most of the range $20 < \pTgen\ < 100$
\gev. Its value is \EffKin\ = 50\% at \pTgen\ = 20 \gev\ for all \rr,
due primarily to detector efficiency, and \EffKin\ = 70\% at \pTgen\ =
100 \gev\ for all \rr, due to the effects of momentum resolution and
background fluctuations. The jet finding efficiency is $95\%$ for
\pTpart\ = 20 \gev\ and 100\% for $\pTpart>40$ \gev, for all \rr.

For the \PbPb\ analysis, the primary unfolding algorithm is an
iterative procedure based on Bayes' Theorem~\cite{D'Agostini:1994zf},
as implemented in the RooUnfold software package\cite{RooUnfold}.
Regularization is imposed by requiring only small variation between
successive iterations, which occurs typically after three iterations.
Closure of the unfolding procedure for \Drecoil\ was tested in model
studies with correlated spectrum and background fluctuations similar
to those of this measurement \cite{deBarros:2012ws}. An alternative
unfolding algorithm, regularized Singular Value Decomposition (SVD)
\cite{Hocker:1995kb}, was used to estimate the systematic
uncertainties. 

Both unfolding algorithms were also used for the \pp\ analysis. In
this case, the SVD algorithm was used to determine the central values,
while Bayesian unfolding is used to estimate the systematics. This was
found to be the optimal approach for the more limited statistics of
the \pp\ distributions.

Both unfolding methods require initial specification of a prior
distribution. For the \PbPb\ analysis, the prior is the \Drecoil\
distribution for \pp\ collisions at \sqrts\ = 2.76 TeV, calculated
using PYTHIA (Perugia 10 tune). For \pp\ collisions at \sqrts\ = 7
TeV, the prior is calculated using PYTHIA (Perugia 10 tune
\cite{Skands:2010ak}).

\subsection{Correction for instrumental response}
\label{sect:InstrCorr}

The procedures to correct the jet energy for instrumental effects
are the same as those described in \cite{Abelev:2013kqa}. The dominant
correction is due to tracking efficiency, with \pT\ resolution generating the
second-largest correction.

Corrections for instrumental effects are determined from simulations
of \pp\ collisions at \sqrts\ = 2.76 TeV generated by PYTHIA, together
with detailed detector simulations generated using GEANT followed by
event reconstruction. The lower tracking efficiency in central \PbPb\
collisions was modeled by randomly discarding additional
detector-level tracks. The additional rejection factor was determined
by comparing Hijing and Pythia efficiencies and corresponds to
$2$-$3\%$, with weak \pT\ dependence \cite{Adam:2015ewa}.

Jet reconstruction is carried out for each event at both the particle
and detector level. The instrumental response matrix, \Rdet, is
generated by associating particle-level and detector-level jets whose
centroids are close in $(\eta,\phi)$, following the procedure
described in \cite{Abelev:2013kqa}.

\subsection{Correction for background fluctuations}
\label{sect:BackgroundFluctuations}

The adjustment of reconstructed jet energy by the estimated background
density \rhoAi\ (Eq. \ref{eq:pTraw}) accounts approximately for
event-wise variation in the background level, which arises from
variation in multiplicity within the 0-10\% centrality percentile bin
\cite{Abelev:2012ej}. The jet energy scale of the \Drecoil\
distribution must still be corrected for
energy smearing, due to local background energy density fluctuations
relative to the median background density $\rho$.

Background fluctuations \dpT\ are measured by two techniques: the
Random Cone method (RC) \cite{Abelev:2012ej}, and a method in which
model jets are embedded into real events \cite{deBarros:2011ph}.  The
distribution of fluctuations in background energy for the RC method
has RMS = 4.35 \gev\ for \rr\ = 0.2, 9.9 \gev\ for \rr\ = 0.4, and 13
\gev\ for \rr\ = 0.5. The RC method is used for the central data
points, with the embedding method used to assess the systematic
uncertainty.

The calculation of $\rho$ (Eq. \ref{eq:rho}) requires algorithmic
choices that are not unique, notably the jet reconstruction algorithm
and the population of jets used for the median calculation. However,
calculation of the response matrix for unfolding of background
fluctuations incorporates the same set of choices. If all jet
candidates are retained, without rejection based on \pTrec, the effect
of any systematic shift in JES due to $\rho$ will be precisely
counterbalanced by a shift of the same magnitude but opposite sign in
the response matrix. This two-step JES correction, with event-by-event
jet energy adjustment for event pedestal \rhoAi\ followed by
ensemble-level unfolding of background fluctuations \dpT, will
consequently be independent of the specific algorithmic choices for
determining $\rho$.

In this analysis, the definition of $\rho$ for the first step excludes
the two hardest jet candidates from the median calculation
(Eq.\ref{eq:rho}), while in the second step only jet candidates with
$\pTrec>20$ \gev\ are used for unfolding. However, this rejection cut in the
second step induces an implicit dependence on the specific definition
of $\rho$. In order to assess this effect, the analysis was repeated
with an alternative definition of $\rho$, in which all jet candidates
are included in the median calculation in the first step.  No
significant differences were observed in the corrected recoil jet
spectra.

\subsection{Cumulative response matrix}
\label{sect:RespMatrix}

\begin{figure}[tbh!f]
\centering
\includegraphics[width=0.49\textwidth]{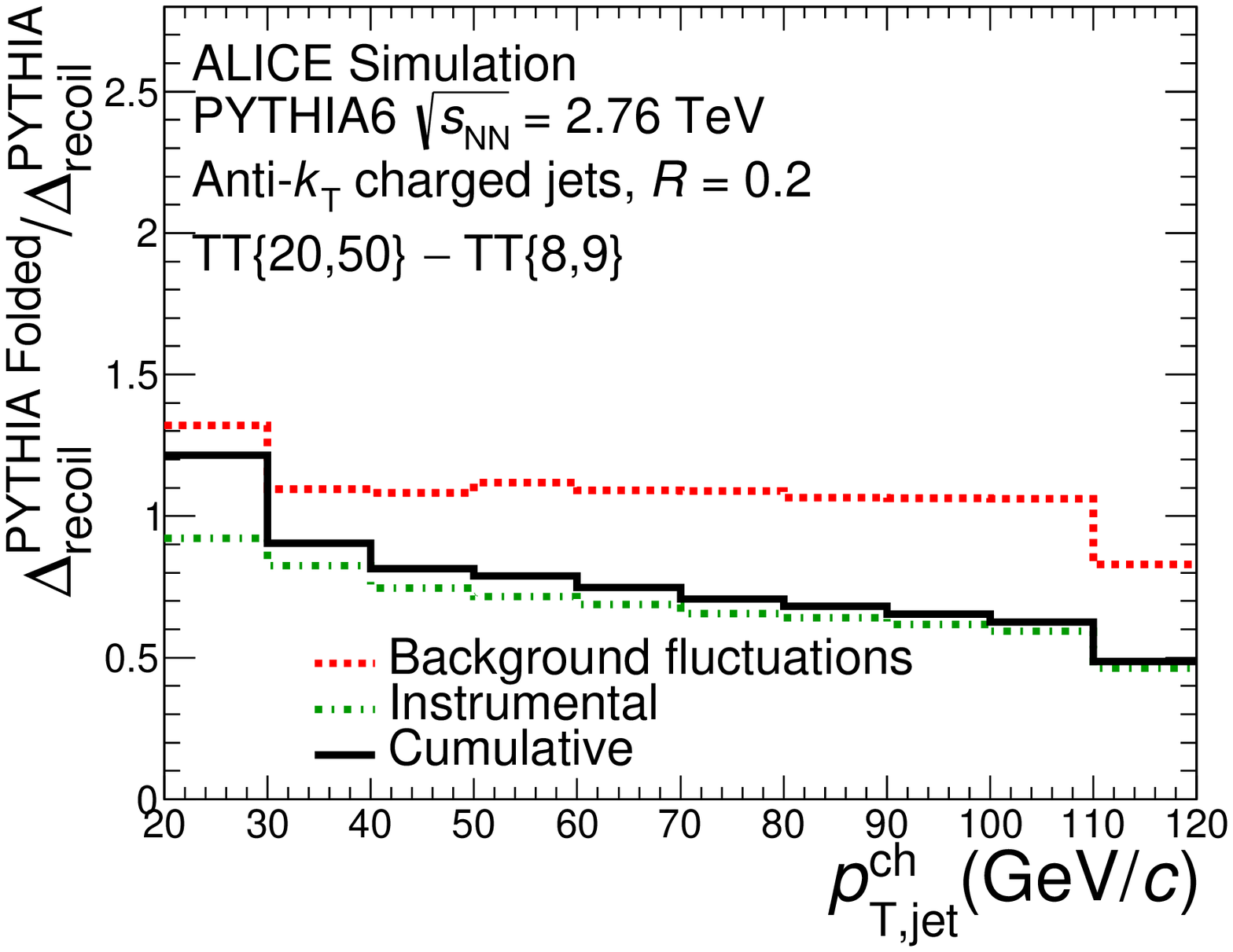}
\includegraphics[width=0.49\textwidth]{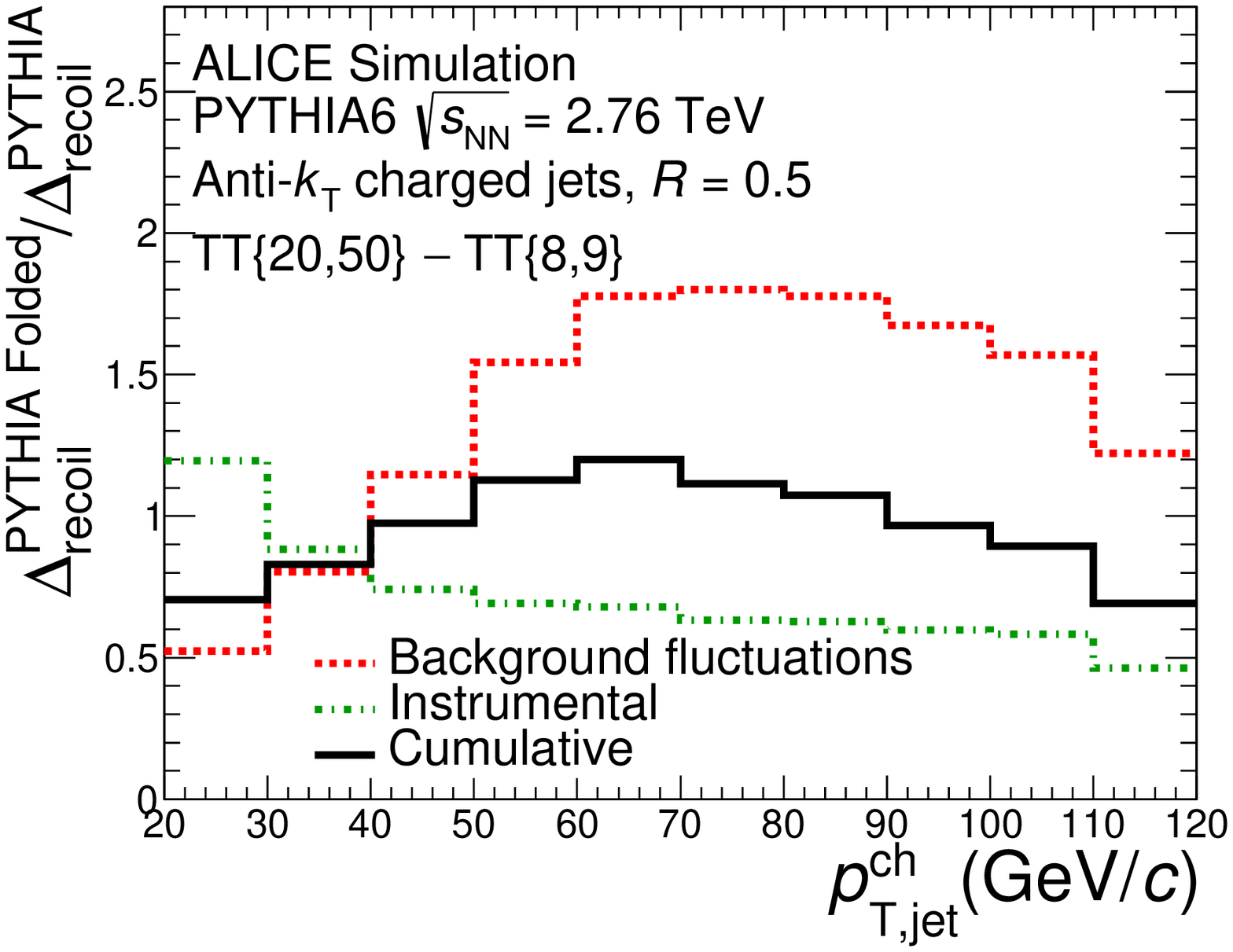}
\caption{Ratio of \Drecoil\ distributions for \pp\ collisions at
  \sqrts\ = 2.76 TeV, for \rr\ = 0.2 (left) and \rr\ = 0.5 (right). The
  numerator is convoluted with \Rdet\ and \dpT\ separately, and with
  \Rtot. The denominator is the unsmeared distribution.}
\label{fig:figfoldex}
\end{figure}

For the \PbPb\ analysis, the cumulative response matrix \Rtot\ is the
product of the response matrices for instrumental response and
background fluctuations. To illustrate the magnitude of corrections to
\Drecoil\ from unfolding the raw distributions with \Rtot, we
calculate the converse effect by convoluting the \Drecoil\
distribution for \pp\ collisions with \Rtot.  Fig. \ref{fig:figfoldex}
compares the particle-level \Drecoil\ distribution calculated using
PYTHIA for \pp\ collisions at \sqrts\ = 2.76 TeV for \rr\ = 0.2 and
0.5 with their convolution with \Rdet\ and \dpT\ separately, and
\Rtot. The figure shows the ratio of the convolution over the
unsmeared distribution. For \rr\ = 0.2, the effects of background
fluctuations are small, and the net effect of \Rtot\ is due primarily
to the instrumental response. For \rr\ = 0.5, the effects of
background fluctuations and instrumental response offset each other to
a large degree, with only a small net effect on the central values of
the distribution. The distributions for \rr\ = 0.4 are similar to
those for \rr\ = 0.5. Since the shape of the \Drecoil\ distribution is
similar in the \pp\ and \PbPb\ analyses, the corrections in the two
analyses will likewise be similar.

For the \pp\ analysis, only the instrumental response was corrected
using unfolding, i.e. \Rtot\ = \Rdet.

\subsection{Other effects}
\label{sect:OtherEffects}

In this section we discuss other effects that do not warrant
correction of the data.

Since this analysis is based on a semi-inclusive observable, with
normalization provided by the number of trigger hadrons measured
offline, correction for online trigger efficiency
(Sect. \ref{sect:Dataset}) is not required. No significant difference
in measured distributions was observed for events in the 0-8\% and
9-10\% centrality intervals.

Tracking efficiency at high-\pT\ is 80\% (Sect. \ref{sect:Dataset}),
so that 20\% of all trigger hadrons will not be observed. However,
this tracking efficiency is uniform over the \pTtrig\ range spanning
both the Reference and Signal TT classes, so the loss in trigger
statistics is unbiased in \pTtrig. Since the measurements are
trigger-normalized semi-inclusive distributions, the reduction in the
observed trigger hadrons corresponds simply to a loss of events, and
correction for this effect is not required.

Section \ref{sect:hJetDiscussion} presented considerations of
trigger-jet energy loss in the interpretation of these measurements. A
related but distinct effect is variation of \RAA, the suppression of
inclusive hadron yield in central \PbPb\ collisions, over the
\pTtrig-interval of the Signal TT bin
\cite{Abelev:2012hxa,CMS:2012aa}. Such a variation can generate
different hard-process selection bias for the same hadron trigger cuts
in \pp\ and \PbPb\ collisions, even if trigger-jet energy loss effects in \PbPb\ are
negligible. Using PYTHIA simulations, we estimate that this variation
may generate an increase in \Drecoil\ of 5\% at \pTjet\ = 20 \gev\ and
15\% at \pTjet\ = 100 \gev, but negligible change in the ratio of
\Drecoil\ in \PbPb\ with different \rr\ (see Fig.\
\ref{fig:RatioYield2050} and discussion below). Such effects will
however be included in theoretical calculations which incorporate
quenching and accurately reproduce the measured \pT-dependence of
inclusive hadron \RAA, and we do not correct the data for them.

High-\pTtrig\ hadron triggers above a fixed \pT\ threshold bias the
event population due to correlation with the Event plane (EP)
orientation, and bias towards more--central events.  Both effects will
bias the underlying event density and its fluctuations in the recoil
jet region. Note, however, that for the \pTtrig\ ranges of the Signal
and Reference TT classes in this analysis, the second-order EP
correlation amplitude \vtwo\ exhibits no significant variation with
hadron \pT\ ($\vtwo\approx{0.01}$ for $\pT>10$ \gev\
\cite{Chatrchyan:2012xq,Abelev:2012di}), and the centrality bias is
also invariant. The subtraction of the two distributions in \Drecoil\
and \Drecoilphi\ thereby removes the effect of such background biases to a
significant extent. Residual effects of these biases are assessed in
Sect.~\ref{sect:SysUncert}, and are included in the systematic
uncertainties.

Multiple, incoherent partonic interactions (MPI) can generate both a
trigger hadron and uncorrelated hard jets in the recoil acceptance, in
the same \PbPb\ collision. A recent analysis of $\gamma$-jet
coincidences corrected for this background using a mixed event
technique \cite{Chatrchyan:2012gt}. Since the rate of uncorrelated
hard interactions is by definition independent of \pTtrig, the
subtraction of the Reference from the Signal distribution in Eq. \ref{eq:DRecoil}
and Eq. \ref{eq:DRecoildphi} suppresses entirely the contribution of jet
candidates from all uncorrelated sources, including jets found in the
recoil acceptance arising from MPI. Correction for MPI effects is
therefore not required, for all observables considered in this
analysis.

\section{Systematic uncertainties}
\label{sect:SysUncert}

\subsection{Systematic uncertainties of {\bf\Drecoil}}
\label{sect:DrecoilUncert}

The systematic uncertainties for the distributions from \PbPb\
collisions are determined by varying parameters and algorithmic
choices in corrections for instrumental response and background
fluctuations. For the \Drecoil\ distribution for \pp\ collisions at
\sqrts\ = 7 TeV, systematic uncertainties are determined by varying
the corrections for the instrumental response.

The significant systematic uncertainties of the \Drecoil\
distributions in \PbPb\ collisions are as follows:

\begin{itemize}

\item Fit range for \cRef\ (Table~\ref{tab:SpecRatioFit}): variation
  of limits for fit generates a variation in \Drecoil\ of less than
  1\%;

\item Tracking efficiency: variation of \Rdet\ by changing the
  tracking efficiency by 5\% generates a
  variation in corrected \Drecoil\ of 4\% at \pTjetch\ $\approx$ 20
  \gev\ and 15\% at \pTjetch\ $\approx$ 100 \gev;

\item Fragmentation model for instrumental response: determination of
  \Rdet\ using PYQUEN rather than PYTHIA. PYQUEN has large-angle
  radiation enabled and was tuned to LHC data
  \cite{Lokhtin:2011qq}. This gives a variation in \Drecoil\ of 2\% at
  $\pTjetch\approx20$ \gev\ and 13\% at $\pTjetch\approx100$ \gev;

\item Event plane and multiplicity bias: the trigger
    hadron yield and background fluctuation distributions are measured
    differentially in bins of azimuthal angle relative to the EP. The
    trigger hadron yield is found to be correlated with EP
    orientation, indicating non-zero elliptic flow. The response
    matrix is then obtained by weighting the azimuth-dependent
    background fluctuation distribution with the azimuth-dependent
    trigger hadron yield. The change in corrected jet yield with and
    without this weighting is less than 5\%. The effects of the
  multiplicity bias are negligible;

\item Background fluctuations: using embedding rather than
  the RC method to measure background fluctuations
  (Sect.~\ref{sect:BackgroundFluctuations}) generates differences in
  the corrected \Drecoil\ distribution of less than 10\%;

\item Variation in unfolding algorithm: termination of Bayesian
  unfolding after five rather than three iterations generates
  variations in \Drecoil\ of $\approx1\%$ over most of the measured
  range. SVD unfolding yields \Drecoil\ distributions that differ from
  the Bayesian-based corrected distributions by 1\%;

\item Choice of unfolding prior: for Bayesian-based unfolding, the
  alternative priors are the \Drecoil\ distribution for \pp\
  collisions at \sqrts\ = 2.76 TeV including a 10\% or 20\% relative
  energy shift, to model jet energy loss. For SVD unfolding, the
  alternative prior is the Bayesian-based unfolded distribution. The
  largest variation in the corrected \Drecoil\ is less than 6\% at all
  \pTjetch;

\item Spectrum binning and limits: variations of upper and lower
  spectrum limits generate variations in corrected \Drecoil\ of less
  than 3\% at low \pTjetch, with negligible variation at high
  \pTjetch. Variation in choice of binning generates changes in
  corrected \Drecoil\ that are less than 4\%.

\end{itemize}

The products of weak decays make negligible contribution to \pTjetch\,
because of the stringent track selection requirements of the analysis
and the low material budget of the ITS and TPC. The systematic
uncertainty of \Drecoil\ due to the contribution of secondary vertex
decays is less than 2\%.

\begin{center} 
  \begin{table}[h]
  \centering
  \begin{tabular}{|l|c|c|} \hline
    {\bf Systematic uncertainty} & \pTjetch\ = 25 \gev  & \pTjetch\ = 75 \gev \\ \hline \hline
    \multicolumn{3}{|l|}{\bf Correlated} \\ \hline
    \hspace{1.0cm}Scale Factor \cRef  & ($-$1,+0)\% & ($-$0,+0.1)\%  \\ 
    \hspace{1.0cm}Tracking efficiency & ($-$6,+6)\% & ($-$16,+16)\% \\ 
    \hspace{1.0cm}Fragmentation model & ($-$0,+2)\% & ($-$0,+14)\% \\ 
    \hspace{1.0cm}EP bias  & ($-$2,+0)\% & ($-$3,+0)\% \\ \hline
    \multicolumn{3}{|l|}{\bf Uncorrelated or shape}\\ \hline
    \hspace{1.0cm}Background fluctuations & ($-$0,+7)\% & ($-$0,+8)\% \\ 
    \hspace{1.0cm}Unfolding algorithm & ($-$0,+1)\% & ($-$0,+2)\% \\ 
    \hspace{1.0cm}Unfolding prior & ($-$5,+0)\% & ($-$11,+6)\%  \\ 
    \hspace{1.0cm}Spectrum limits and binning & ($-$1,+0)\% & ($-$2,+3)\% \\ \hline \hline
    {\bf Cumulative correlated uncertainty} & ($-$7,+6)\%  & ($-$17,+22)\%  \\ \hline
    {\bf Cumulative uncorrelated uncertainty} & ($-$5,+7)\%  & ($-$11, +11)\% \\ \hline
\end{tabular}
\caption{Relative systematic uncertainties of
  \Drecoil, for \rr\ = 0.4 jets and two values of \pTjetch\ for
  central \PbPb\ collisions at \sqrtsNN\ = 2.76 TeV,
  TT\{20,50\}-TT\{8,9\}. Uncertainties are expressed as negative and
  positive differences from the central values, with an entry of zero indicating negligible
  contribution. Uncertainties are classified as correlated and
  uncorrelated, as described in the text.}
\label{table:SysUncert}
\end{table}
\end{center}

Table \ref{table:SysUncert} presents the significant systematic
uncertainties for \rr\ = 0.4, at two values of \pTjetch. Uncertainties
are presented as the relative difference to the central values of the
corrected \Drecoil.  We distinguish between correlated systematic
uncertainties, arising from variations that generate a correlated
change in the magnitude of the spectrum, and uncorrelated (or
shape) uncertainties, arising from variations that preserve the
integral but generate a change in the shape of the spectrum. Cumulative
uncertainties are the quadratic sum of all correlated or
uncorrelated uncertainties. Uncertainties for \rr\ = 0.2 and 0.5
are evaluated in a similar way.

Similar systematic uncertainties were considered for the \Drecoil\ distribution in \pp\ collisions at
\sqrts\ = 7 TeV  as those discussed for \PbPb\
collisions. Uncertainty in tracking efficiency causes variation in
\Drecoil\ by $2\%$ at $\pTjet\ \approx 20$~\gev\ and $9\%$ at $\pTjet\
\approx 60$ \gev. The systematic uncertainty due to track momentum resolution is
 estimated to be 4\% in the entire \pT\ range. The shift of jet energy
 scale due to contamination by secondary particles and fake tracks
 causes a variation 
in \Drecoil\ of less than $2\%$. Variations in the unfolding procedure,
including change in the choice of unfolding algorithm, prior, and
spectrum 
binning, result  in  \Drecoil\ changes of $\approx 5\%$.  
The cumulative systematic uncertainty is given by the quadratic sum of all individual uncertainties.

\subsection{Systematic uncertainties of \Drecoilphi\ and \Dyieldthresh}
\label{sect:AngleDepSysUncert}
  
The systematic uncertainties for the measurement of \Drecoilphi\ and \Dyieldthresh\ are presented in this section.

\begin{itemize}
\item Since the yield of correlated hard jets decreases with
  increasing acoplanarity (i.e. increasing $|\dphi-\pi|$), the scale
  factor \cRef\ in this region should approach unity. To assess this
  effect, \cRef\ is varied from its nominal values in Table
  \ref{tab:SpecRatioFit} to unity. The resulting change in width of
  the uncorrected \Drecoilphi\ distribution (Eq.\ref{eq:dphifit}) is
  0.001 and the change in slope of the \Dyieldthresh\ ratio
  (Fig.~\ref{fig:paper-dphi-integral}) is 0.35, which
  are taken as the systematic uncertainties.

\item Tracking efficiency less than unity will result in jets that are
  reconstructed from a subset of their charged track constituents,
  with consequent variation of jet centroid. However, the 5\% relative
  uncertainty of tracking efficiency generates negligible variation in
  the width of the \Drecoilphi\ distribution and in the slope of the
  \Dyieldthresh\ ratio.

\item The EP bias due to the hadron trigger, discussed in
Sect.~\ref{sect:DrecoilUncert}, generates a change of $0.005$ in the width of
the \Drecoilphi\, and 0.07 in the slope of the \Dyieldthresh\ ratio.

\end{itemize}

\begin{table}[h] 
\centering
\begin{tabular}{|c|c|c|}\hline
Systematic uncertainty & Width of \Drecoilphi\ & Slope of \Dyieldthresh\ ratio\\
\hline \hline
Scale factor \cRef & $\pm$0.001 & $\pm$0.35 \\
Tracking efficiency & negligible & negligible \\
EP bias & $\pm$0.005 & $\pm$0.07 \\
\hline \hline
{\bf Cumulative uncertainty} & $\pm$0.005 & $\pm$0.36 \\
\hline
\end{tabular}
\caption{Systematic uncertainties for the width of
    the \Drecoilphi\ distribution (Eq.\ref{eq:dphifit}) and the slope
    of the \Dyieldthresh\ ratio (Fig.\ \ref{fig:paper-dphi-integral},
    right panel). The cumulative
    uncertainty is the quadratic sum of all contributions.}
\label{tab:sys-dphi}
\end{table}

Table \ref{tab:sys-dphi} shows all sources of systematic uncertainty
for \Drecoilphi\ and \Dyieldthresh, with the cumulative uncertainty
given by their quadratic sum. Instrumental effects generate
negligible uncertainty in the azimuthal correlations.

\section{Distributions for \pp\ collisions at \sqrts = 2.76 TeV}
\label{sect:ppReference}

As noted above, comparison of \PbPb\ measurements to similar
distributions in \pp\ collisions at \sqrts\ = 2.76 TeV requires
calculations based on PYTHIA and NLO pQCD. In order to validate this
approach, we compare PYTHIA and NLO pQCD-based calculations to ALICE
measurements of \Drecoil\ distributions in \pp\ collisions at \sqrts\
= 7 TeV, using the data shown in Fig. \ref{fig:hJetRecoil}.

The NLO pQCD-based framework was developed initially to calculate the
spin-dependent hadron-jet coincidence cross section at \sqrts\ = 200
GeV \cite{deFlorian:2009fw}. The calculation uses the DSS
fragmentation function \cite{deFlorian:2007hc} and the CT10 NLO parton
distribution function \cite{Lai:2010vv}.  We model hadronization by a
shift in \pT\ for parton-level jets \cite{Dasgupta:2007wa}, with the
magnitude of the shift determined by a fit to inclusive jet
distributions \cite{Soyez:2011np}. The resulting particle-level jet
distribution is transformed to a charged-jet distribution by applying
a response matrix calculated using PYTHIA. The systematic uncertainty
of the resulting spectrum is estimated by independently varying the
parton distribution function and the factorization and renormalization
scales by a factor two.

\begin{figure}[tbh!f]
\centering
\includegraphics[width=0.49\textwidth]{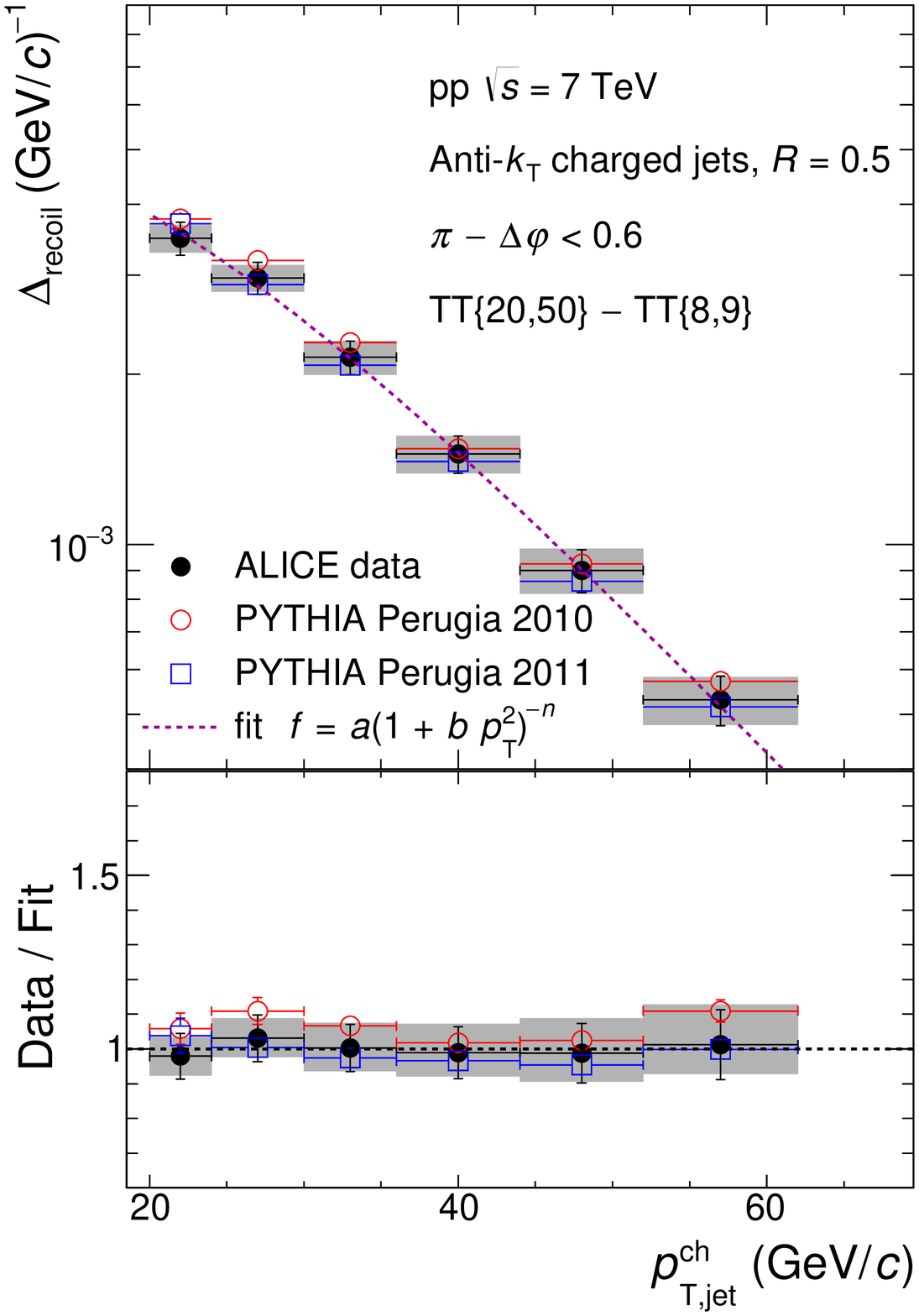}
\includegraphics[width=0.49\textwidth]{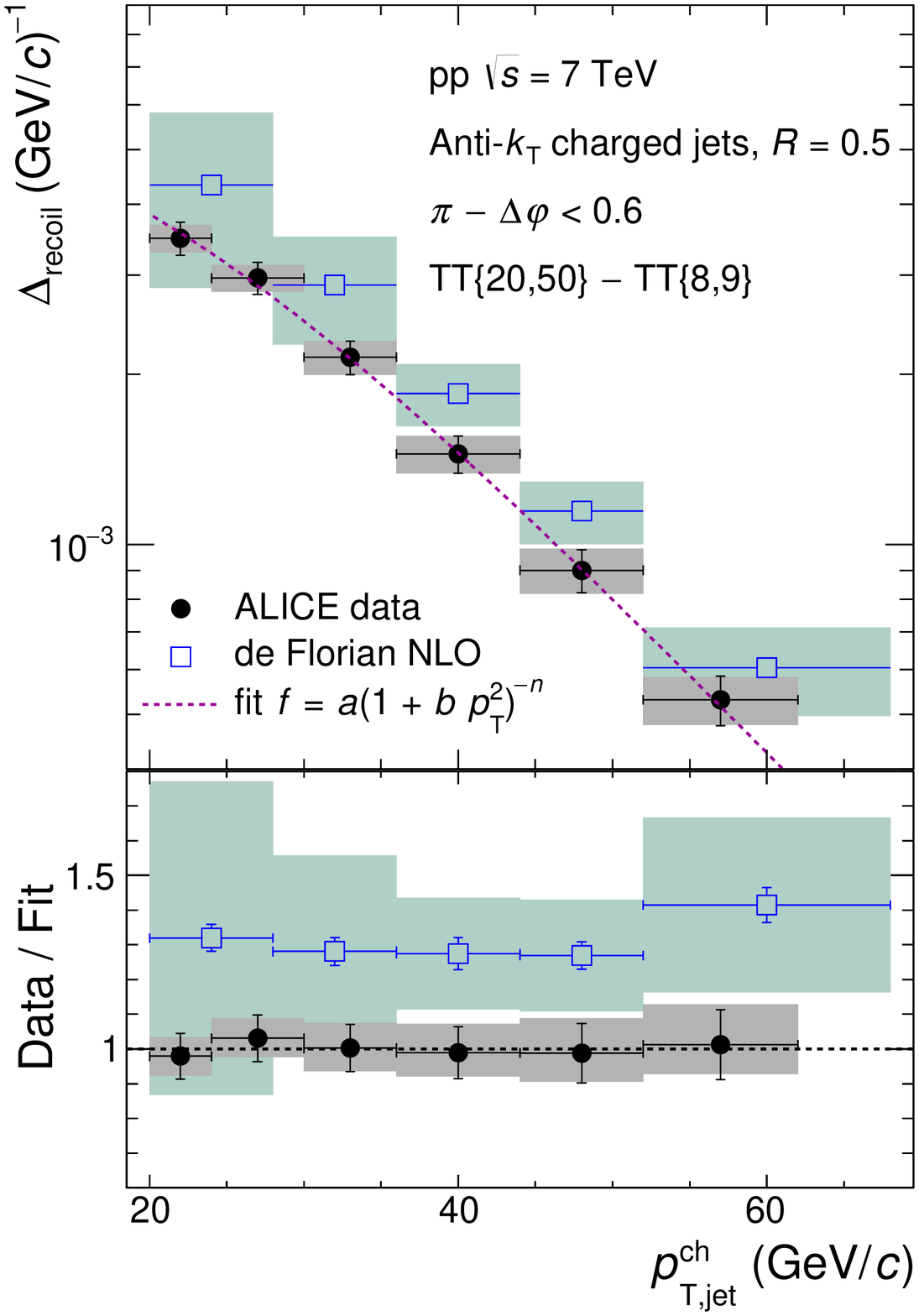}
\caption{\Drecoil\ distributions for \rr\ = 0.5, for ALICE data and
  calculations of \pp\ collisions \sqrts\ = 7 TeV.  ALICE data, which
  are the same in both panels, are compared with calculations based on
  PYTHIA (left) and NLO pQCD (right). The green boxes in the right
  panel show the systematic uncertainty of the NLO calculation. The
  lower panels show the ratios of data and calculations to a smooth
  function fit to the data.}
\label{fig:ppSpectraDataAndPythia}
\end{figure}

Figure~\ref{fig:ppSpectraDataAndPythia}, upper panels, show \Drecoil\
distributions for \rr\ = 0.5, from ALICE data and calculations for
\pp\ collisions at \sqrts\ = 7~TeV. The lower panels show the ratios
of these distributions to a function which parameterizes the ALICE
data. The PYTHIA calculations for both tunes agree with the
measurement within uncertainties; similar agreement is found for \rr\
= 0.2 and 0.4. The central values of the NLO calculation are above the
measured data by about $20\%$, though the calculation is consistent
with data within uncertainties for \rr\ = 0.5. The discrepancy in
central values is larger for smaller \rr, reaching about $50\%$ for
\rr\ = 0.2, which is not consistent within systematic
uncertainties. In pQCD calculations the difference between the parton
and jet momenta involves an expansion in terms of $\log(R)$, whose
contribution may be significant for small \rr\
\cite{Dasgupta:2014yra}. Improved agreement between the NLO
calculation and data for \rr\ = 0.2 may therefore be achievable using
resummation techniques \cite{Dasgupta:2014yra}.

\begin{figure}[tbh!f]
\centering
\includegraphics[width=0.49\textwidth]{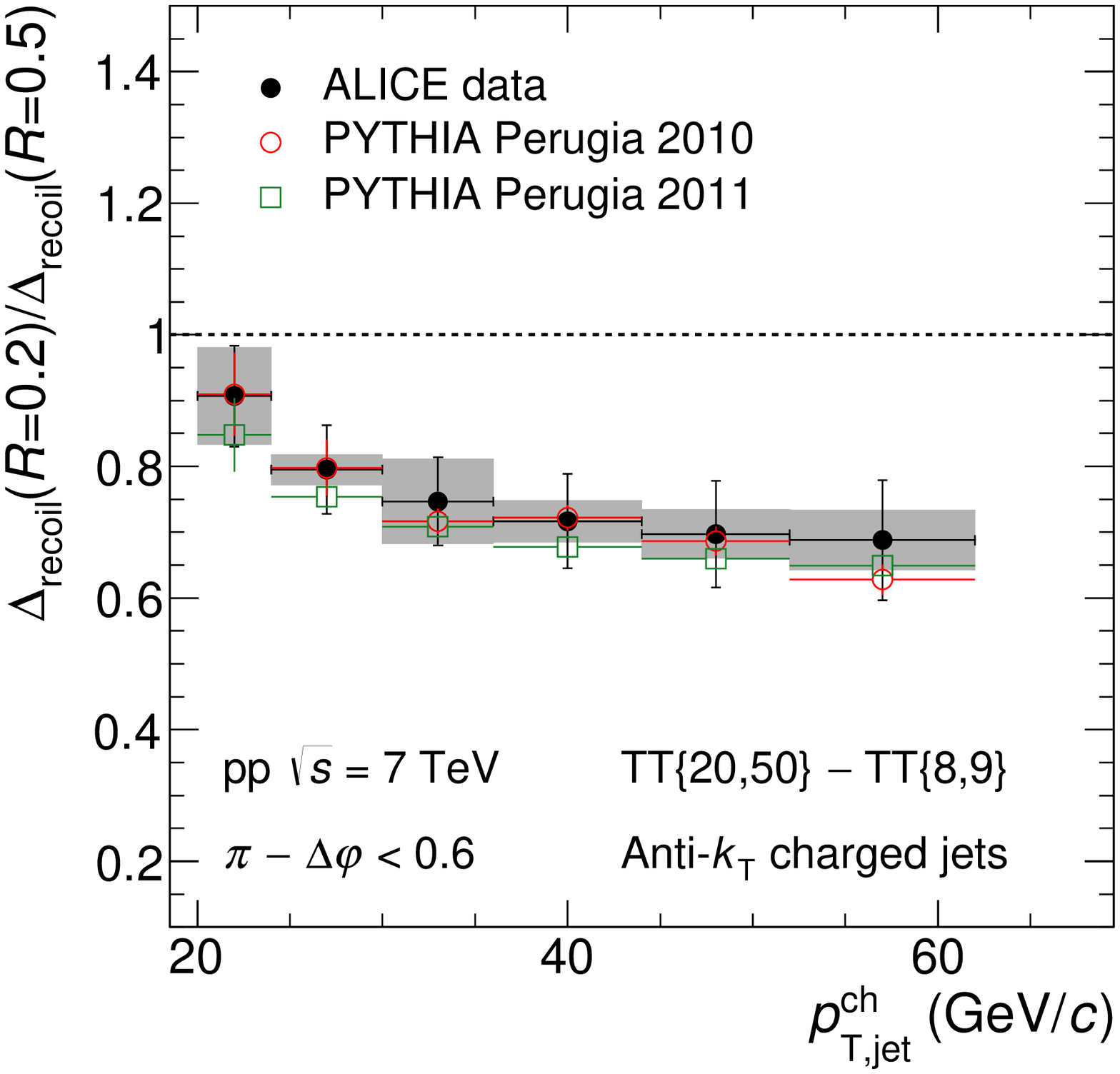}
\includegraphics[width=0.49\textwidth]{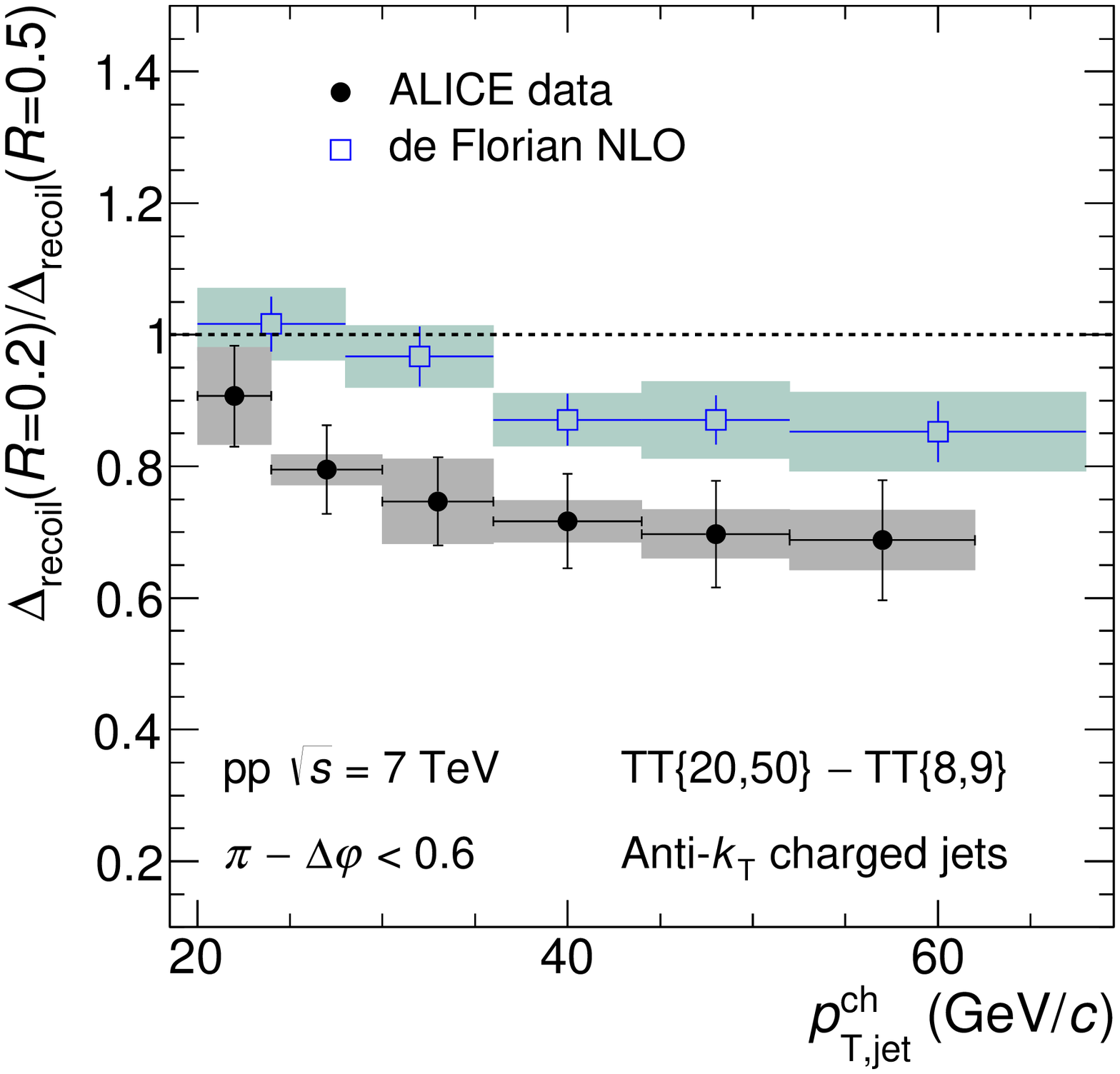}
\caption{Ratio of \Drecoil\ distributions with \rr\ = 0.2 and 0.5, for
  ALICE data and calculations from \pp\ collisions at \sqrts\ = 7
  TeV. ALICE data, which are the same in both panels, are compared
  with calculations using PYTHIA (left) and NLO pQCD (right).}
\label{fig:ppSpectraRatioAndPythia}
\end{figure}

Figure~\ref{fig:ppSpectraRatioAndPythia} shows the ratio of \Drecoil\
distributions for \rr\ = 0.2 and 0.5 in \pp\ collisions at \sqrts\ =
7~TeV.  The measured ratio is compared with PYTHIA and NLO pQCD-based
calculations. The grey boxes show the systematic uncertainty of the
measured ratio, taking into account correlations of numerator and
denominator.

The NLO calculation generates larger ratios than those observed in the
data. A related observable, the ratio of inclusive jet production
cross sections for \rr\ = 0.2 and 0.4 in \pp\ collisions at \sqrts\ =
2.76 TeV, has also been compared with pQCD calculations
\cite{Abelev:2013fn}.  This comparison shows that both hadronization
corrections and perturbative effects that are effectively
next-to-next-to-leading order (NNLO) in the individual cross sections
\cite{Soyez:2011np} are required for agreement. Perturbative QCD
calculations to higher order than NLO are also needed to describe the
ratio of \Drecoil\ distributions presented here.

In contrast, PYTHIA simulations agree within uncertainties with data,
both for \Drecoil\ at fixed \rr\ and the \Drecoil\ ratio for two
different values of \rr.  These comparisons therefore favor PYTHIA
calculations for the reference distributions at \sqrts\ = 2.76
TeV. PYTHIA combines a LO matrix element with a parton shower
resummation of leading logarithmic terms of soft gluon radiation at all
orders, leading to an improved description of data compared to a fixed
order analytic calculation.

\section{Results}
\label{sect:Results}

\subsection{\Drecoil}
\label{sect:ResultsDrecoil}

\begin{figure}[tbh!f]
\centering
\includegraphics[width=0.6\textwidth]{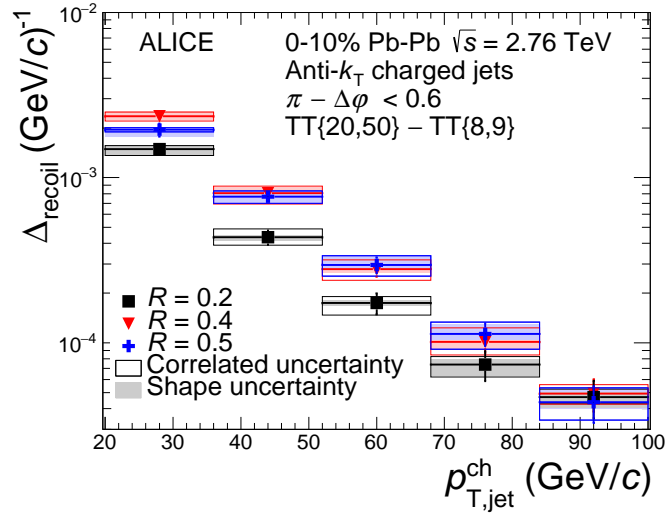}
\caption{\Drecoil\ for 0-10\% central \PbPb\ collisions at \sqrtsNN\ = 2.76
  TeV for \antikT\ jets with \rr\ = 0.2, 0.4 and 0.5. The vertical
  error bars are the square root of the diagonal elements of the unfolding covariance
  matrix, with the boxes indicating correlated and uncorrelated
  (shape) systematic uncertainties. }
\label{fig:DrecoilCorrected}
\end{figure}

Figure \ref{fig:DrecoilCorrected} shows corrected \Drecoil\
distributions for central \PbPb\ collisions, for \rr\ = 0.2, 0.4 and
0.5. The shape of the distributions is approximately exponential, with
larger per-trigger yield for \rr\ = 0.4 and 0.5 than for \rr\ = 0.2.

The \rr\ dependence of \Drecoil\ is related to the distribution of jet
energy transverse to the jet axis. Scattering of the parton shower
within the hot QCD medium may broaden this distribution
\cite{Wang:2013cia,Kurkela:2014tla}. The magnitude of intra-jet
broadening can be measured by comparing \Drecoil\ distributions for
\PbPb\ collisions with those for \pp\ collisions, in which jets are
generated in vacuum. We utilize two related observables for this
purpose: (i) \DIAA, which is the ratio of \Drecoil\ for \PbPb\ to that
for \pp\ collisions simulated using PYTHIA, for fixed \rr, and (ii)
the ratio of \Drecoil\ at two different \rr\ in \PbPb, compared with
that in \pp\ collisions.

\begin{figure}[tbh!f]
\centering
\includegraphics[width=0.6\textwidth]{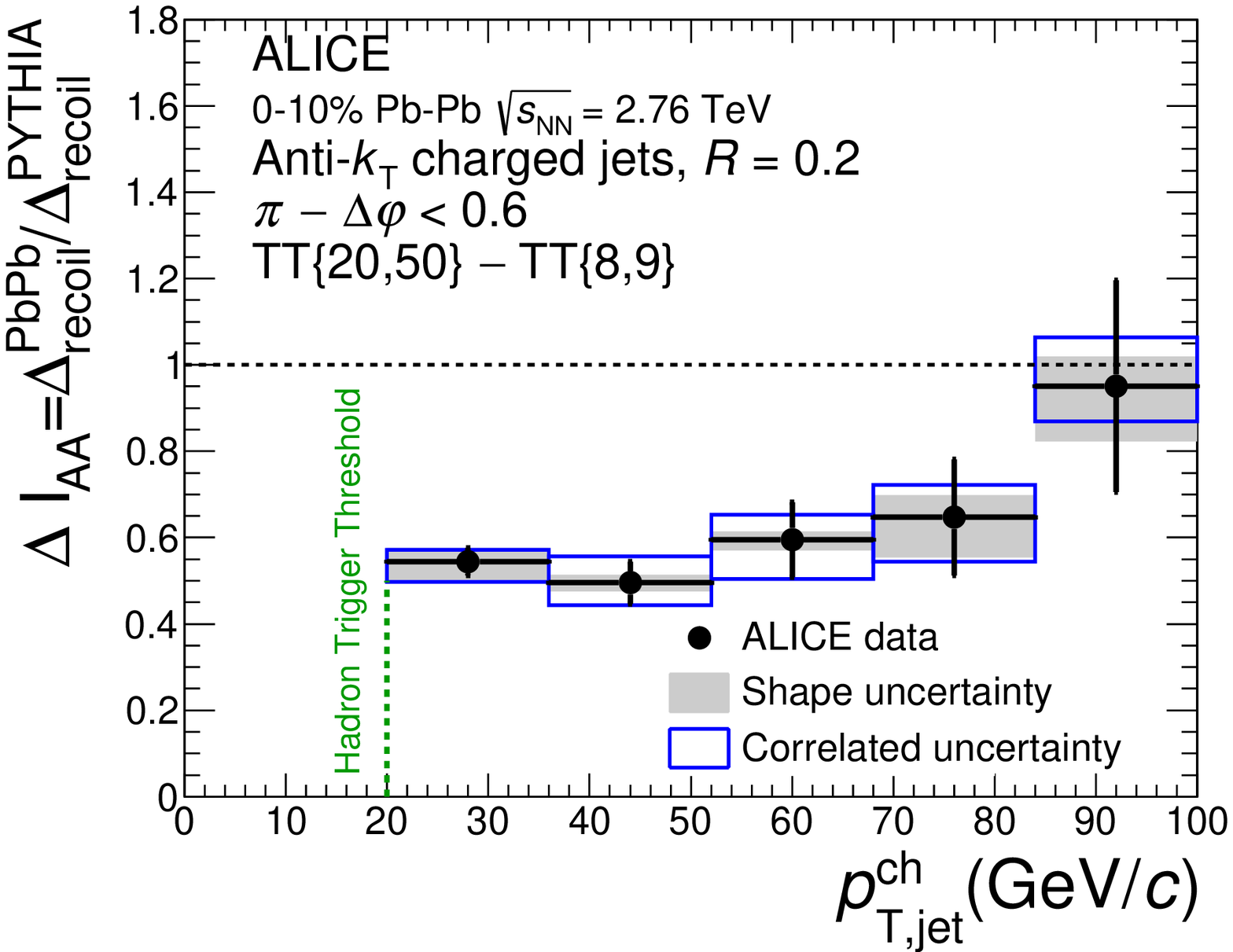}
\includegraphics[width=0.6\textwidth]{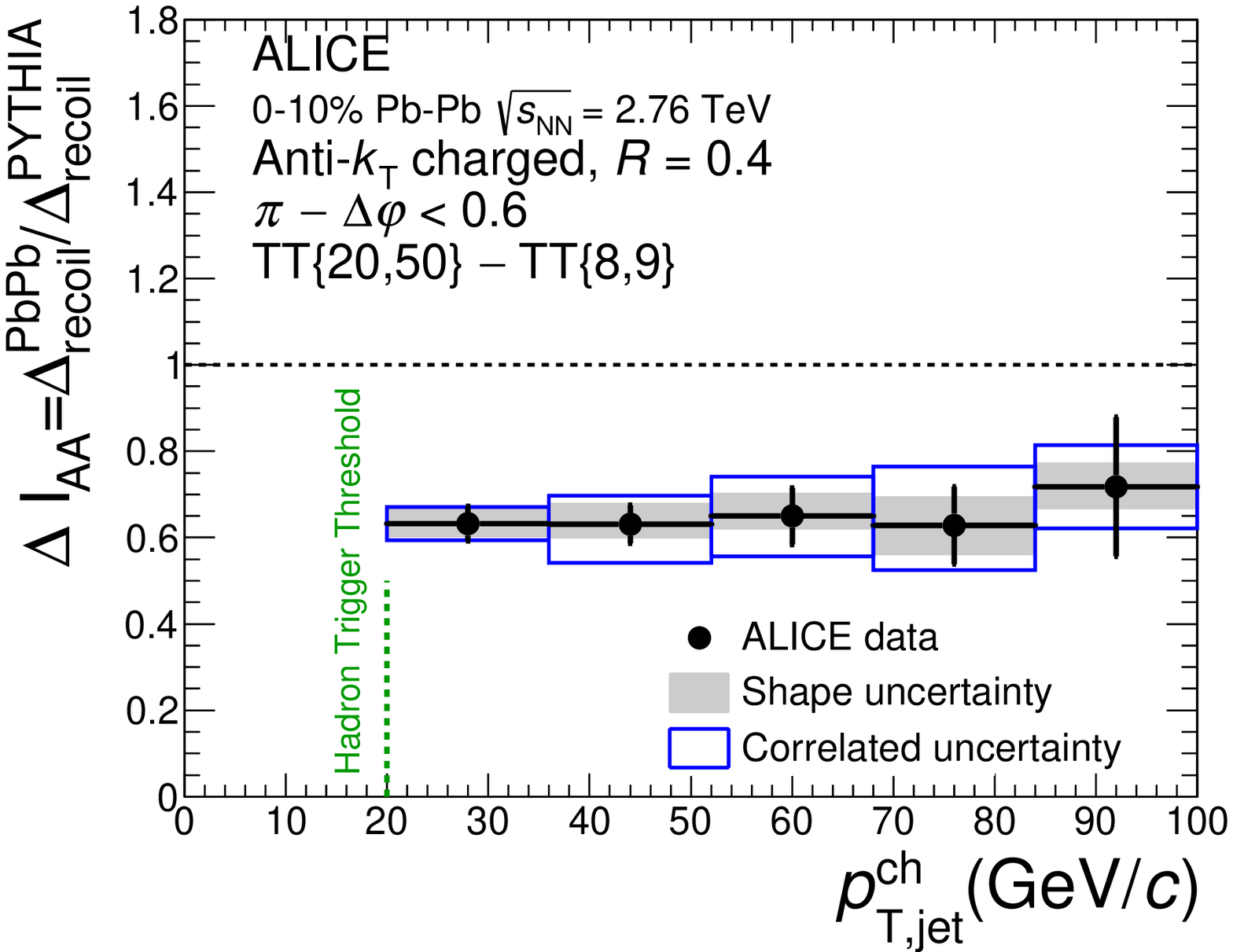}
\includegraphics[width=0.6\textwidth]{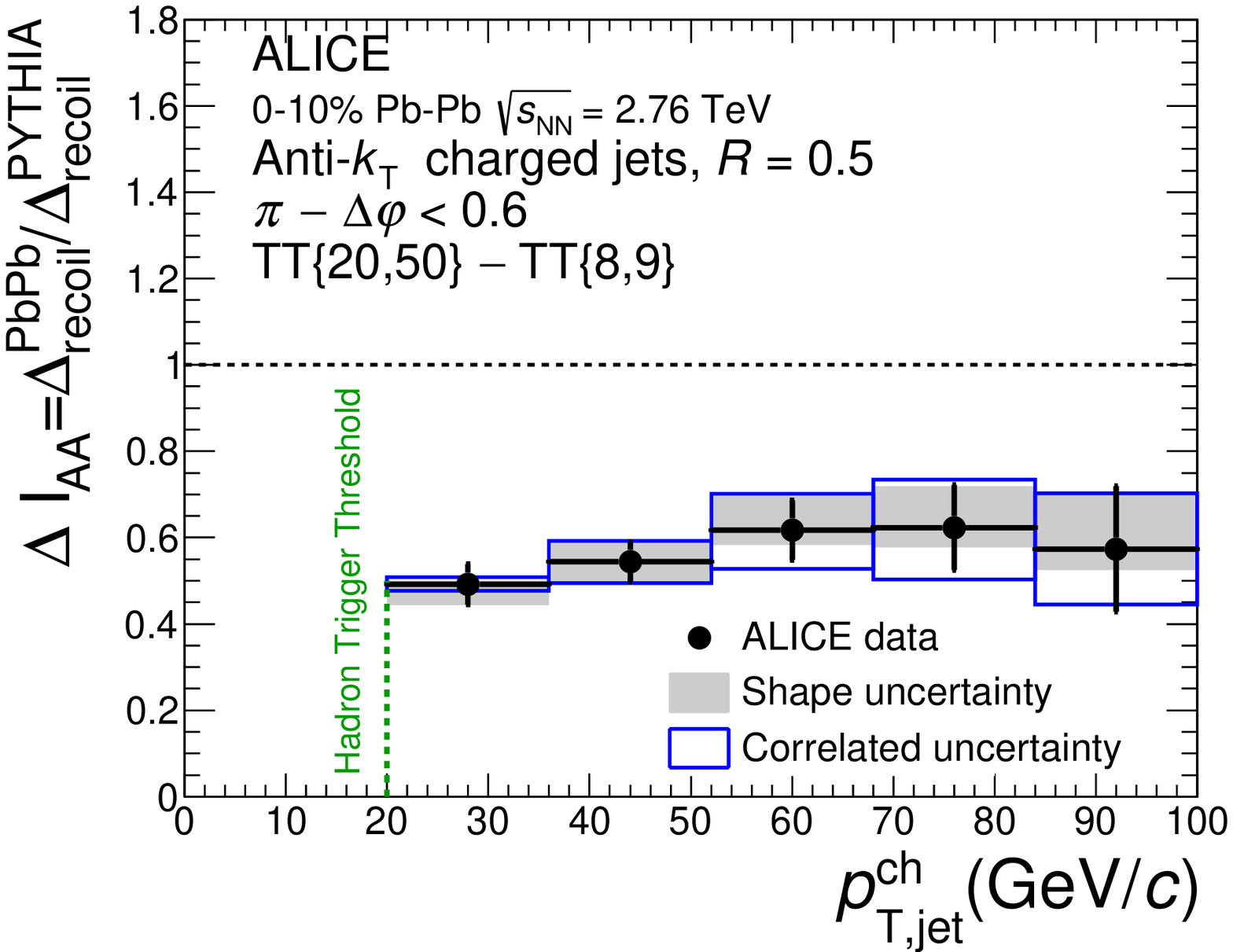}
\caption{\DIAA, the ratio of \Drecoil\ in central \PbPb\ and \pp\
  collisions at \sqrts\ = 2.76 TeV, for \rr\ = 0.2, 0.4 and
  0.5. \Drecoil\ for \pp\ collisions are calculated using
  PYTHIA.}
\label{fig:DeltaIAATT2050}
\end{figure}
 
Figure \ref{fig:DeltaIAATT2050} shows \DIAA\ for \rr\ = 0.2, 0.4 and
0.5. Suppression of the yield of recoil jets in \PbPb\ collisions is observed,
with similar magnitude for all \rr. 

\begin{figure}[tbh!f]
\centering
\includegraphics[width=0.6\textwidth]{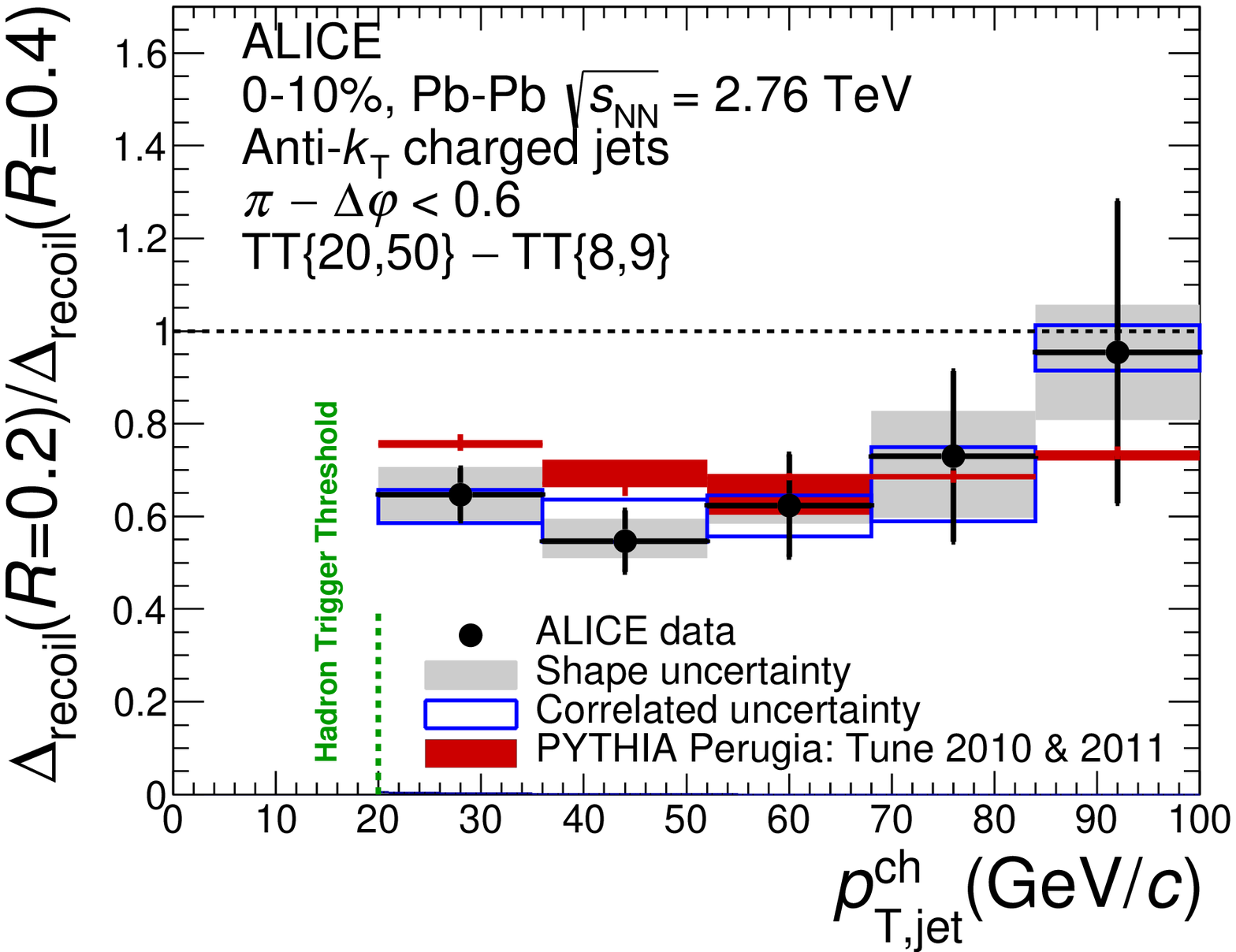}
\includegraphics[width=0.6\textwidth]{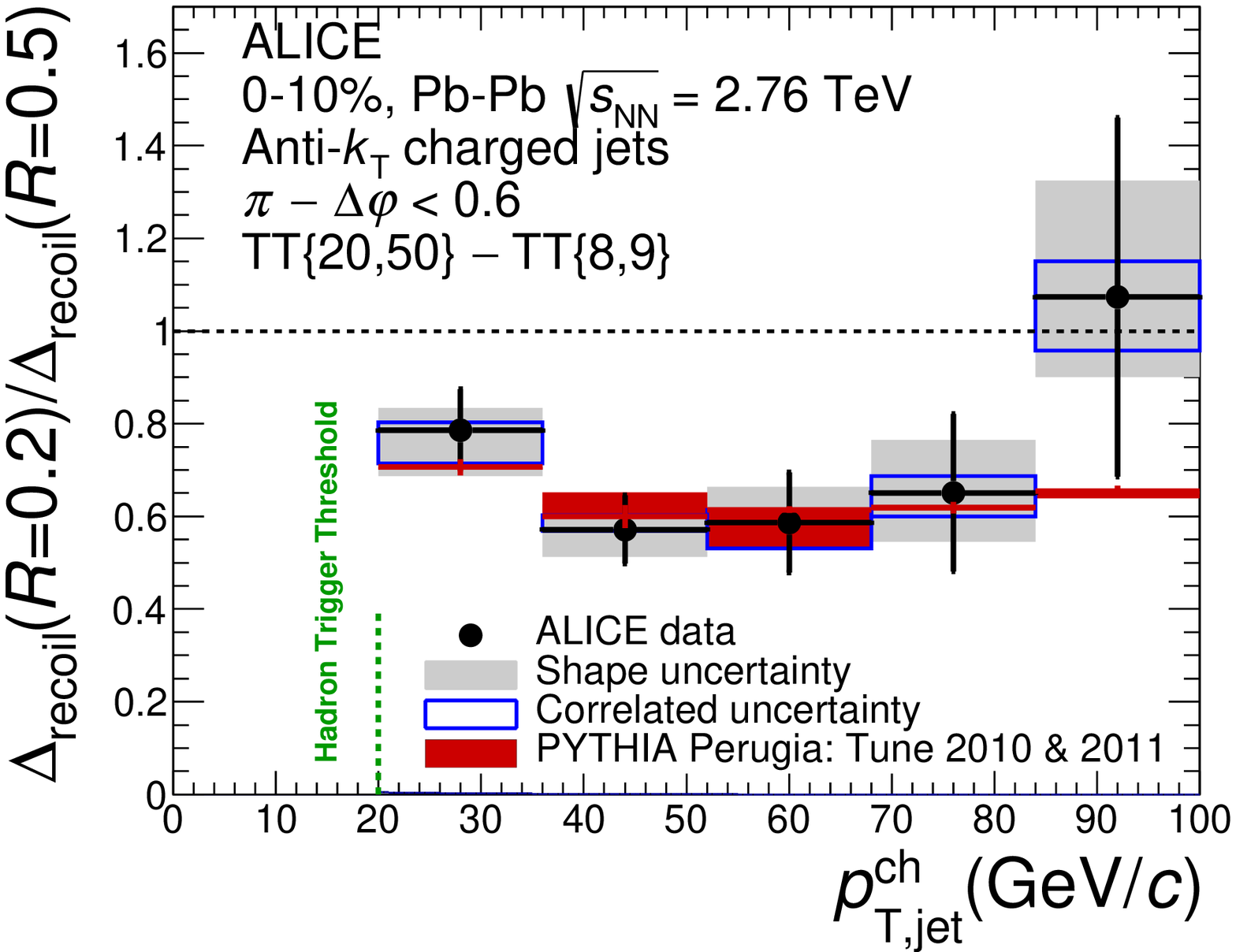}
\caption{Ratio of \Drecoil\ for \rr\ = 0.2 relative to \rr\ = 0.4
  (top) and to \rr\ = 0.5 (bottom), for central \PbPb\ (black) and
  \pp\ collisions simulated using PYTHIA (red) at \sqrts\ = 2.76
  TeV.}
\label{fig:RatioYield2050}
\end{figure}

Figure ~\ref{fig:RatioYield2050} shows the ratio of \Drecoil\ for \rr\
= 0.2 relative to \Drecoil\ for \rr\ = 0.4 and 0.5, in central \PbPb\ and
\pp\ collisions. The systematic uncertainties of the \PbPb\ ratios
take into account the correlated systematic uncertainties in numerator
and denominator. The shape uncertainties are propagated independently
in the ratio, since variation in their components induces different
effects as a function of \rr. The distributions for \PbPb\ and \pp\
collisions are seen to be similar, with no evidence for
intra-jet broadening in central \PbPb\ collisions within the
uncertainties.

The CMS collaboration has reported a significant redistribution of
energy within $\rr<0.3$ for jets in central \PbPb\
collisions\cite{Chatrchyan:2013kwa}, potentially in contrast to
Fig.~\ref{fig:RatioYield2050}. However, that measurement and the one
reported here cannot be compared directly. Modeling of the two
measurements within the same theoretical framework is required for
their comparison.

Figures \ref{fig:DeltaIAATT2050} and \ref{fig:RatioYield2050} show
that the recoil jet yield is suppressed, while the intra-jet energy
profile is not changed significantly for $\rr \le 0.5$. We note in
addition that the infrared cutoff for jet constituents (tracks) in
this measurement is \pTconst\ = 0.15 \gev, which strongly constrains
the correlated energy within the jet cone that would not be detected
by this measurement.

Taken together, these observations are consistent with a picture in
which there is significant in-medium transport of radiation to angles
larger than 0.5 radians. This picture was initially suggested by a
measurement showing that the energy imbalance of highly asymmetric jet
pairs is compensated, on an ensemble-averaged basis, by the energy
carried by soft particles at large angles relative to the jet axis
\cite{Chatrchyan:2011sx}. Also in this case, however, quantitative
comparison of these measurements requires their calculation in a
common theoretical framework.

The \Drecoil\ distributions in both \pp\ and \PbPb\ collisions are
well-described by an exponential distribution
$\propto{e}^{-\pTjetch/b}$, with values of $b$ around 16
\gev. Fig. \ref{fig:DeltaIAATT2050} shows that \DIAA\ has negligible
dependence on \pTjetch\ for \rr\ = 0.4 and 0.5 within
$60<\pTjetch<100$ \gev, which indicates that the values of $b$ are
similar within this \pTjet\ range for the \pp\ and \PbPb\
distributions. The value of \DIAA\ in this region can therefore be
expressed as the horizontal shift of an exponential distribution of
fixed slope. For \rr\ = 0.5 in the range $60<\pTjetch<100$ \gev, the
suppression in \DIAA\ corresponds to a shift in \pTjetch\ of $-8\pm2$
(stat) \gev. In the scenario of negligible trigger-jet energy loss, this
shift corresponds to the average partonic energy loss of the recoil
jet population via energy transport to large angles, outside the jet
cone.

\subsection{Azimuthal correlations}
\label{sect:AngDep}

\begin{figure}[tbh!f]
\centering
\includegraphics[width=0.60\textwidth]{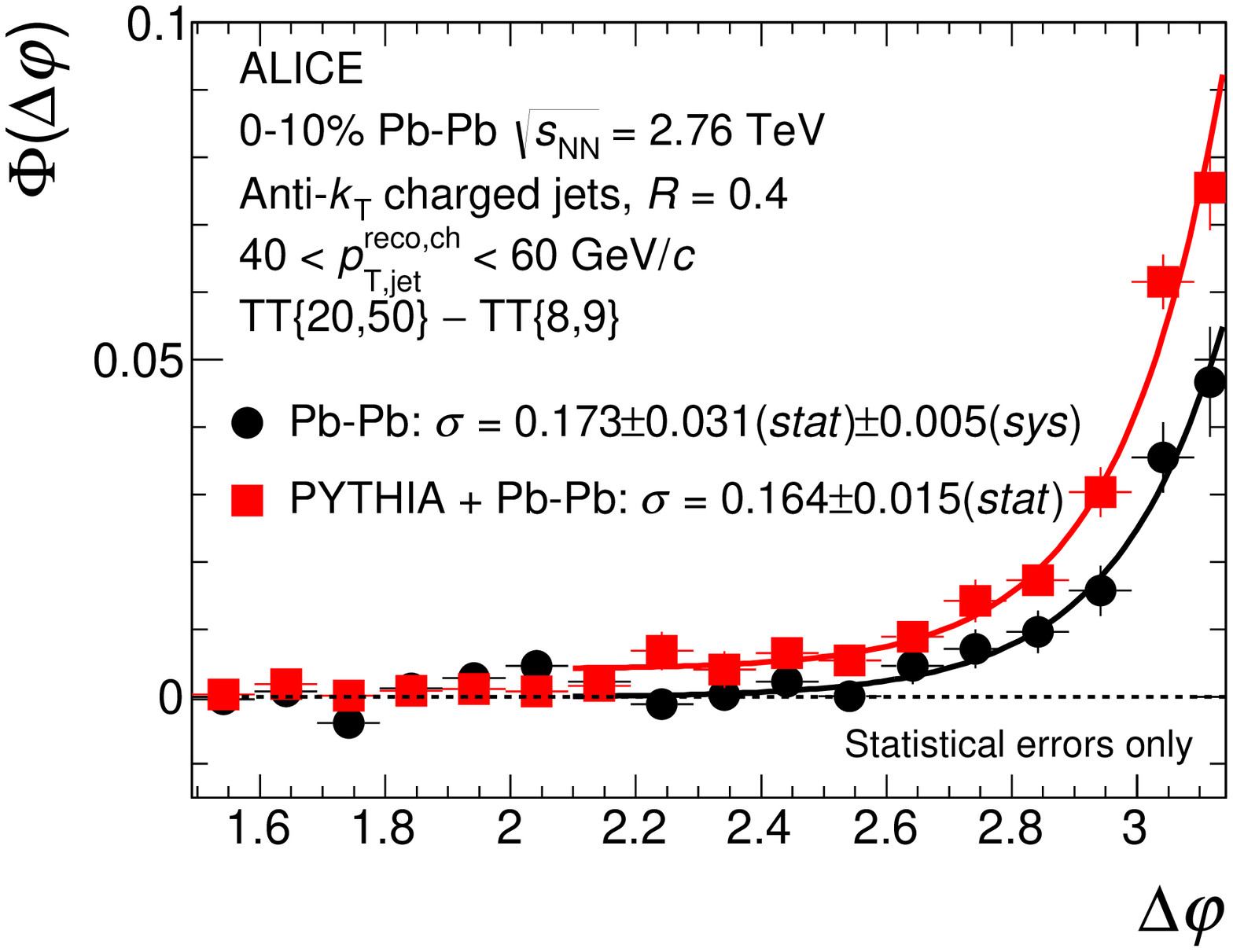}
\caption{\Drecoilphi\ distributions for 0-10\% central \PbPb\ data
  (black circles) and \pp\ collisions simulated by detector-level
  PYTHIA events embedded into central \PbPb\ events (red squares), at
  \sqrtsNN\ = 2.76 TeV. Jets have $40<\pTreco<60$, with \pTreco\ not
  corrected for background fluctuations and instrumental effects.  The
  lines show the result of fitting Eq. \ref{eq:dphifit} to the
  distributions, with the value of $\sigma$ from the fit as
  indicated. The error bars show statistical errors only. The \PbPb\
  data points are the same as the solid circles shown in the left
  panel of Fig. \ref{fig:data-dphi-raw}.}
\label{fig:paper-dphi-data-vs-embed}
\end{figure}

Figure \ref{fig:paper-dphi-data-vs-embed} shows the uncorrected
\Drecoilphi\ distributions for central \PbPb\ data and \pp\
simulations. As noted in Sect.~\ref{sect:AngleDepRaw}, we compare the
uncorrected \Drecoilphi\ distribution of \PbPb\ data to a reference
distribution for \pp\ collisions (PYTHIA, Perugia 2010 tune), modified
by the background and instrumental effects expected for central \PbPb\
collisions. We recall that \Drecoilphi\ suppresses the
uncorrelated contribution from MPI, which otherwise would provide a
significant background at large $\pi-\dphi$.

The absolute yield of the \PbPb\ distribution is seen to be smaller
than that of the \pp\ reference. This is consistent with the
suppression observed for \DIAA\ (Fig.\ \ref{fig:DeltaIAATT2050}),
which is the ratio of the integrals of the \Drecoilphi\ distributions
over the range $\pi-\dphi<0.6$.

The \Drecoilphi\ distributions for \PbPb\ and \pp\ collisions are characterized
by fitting a function corresponding to an exponential plus a
pedestal term \cite{Chatrchyan:2012gt},

\begin{equation}
f(\dphi)=p_0\times e^{(\dphi-\pi)/\sigma} + p_1 ,
\label{eq:dphifit}
\end{equation}

\noindent
where the parameter $\sigma$ reflects the width of the
distribution. The fit range is $2\pi/3 <\dphi<\pi$.  The fitted values
are
$\sigma_{\rm{Pb-Pb}}=0.173\pm0.031\mathrm{(stat.)}\pm0.005\mathrm{(sys.)}$
and $\sigma_{\rm{PYTHIA}}=0.164\pm0.015\mathrm{(stat.)}$, which are
consistent within uncertainties. We find no evidence from this
comparison for medium-induced acoplanarity of recoil jets with
uncorrected energy in the range $40<\pTreco<60$ \gev.

The azimuthal distribution between a direct photon ($p_{T,\gamma}>60$
\gev) and a recoil jet ($p_{T,\mathrm{jet}}>30$ \gev) has been
measured in central \PbPb\ collisions and compared to that from PYTHIA
events embedded in a simulation of \PbPb\ collisions
\cite{Chatrchyan:2012gt}. Fits of an exponential function to these
distributions give similar values of $\sigma$ for central \PbPb\ and
embedded PYTHIA, likewise indicating no evidence for medium-induced
acoplanarity, though the values of $\sigma$ are larger than those for
the analysis reported here. Comparison of the shape of the azimuthal
distribution of di-jet pairs in central \PbPb\ data and embedded
PYTHIA events has been reported \cite{Aad:2010bu,Chatrchyan:2011sx},
with indication of an enhancement in the tail of the distribution for
central \PbPb\ collisions\cite{Aad:2010bu}.

\begin{figure}[tbh!f]
\centering
\includegraphics[width=0.49\textwidth]{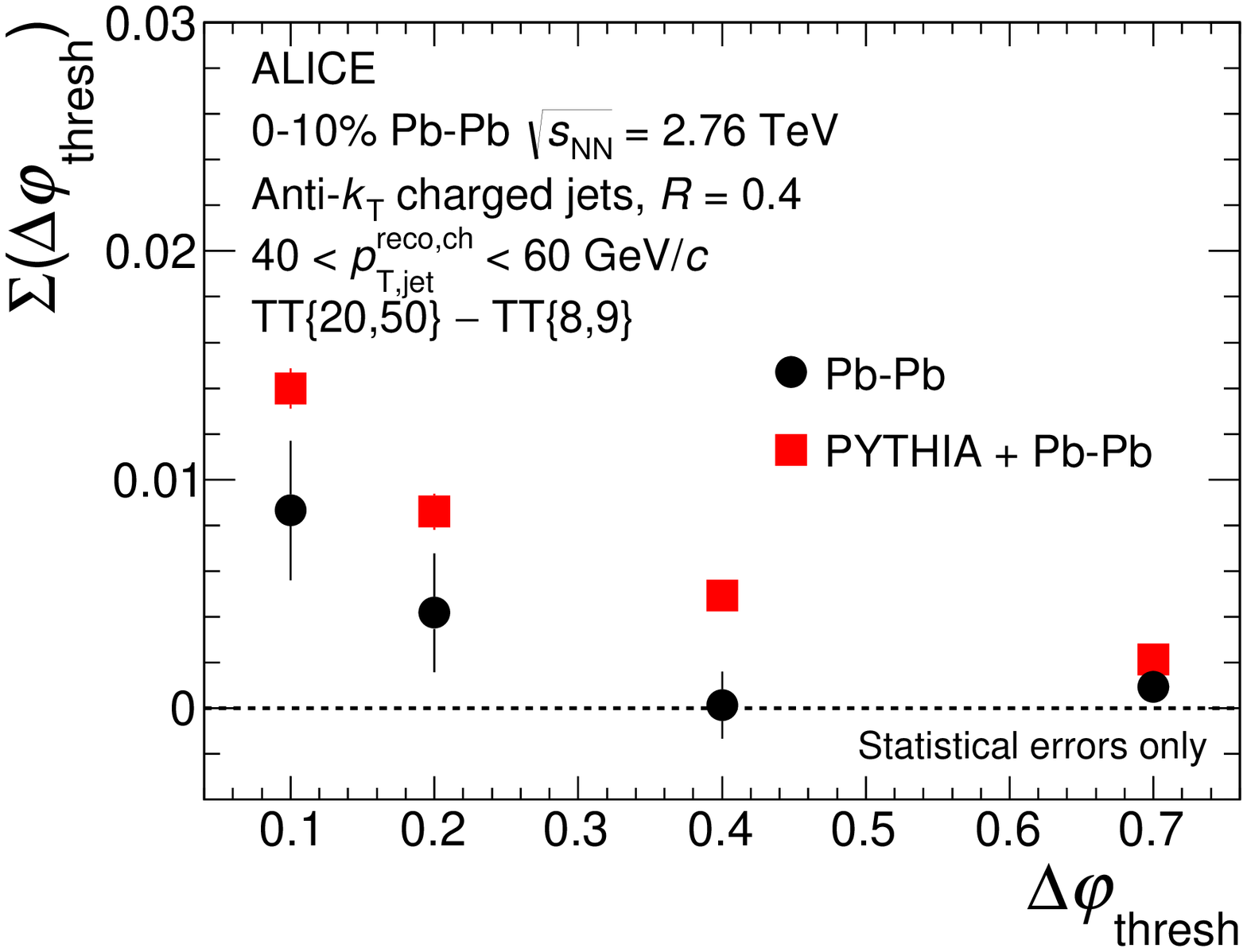}
\includegraphics[width=0.49\textwidth]{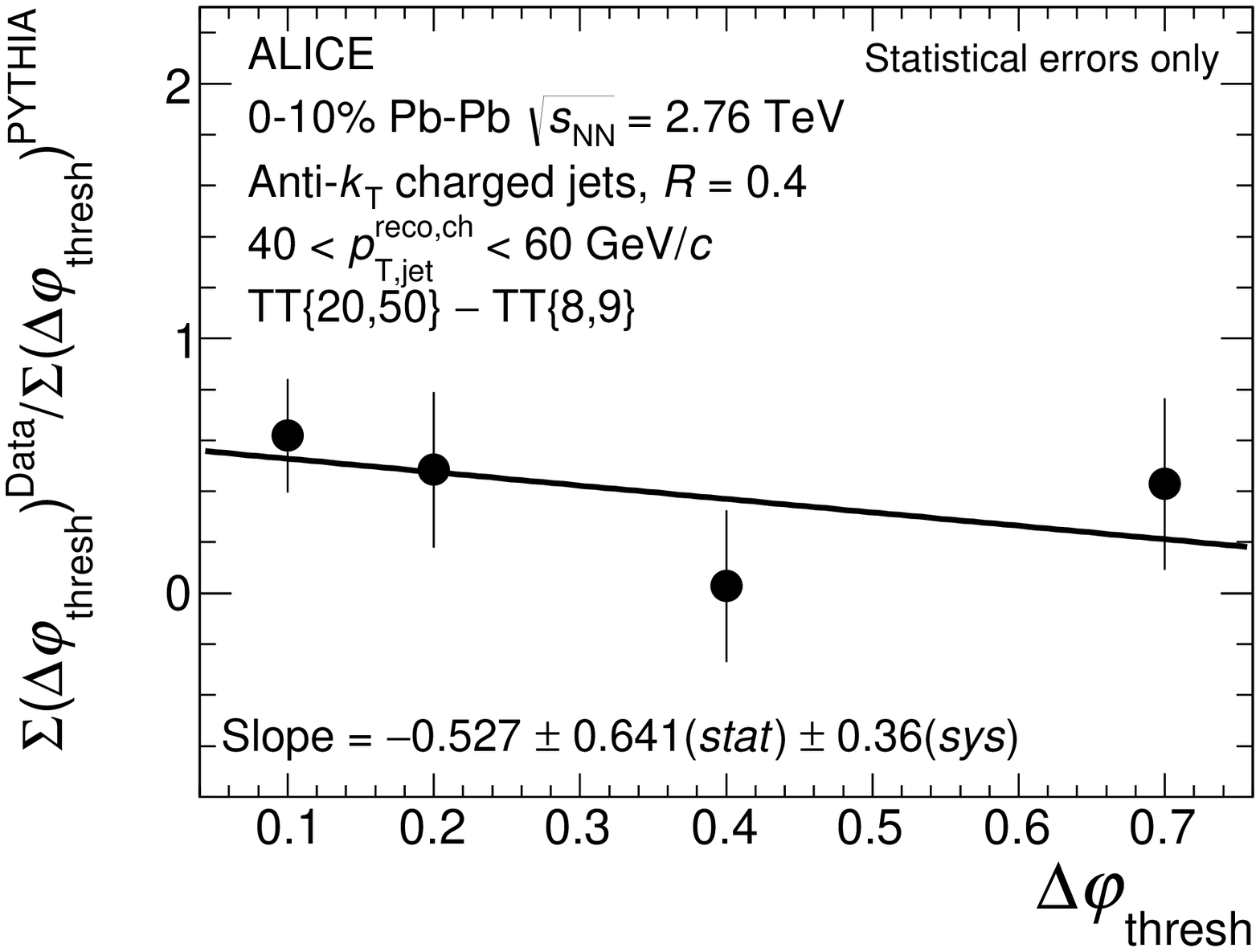}
\caption{Left: \Dyieldthresh\ distributions for central \PbPb\ data
  (black circles) and the \pp\ reference distribution (red squares),
  obtained by embedding detector-level PYTHIA events into real data. Right: ratio of
  the \Dyieldthresh\ distribution between \PbPb\ data and the PYTHIA
  reference shown in the left along with a first-order polynomial
  fit. The error bars in both panels show statistical errors only.}
\label{fig:paper-dphi-integral}
\end{figure}

More detailed characterization of the change in the angular
distribution with change in TT interval is provided by \Dyieldthresh\
(Eq. \ref{eq:Dyield}), which measures the yield in the tail
  of the distribution beyond a threshold, $\pi/2<\dphi<\pi-\phithresh$.
Fig. \ref{fig:paper-dphi-integral}, left panel, shows the
\Dyieldthresh\ distributions for central \PbPb\ data and the embedded
PYTHIA reference. As discussed in Sect.~\ref{sect:AngleDepRaw}, for
the measurement of \Dyieldthresh\ the dataset is divided into
exclusive subsets, with each subset used for only one value of
threshold, so that the data points in Fig.\
\ref{fig:paper-dphi-integral} are statistically uncorrelated.

The relative contribution to \Dyieldthresh\ of different physics
processes may vary with \phithresh. At sufficiently large angular
deflection of the jet centroid, at a value of \phithresh\ which has
not yet been determined, the yield is expected to arise predominantly
from single hard (Moli{\`e}re) scattering in the hot QCD medium
\cite{D'Eramo:2012jh,Wang:2013cia}. Figure
\ref{fig:paper-dphi-integral}, right panel, shows the ratio of the two
distributions in the left panel. We utilize this ratio of absolutely
normalized distributions in \pp\ and \PbPb\ collisions to search for
effects due to Moli{\`e}re scattering.

The value of the ratio at small \phithresh\ corresponds approximately
to the \dphi-integrated suppression in recoil yield in
Fig.~\ref{fig:DeltaIAATT2050}, while its dependence on \phithresh\
provides a comparison of the shapes of the distributions. This
comparison is quantified by fitting a first-order polynomial function
to the ratio of \Dyieldthresh\ in the right panel.  The fit gives a
slope of $-0.527\pm0.641\mathrm{(stat.)}\pm0.36\mathrm{(sys.)}$, which
is consistent with zero within uncertainties. If Moliere scattering
were the only mechanism modifying the \PbPb\ distribution relative to
that of \pp\ collisions the ratio at large \phithresh\ should be
larger than unity; however, the ratio is seen to be below unity at the
largest measured \phithresh, indicating that other mechanisms have
large effect in this region. We find no evidence from this measurement
for medium-induced Moli{\`e}re scattering. The uncertainty in this
measurement is dominated by the statistical error, however, meaning
that additional data, together with measurements at other jet energies
and larger angular deviations, will provide more precise constraints
on the rate of large-angle scattering in the hot QCD medium.

\section{Summary}
\label{sect:Summary}

We have reported measurements of jet quenching in central \PbPb\
collisions at the LHC, using a new analysis method based on the
semi-inclusive distribution of jets recoiling from a high-\pT\ trigger
hadron. Discrimination of coincident jet yield from background is
carried out at the level of ensemble-averaged distributions with a
trigger-difference technique, with no selection bias imposed on the
recoil jet population. This approach enables measurement of the
distribution of jets with large \rr\ and low infrared cutoff for jet
constituents, over a broad range of jet energy. Distributions are
reported for charged jets in the range $20<\pTjetch<100$ \gev, for
\rr\ = 0.2, 0.4 and 0.5.

The differential recoil jet yield in central \PbPb\ collisions is
suppressed relative to that in \pp\ collisions by up to a factor two
for $0.2\leq\rr\leq0.5$. Together with the low infrared cutoff of this
measurement, this indicates that medium-induced energy loss arises
predominantly from radiation at angles larger than 0.5 relative to the
jet axis. The energy carried by this radiation, which is reflected in
the magnitude of the spectrum shift under the assumption of negligible
trigger-jet energy loss, is estimated to be $8\pm2$ \gev\ for charged
jets with \rr\ = 0.5, in the range $60<\pTjetch<100$ \gev.

The ratio of differential recoil jet yields with different \rr\ is
similar for \PbPb\ and \pp\ collisions. No significant medium-induced
modification of the intra-jet energy distribution for angles
$\rr\leq0.5$ relative to the jet axis is thereby observed.
 
The width of the azimuthal distribution of recoil jets relative to the
trigger axis is measured to be similar in \PbPb\ and \pp\ collisions
for jets of $40 < \pTreco < 60$ \gev. No significant medium-induced
acoplanarity is therefore observed, consistent with findings from
di-jet and direct photon-jet measurements. Large angular deflection of
the recoil jet may be sensitive to the rate of single Moli{\`e}re
scattering in the hot QCD medium and provide a direct probe of its
quasi-particle degrees of freedom. We observe no significant rate of
such large-angle scatterings, though with limited statistical
precision at present. These data, when combined with theoretical
calculations, will provide guidance for the necessary precision to
achieve discriminating measurements in the future.

\newenvironment{acknowledgement}{\relax}{\relax}
\begin{acknowledgement}
\section*{Acknowledgements}
\label{sect:Acknowledgments}
We thank Daniel de Florian, Francesco d'Eramo and Krishna Rajagopal
for valuable discussions and calculations.
The ALICE Collaboration would like to thank all its engineers and technicians for their invaluable contributions to the construction of the experiment and the CERN accelerator teams for the outstanding performance of the LHC complex.
The ALICE Collaboration gratefully acknowledges the resources and support provided by all Grid centres and the Worldwide LHC Computing Grid (WLCG) collaboration.
The ALICE Collaboration acknowledges the following funding agencies for their support in building and
running the ALICE detector:
State Committee of Science,  World Federation of Scientists (WFS)
and Swiss Fonds Kidagan, Armenia,
Conselho Nacional de Desenvolvimento Cient\'{\i}fico e Tecnol\'{o}gico (CNPq), Financiadora de Estudos e Projetos (FINEP),
Funda\c{c}\~{a}o de Amparo \`{a} Pesquisa do Estado de S\~{a}o Paulo (FAPESP);
National Natural Science Foundation of China (NSFC), the Chinese Ministry of Education (CMOE)
and the Ministry of Science and Technology of China (MSTC);
Ministry of Education and Youth of the Czech Republic;
Danish Natural Science Research Council, the Carlsberg Foundation and the Danish National Research Foundation;
The European Research Council under the European Community's Seventh Framework Programme;
Helsinki Institute of Physics and the Academy of Finland;
French CNRS-IN2P3, the `Region Pays de Loire', `Region Alsace', `Region Auvergne' and CEA, France;
German Bundesministerium fur Bildung, Wissenschaft, Forschung und Technologie (BMBF) and the Helmholtz Association;
General Secretariat for Research and Technology, Ministry of
Development, Greece;
Hungarian Orszagos Tudomanyos Kutatasi Alappgrammok (OTKA) and National Office for Research and Technology (NKTH);
Department of Atomic Energy and Department of Science and Technology of the Government of India;
Istituto Nazionale di Fisica Nucleare (INFN) and Centro Fermi -
Museo Storico della Fisica e Centro Studi e Ricerche "Enrico
Fermi", Italy;
MEXT Grant-in-Aid for Specially Promoted Research, Ja\-pan;
Joint Institute for Nuclear Research, Dubna;
National Research Foundation of Korea (NRF);
Consejo Nacional de Cienca y Tecnologia (CONACYT), Direccion General de Asuntos del Personal Academico(DGAPA), M\'{e}xico, 
Amerique Latine Formation academique - European Commission(ALFA-EC) and the EPLANET Program (European Particle Physics Latin American Network);
Stichting voor Fundamenteel Onderzoek der Materie (FOM) and the Nederlandse Organisatie voor Wetenschappelijk Onderzoek (NWO), Netherlands;
Research Council of Norway (NFR);
National Science Centre, Poland;
Ministry of National Education/Institute for Atomic Physics and Consiliul Naţional al Cercetării Ştiinţifice - Executive Agency for Higher Education Research Development and Innovation Funding (CNCS-UEFISCDI) - Romania;
Ministry of Education and Science of Russian Federation, Russian
Academy of Sciences, Russian Federal Agency of Atomic Energy,
Russian Federal Agency for Science and Innovations and The Russian
Foundation for Basic Research;
Ministry of Education of Slovakia;
Department of Science and Technology, South Africa;
Centro de Investigaciones Energeticas, Medioambientales y Tecnologicas (CIEMAT), E-Infrastructure shared between Europe and Latin America (EELA), Ministerio de Econom\'{i}a y Competitividad (MINECO) of Spain, Xunta de Galicia (Conseller\'{\i}a de Educaci\'{o}n),
Centro de Aplicaciones Tecnológicas y Desarrollo Nuclear (CEA\-DEN), Cubaenerg\'{\i}a, Cuba, and IAEA (International Atomic Energy Agency);
Swedish Research Council (VR) and Knut $\&$ Alice Wallenberg
Foundation (KAW);
Ukraine Ministry of Education and Science;
United Kingdom Science and Technology Facilities Council (STFC);
The United States Department of Energy, the United States National
Science Foundation, the State of Texas, and the State of Ohio;
Ministry of Science, Education and Sports of Croatia and  Unity through Knowledge Fund, Croatia.
Council of Scientific and Industrial Research (CSIR), New Delhi, India

\end{acknowledgement}

\bibliographystyle{utphys}
\bibliography{references}

\newpage
\appendix
\section{The ALICE Collaboration}
\label{app:collab}



\begingroup
\small
\begin{flushleft}
J.~Adam\Irefn{org40}\And
D.~Adamov\'{a}\Irefn{org83}\And
M.M.~Aggarwal\Irefn{org87}\And
G.~Aglieri Rinella\Irefn{org36}\And
M.~Agnello\Irefn{org111}\And
N.~Agrawal\Irefn{org48}\And
Z.~Ahammed\Irefn{org132}\And
S.U.~Ahn\Irefn{org68}\And
I.~Aimo\Irefn{org94}\textsuperscript{,}\Irefn{org111}\And
S.~Aiola\Irefn{org137}
% ...
\renewcommand\labelenumi{\textsuperscript{\theenumi}~}

\section*{Affiliation notes}
\renewcommand\theenumi{\roman{enumi}}
\begin{Authlist}
\item \Adef{0}Deceased
\item \Adef{idp3802240}{Also at: M.V. Lomonosov Moscow State University, D.V. Skobeltsyn Institute of Nuclear, Physics, Moscow, Russia}
\item \Adef{idp5937984}{Also at: University of Kansas, Lawrence, Kansas, United States}
\end{Authlist}

\section*{Collaboration Institutes}
\renewcommand\theenumi{\arabic{enumi}~}
\begin{Authlist}

\item \Idef{org1}A.I. Alikhanyan National Science Laboratory (Yerevan Physics Institute) Foundation, Yerevan, Armenia
\item \Idef{org2}Benem\'{e}rita Universidad Aut\'{o}noma de Puebla, Puebla, Mexico
\item \Idef{org3}Bogolyubov Institute for Theoretical Physics, Kiev, Ukraine
\item \Idef{org4}Bose Institute, Department of Physics and Centre for Astroparticle Physics and Space Science (CAPSS), Kolkata, India
\item \Idef{org5}Budker Institute for Nuclear Physics, Novosibirsk, Russia
\item \Idef{org6}California Polytechnic State University, San Luis Obispo, California, United States
\item \Idef{org7}Central China Normal University, Wuhan, China
\item \Idef{org8}Centre de Calcul de l'IN2P3, Villeurbanne, France
\item \Idef{org9}Centro de Aplicaciones Tecnol\'{o}gicas y Desarrollo Nuclear (CEADEN), Havana, Cuba
\item \Idef{org10}Centro de Investigaciones Energ\'{e}ticas Medioambientales y Tecnol\'{o}gicas (CIEMAT), Madrid, Spain
\item \Idef{org11}Centro de Investigaci\'{o}n y de Estudios Avanzados (CINVESTAV), Mexico City and M\'{e}rida, Mexico
\item \Idef{org12}Centro Fermi - Museo Storico della Fisica e Centro Studi e Ricerche ``Enrico Fermi'', Rome, Italy
\item \Idef{org13}Chicago State University, Chicago, Illinois, USA
\item \Idef{org14}China Institute of Atomic Energy, Beijing, China
\item \Idef{org15}Commissariat \`{a} l'Energie Atomique, IRFU, Saclay, France
\item \Idef{org16}COMSATS Institute of Information Technology (CIIT), Islamabad, Pakistan
\item \Idef{org17}Departamento de F\'{\i}sica de Part\'{\i}culas and IGFAE, Universidad de Santiago de Compostela, Santiago de Compostela, Spain
\item \Idef{org18}Department of Physics and Technology, University of Bergen, Bergen, Norway
\item \Idef{org19}Department of Physics, Aligarh Muslim University, Aligarh, India
\item \Idef{org20}Department of Physics, Ohio State University, Columbus, Ohio, United States
\item \Idef{org21}Department of Physics, Sejong University, Seoul, South Korea
\item \Idef{org22}Department of Physics, University of Oslo, Oslo, Norway
\item \Idef{org23}Dipartimento di Elettrotecnica ed Elettronica del Politecnico, Bari, Italy
\item \Idef{org24}Dipartimento di Fisica dell'Universit\`{a} 'La Sapienza' and Sezione INFN Rome, Italy
\item \Idef{org25}Dipartimento di Fisica dell'Universit\`{a} and Sezione INFN, Cagliari, Italy
\item \Idef{org26}Dipartimento di Fisica dell'Universit\`{a} and Sezione INFN, Trieste, Italy
\item \Idef{org27}Dipartimento di Fisica dell'Universit\`{a} and Sezione INFN, Turin, Italy
\item \Idef{org28}Dipartimento di Fisica e Astronomia dell'Universit\`{a} and Sezione INFN, Bologna, Italy
\item \Idef{org29}Dipartimento di Fisica e Astronomia dell'Universit\`{a} and Sezione INFN, Catania, Italy
\item \Idef{org30}Dipartimento di Fisica e Astronomia dell'Universit\`{a} and Sezione INFN, Padova, Italy
\item \Idef{org31}Dipartimento di Fisica `E.R.~Caianiello' dell'Universit\`{a} and Gruppo Collegato INFN, Salerno, Italy
\item \Idef{org32}Dipartimento di Scienze e Innovazione Tecnologica dell'Universit\`{a} del  Piemonte Orientale and Gruppo Collegato INFN, Alessandria, Italy
\item \Idef{org33}Dipartimento Interateneo di Fisica `M.~Merlin' and Sezione INFN, Bari, Italy
\item \Idef{org34}Division of Experimental High Energy Physics, University of Lund, Lund, Sweden
\item \Idef{org35}Eberhard Karls Universit\"{a}t T\"{u}bingen, T\"{u}bingen, Germany
\item \Idef{org36}European Organization for Nuclear Research (CERN), Geneva, Switzerland
\item \Idef{org37}Excellence Cluster Universe, Technische Universit\"{a}t M\"{u}nchen, Munich, Germany
\item \Idef{org38}Faculty of Engineering, Bergen University College, Bergen, Norway
\item \Idef{org39}Faculty of Mathematics, Physics and Informatics, Comenius University, Bratislava, Slovakia
\item \Idef{org40}Faculty of Nuclear Sciences and Physical Engineering, Czech Technical University in Prague, Prague, Czech Republic
\item \Idef{org41}Faculty of Science, P.J.~\v{S}af\'{a}rik University, Ko\v{s}ice, Slovakia
\item \Idef{org42}Faculty of Technology, Buskerud and Vestfold University College, Vestfold, Norway
\item \Idef{org43}Frankfurt Institute for Advanced Studies, Johann Wolfgang Goethe-Universit\"{a}t Frankfurt, Frankfurt, Germany
\item \Idef{org44}Gangneung-Wonju National University, Gangneung, South Korea
\item \Idef{org45}Gauhati University, Department of Physics, Guwahati, India
\item \Idef{org46}Helsinki Institute of Physics (HIP), Helsinki, Finland
\item \Idef{org47}Hiroshima University, Hiroshima, Japan
\item \Idef{org48}Indian Institute of Technology Bombay (IIT), Mumbai, India
\item \Idef{org49}Indian Institute of Technology Indore, Indore (IITI), India
\item \Idef{org50}Inha University, Incheon, South Korea
\item \Idef{org51}Institut de Physique Nucl\'eaire d'Orsay (IPNO), Universit\'e Paris-Sud, CNRS-IN2P3, Orsay, France
\item \Idef{org52}Institut f\"{u}r Informatik, Johann Wolfgang Goethe-Universit\"{a}t Frankfurt, Frankfurt, Germany
\item \Idef{org53}Institut f\"{u}r Kernphysik, Johann Wolfgang Goethe-Universit\"{a}t Frankfurt, Frankfurt, Germany
\item \Idef{org54}Institut f\"{u}r Kernphysik, Westf\"{a}lische Wilhelms-Universit\"{a}t M\"{u}nster, M\"{u}nster, Germany
\item \Idef{org55}Institut Pluridisciplinaire Hubert Curien (IPHC), Universit\'{e} de Strasbourg, CNRS-IN2P3, Strasbourg, France
\item \Idef{org56}Institute for Nuclear Research, Academy of Sciences, Moscow, Russia
\item \Idef{org57}Institute for Subatomic Physics of Utrecht University, Utrecht, Netherlands
\item \Idef{org58}Institute for Theoretical and Experimental Physics, Moscow, Russia
\item \Idef{org59}Institute of Experimental Physics, Slovak Academy of Sciences, Ko\v{s}ice, Slovakia
\item \Idef{org60}Institute of Physics, Academy of Sciences of the Czech Republic, Prague, Czech Republic
\item \Idef{org61}Institute of Physics, Bhubaneswar, India
\item \Idef{org62}Institute of Space Science (ISS), Bucharest, Romania
\item \Idef{org63}Instituto de Ciencias Nucleares, Universidad Nacional Aut\'{o}noma de M\'{e}xico, Mexico City, Mexico
\item \Idef{org64}Instituto de F\'{\i}sica, Universidad Nacional Aut\'{o}noma de M\'{e}xico, Mexico City, Mexico
\item \Idef{org65}iThemba LABS, National Research Foundation, Somerset West, South Africa
\item \Idef{org66}Joint Institute for Nuclear Research (JINR), Dubna, Russia
\item \Idef{org67}Konkuk University, Seoul, South Korea
\item \Idef{org68}Korea Institute of Science and Technology Information, Daejeon, South Korea
\item \Idef{org69}KTO Karatay University, Konya, Turkey
\item \Idef{org70}Laboratoire de Physique Corpusculaire (LPC), Clermont Universit\'{e}, Universit\'{e} Blaise Pascal, CNRS--IN2P3, Clermont-Ferrand, France
\item \Idef{org71}Laboratoire de Physique Subatomique et de Cosmologie, Universit\'{e} Grenoble-Alpes, CNRS-IN2P3, Grenoble, France
\item \Idef{org72}Laboratori Nazionali di Frascati, INFN, Frascati, Italy
\item \Idef{org73}Laboratori Nazionali di Legnaro, INFN, Legnaro, Italy
\item \Idef{org74}Lawrence Berkeley National Laboratory, Berkeley, California, United States
\item \Idef{org75}Lawrence Livermore National Laboratory, Livermore, California, United States
\item \Idef{org76}Moscow Engineering Physics Institute, Moscow, Russia
\item \Idef{org77}National Centre for Nuclear Studies, Warsaw, Poland
\item \Idef{org78}National Institute for Physics and Nuclear Engineering, Bucharest, Romania
\item \Idef{org79}National Institute of Science Education and Research, Bhubaneswar, India
\item \Idef{org80}Niels Bohr Institute, University of Copenhagen, Copenhagen, Denmark
\item \Idef{org81}Nikhef, Nationaal instituut voor subatomaire fysica, Amsterdam, Netherlands
\item \Idef{org82}Nuclear Physics Group, STFC Daresbury Laboratory, Daresbury, United Kingdom
\item \Idef{org83}Nuclear Physics Institute, Academy of Sciences of the Czech Republic, \v{R}e\v{z} u Prahy, Czech Republic
\item \Idef{org84}Oak Ridge National Laboratory, Oak Ridge, Tennessee, United States
\item \Idef{org85}Petersburg Nuclear Physics Institute, Gatchina, Russia
\item \Idef{org86}Physics Department, Creighton University, Omaha, Nebraska, United States
\item \Idef{org87}Physics Department, Panjab University, Chandigarh, India
\item \Idef{org88}Physics Department, University of Athens, Athens, Greece
\item \Idef{org89}Physics Department, University of Cape Town, Cape Town, South Africa
\item \Idef{org90}Physics Department, University of Jammu, Jammu, India
\item \Idef{org91}Physics Department, University of Rajasthan, Jaipur, India
\item \Idef{org92}Physik Department, Technische Universit\"{a}t M\"{u}nchen, Munich, Germany
\item \Idef{org93}Physikalisches Institut, Ruprecht-Karls-Universit\"{a}t Heidelberg, Heidelberg, Germany
\item \Idef{org94}Politecnico di Torino, Turin, Italy
\item \Idef{org95}Purdue University, West Lafayette, Indiana, United States
\item \Idef{org96}Pusan National University, Pusan, South Korea
\item \Idef{org97}Research Division and ExtreMe Matter Institute EMMI, GSI Helmholtzzentrum f\"ur Schwerionenforschung, Darmstadt, Germany
\item \Idef{org98}Rudjer Bo\v{s}kovi\'{c} Institute, Zagreb, Croatia
\item \Idef{org99}Russian Federal Nuclear Center (VNIIEF), Sarov, Russia
\item \Idef{org100}Russian Research Centre Kurchatov Institute, Moscow, Russia
\item \Idef{org101}Saha Institute of Nuclear Physics, Kolkata, India
\item \Idef{org102}School of Physics and Astronomy, University of Birmingham, Birmingham, United Kingdom
\item \Idef{org103}Secci\'{o}n F\'{\i}sica, Departamento de Ciencias, Pontificia Universidad Cat\'{o}lica del Per\'{u}, Lima, Peru
\item \Idef{org104}Sezione INFN, Bari, Italy
\item \Idef{org105}Sezione INFN, Bologna, Italy
\item \Idef{org106}Sezione INFN, Cagliari, Italy
\item \Idef{org107}Sezione INFN, Catania, Italy
\item \Idef{org108}Sezione INFN, Padova, Italy
\item \Idef{org109}Sezione INFN, Rome, Italy
\item \Idef{org110}Sezione INFN, Trieste, Italy
\item \Idef{org111}Sezione INFN, Turin, Italy
\item \Idef{org112}SSC IHEP of NRC Kurchatov institute, Protvino, Russia
\item \Idef{org113}SUBATECH, Ecole des Mines de Nantes, Universit\'{e} de Nantes, CNRS-IN2P3, Nantes, France
\item \Idef{org114}Suranaree University of Technology, Nakhon Ratchasima, Thailand
\item \Idef{org115}Technical University of Ko\v{s}ice, Ko\v{s}ice, Slovakia
\item \Idef{org116}Technical University of Split FESB, Split, Croatia
\item \Idef{org117}The Henryk Niewodniczanski Institute of Nuclear Physics, Polish Academy of Sciences, Cracow, Poland
\item \Idef{org118}The University of Texas at Austin, Physics Department, Austin, Texas, USA
\item \Idef{org119}Universidad Aut\'{o}noma de Sinaloa, Culiac\'{a}n, Mexico
\item \Idef{org120}Universidade de S\~{a}o Paulo (USP), S\~{a}o Paulo, Brazil
\item \Idef{org121}Universidade Estadual de Campinas (UNICAMP), Campinas, Brazil
\item \Idef{org122}University of Houston, Houston, Texas, United States
\item \Idef{org123}University of Jyv\"{a}skyl\"{a}, Jyv\"{a}skyl\"{a}, Finland
\item \Idef{org124}University of Liverpool, Liverpool, United Kingdom
\item \Idef{org125}University of Tennessee, Knoxville, Tennessee, United States
\item \Idef{org126}University of the Witwatersrand, Johannesburg, South Africa
\item \Idef{org127}University of Tokyo, Tokyo, Japan
\item \Idef{org128}University of Tsukuba, Tsukuba, Japan
\item \Idef{org129}University of Zagreb, Zagreb, Croatia
\item \Idef{org130}Universit\'{e} de Lyon, Universit\'{e} Lyon 1, CNRS/IN2P3, IPN-Lyon, Villeurbanne, France
\item \Idef{org131}V.~Fock Institute for Physics, St. Petersburg State University, St. Petersburg, Russia
\item \Idef{org132}Variable Energy Cyclotron Centre, Kolkata, India
\item \Idef{org133}Vin\v{c}a Institute of Nuclear Sciences, Belgrade, Serbia
\item \Idef{org134}Warsaw University of Technology, Warsaw, Poland
\item \Idef{org135}Wayne State University, Detroit, Michigan, United States
\item \Idef{org136}Wigner Research Centre for Physics, Hungarian Academy of Sciences, Budapest, Hungary
\item \Idef{org137}Yale University, New Haven, Connecticut, United States
\item \Idef{org138}Yonsei University, Seoul, South Korea
\item \Idef{org139}Zentrum f\"{u}r Technologietransfer und Telekommunikation (ZTT), Fachhochschule Worms, Worms, Germany
\end{Authlist}
\endgroup

\end{document}